\documentstyle[12pt]{article}

\begin{document}
{}
\vskip.5in
\centerline{\Large {\bf Renormalization of Hamiltonians }}
\vskip.1in
\centerline{\Large {\bf in the Light-Front Fock Space}}
\vskip .2in
\centerline{ June, 1997}

\vskip .3in
\centerline{Stanis{\l}aw D. G{\l}azek}
\vskip .1in
\centerline{Institute of Theoretical Physics, Warsaw University}
\centerline{ul. Ho{\.z}a 69, 00-681 Warsaw}

\vskip.3in
\centerline{\bf Abstract}
\vskip.1in

We outline an ultraviolet renormalization procedure for hamiltonians 
acting in the light-front Fock space. The hamiltonians are defined 
and calculated using creation and annihilation operators with
no limitation of the space of states.  Both, the regularization of 
the initial hamiltonian and the definition of the renormalized 
effective hamiltonians, preserve the light-front frame kinematical 
symmetries. The general equations for the effective hamiltonians are 
illustrated by second order calculations of the self-energy and 
two-particle interaction terms in Yukawa theory, QED and QCD. 
Infrared singularities are regulated but not renormalized.


\vskip1.5in

\newpage

{\bf 1. INTRODUCTION}
\vskip.1in

This paper describes elements of a renormalization theory for
hamiltonians.  The two main elements are the similarity renormalization
group for deriving the hamiltonians and the Schr\"odinger equation for
their eigenstates.  The hamiltonians are defined in the Fock space.

The Fock space is constructed using light-front creation operators
acting on the vacuum state.  The vacuum is annihilated by the
corresponding annihilation operators.  We use this representation of
states because, due to the boost symmetry, it is easier than the usual
equal-time representation to work with in a relativistic theory.

The similarity renormalization group is defined in terms of effective
creation and annihilation operators which appear in the effective
hamiltonians. The effective hamiltonians can be calculated using the
method of successive approximations and perturbation theory.  Eventually,
one obtains a large number of terms and one must study their contribution to
the effective Schr\"odinger equation.  The space of solutions of the
Schr\"odinger equation includes bound states.

The goal of this paper is to present the general framework.  In
addition, the general presentation is illustrated by second order
perturbative calculations of effective hamiltonians in Yukawa theory,
QED and QCD.  The effective hamiltonians lead to the on-shell scattering
matrices which agree with standard results.  The bound state eigenvalue
problems for the effective hamiltonians are only briefly described
because little is known about them beyond the nonrelativistic limit.
This article is meant to help in facilitating further studies by
providing a comprehensive introduction and showing examples of a
systematic derivation of the effective hamiltonians.  From this point of
view, details of the approach are less important than the general
framework and they are forseen to change in future.

One reason for renormalization to play an important role in the
hamiltonian approach is divergences.  Initial expressions for bare
hamiltonians are divergent because they are obtained from formal
expressions in local field theory.  Nevertheless, such hamiltonians are
known to closely reproduce experimental data through tree diagrams.  The
most prominent example is the canonical hamiltonian of QED.
Renormalization problems appear only when one is forced to sum over
intermediate states and the sum diverges.  The divergences correspond to
diverging loop integrals in Feynman diagrams.  However, the hamiltonian
approach greately differs from the lagrangian diagrammatic approaches.

One apparent difference is that the hamiltonian sums over intermediate
states involve integrals over the three-dimensional momentum space while
the integrals in the lagrangian calculus are four-dimensional.  Although
a connection exists for finite integrals which are not sensitive to
cutoffs, when the integrals diverge the connection is broken.  In the
hamiltonian approach, we introduce the effective Fock space basis and we
construct the hamiltonians using the three-dimensional regularization
and renormalization.  In the lagrangian approach, one directly
calculates Green functions using four-dimensional regularization and
renormalization techniques.  Equivalence of the two approaches in
diverging cases remains to be shown, especially when the bound states
are taken into account.

Besides removing divergences, the renormalization group is useful in the
hamiltonian approach because it introduces a hierarchy of scales.
Phenomena of different scales are dealt with in a certain order.  This
enables us to solve problems involving many scales.  Particle theories
contain many, possibly even infinitely many different scales.  Using the
renormalization group approach, we can start from the hamiltonian of a
basic theory that couples all degrees of freedom of all scales and we
can reduce the initial hamiltonian to a renormalized hamiltonian in
which couplings between degrees of freedom of vastly different scales,
vanish.  Eventually, the goal is to obtain an effective hamiltonian for
such a small range of scales that the spectrum of solutions can be found
in practice.

In the hamiltonian renormalization group approach, one introduces
different scales of momentum.  We define the momentum scales for
particles using their relative momenta.  For example, consider the case
where some dressing mechanisms for particles are expected to work at
different scales.  One can think of the constituent quarks as built from
the smaller current quarks and gluons.  The relative momenta of the
latter determine the resolution scale which is required to see them.
For addressing such issues, the renormalization group evolution in the
relative momentum scales is a natural choice.  But, so far, it has not
been exactly understood in a hamiltonian approach how the constituent
structure appears in QCD.

A number of model subspaces of the Fock space need to be
considered when solving for the spectrum of a field-theoretic effective
hamiltonian because the full space of states is too large for
computations.  Also, different physical problems require different model
subspaces.  Nevertheless, working within a subspace of interest, one
should secure that the results for physical quantities are independent
of the renormalization group cutoff which limits the relative energy
transfers. The cutoff independence can appear only in a certain range of
cutoffs which corresponds to the model subspace.  However, once
the cutoff independence in the finite range is achieved, one expects to
have solved the theory in this range.

The following diagram will illustrate the situation.
\vskip.2in
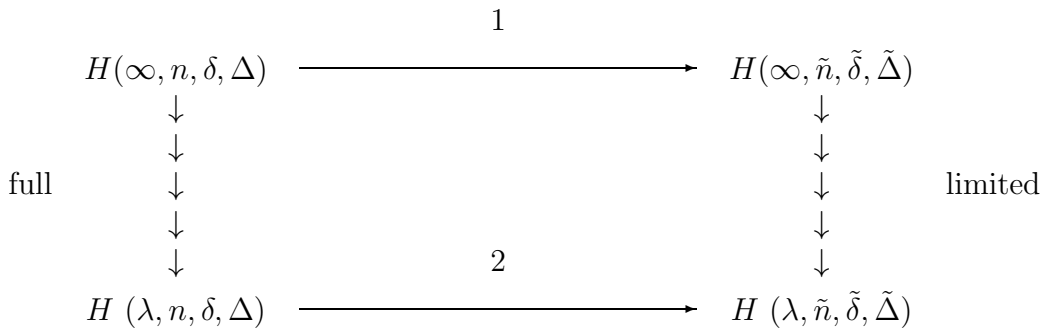
\begin{tabular}{rcccl}
&                 &            1           &                                &\\
& $H(\infty,n,\delta,\Delta)$ & $^{\vector(3,0){150}}$ & $H(\infty,\tilde n,\tilde \delta,\tilde \Delta)$ &\\
& $\downarrow$    &                        & $\downarrow$                   &\\
& $\downarrow$    &                        & $\downarrow$                   &\\
full & $\downarrow$&                       & $\downarrow$ & limited          \\
& $\downarrow$    &                        & $\downarrow$                   &\\
& $\downarrow$    &            2           & $\downarrow$                   &\\
&$H~(\lambda,n,\delta,\Delta)$& $^{\vector(3,0){150}}$ & $H~(\lambda,\tilde n,\tilde \delta, \tilde \Delta)$&
\end{tabular}
\vskip.3in

In this diagram, the vertical arrows indicate evolution in the
renormalization group parameter $\lambda$.  $\lambda$ limits the
relative energy transfers. It will be explained in detail in next
Sections how the limits are imposed. $\lambda$ ranges, in principle, 
from infinity to
zero.  The hamiltonians depend on additional parameters $n$, $\delta$
and $\Delta$.  This notation is used here for the purpose of the
introductory discussion and it will be partly altered in the next
Sections.

$n$ stands for the cutoff on the {\it change} of the particle number.
It defines the limits on the numbers of creation and annihilation
operators that can appear in a single hamiltonian term.  For example,
the canonical expressions for light-front hamiltonians in local field
theories of physical interest limit the number of creation and
annihilation operators in a single term to 4, and the particle number
can change at most by $n = 2$. One can set similar or less stringent
restrictions on the effective hamiltonians with finite width $\lambda$.

$\delta$ stands for the infrared cutoff.  For example, it may define the
lower bound on the longitudinal momentum carried by particles that can
appear or disappear as a result of a single interaction.

$\Delta$ stands for the ultraviolet cutoff which defines the upper
limit on the relative transverse momentum of particles which can appear
or disappear in a single interaction.

The left branch of the diagram is marked ``full'' because it corresponds
to the renormalization group evolution in the full Fock space.  The
parameters $n$, $\delta$ and $\Delta$ do not limit the particle number
in the space of states.

Counterterms in the initial $H(\infty,n,\delta,\Delta)$ are constructed
to make the physical results have well defined limits when the cutoffs
$n$, $\delta$ and $\Delta$ are relaxed.  The construction of
counterterms in perturbation theory will be discussed in detail in the
next Sections.  Once the counterterms remove the regularization
dependence from the effective dynamics the arguments $n$, $\delta$ and
$\Delta$ in $H(\lambda,n,\delta,\Delta)$ in the lower left corner of the
diagram are equivalent to their limiting values, $n = \infty$, $\delta =
0$ and $\Delta = \infty$.  Then, $H(\lambda,n,\delta,\Delta) \equiv
H(\lambda)$ which is independent of $n$, $\delta$ and $\Delta$ for their
extreme values.  The infrared regulator $\delta$ may still appear in the
effective hamiltonian if there are massless vector particles in the
theory.

The horizontal arrow marked 1, denotes simplifications which can be made
in the initial hamiltonian if some of its matrix elements are small
and/or one can drop certain couplings without losing accuracy in
representing solutions of the theory.  The model parameters $\tilde n$,
$\tilde \delta$ and $\tilde \Delta$ are introduced to limit dynamics to
the selected part.  For example, $\tilde n$ can limit the number of bare
particles in the initial model space.  Also, $\tilde \delta$ can limit
the bare particle momenta from below when it is known that only momenta
in a certain range matter.  Similarly, $\tilde \Delta$ can limit from
above the free energies of bare particle states which are taken into
account in a calculation when it is known that some range of the free
energies dominates the dynamics and the introduction of the upper limit
does not lead to significant errors.

Accuracy of the step denoted by the arrow 1 has to be checked by
relaxing the model cutoffs and measuring the resulting changes in the
spectrum of $H(\lambda,\tilde n, \tilde \delta,\tilde \Delta)$.
Naturally, these cutoffs may have to be varied in a big range because
they are introduced along the arrow 1 for bare particles.

The right branch of the figure is marked ``limited'' because it
describes the renormalization group flow in the limited bare model
space selected by parameters $\tilde n$, $\tilde \delta$ and $\tilde \Delta$.

The effective hamiltonians at the bottom of the diagram, namely,
$H(\lambda,n,\delta,\Delta)$ and $H(\lambda,\tilde n,\tilde \delta,
\tilde \Delta)$ with energy transfers limited by finite $\lambda$, are
connected by the arrow marked 2. This arrow denotes the procedure of
introducing helpful cutoffs which enable us to approximately solve for
the spectrum of the effective hamiltonian $H(\lambda,n,\delta,\Delta)$.
This time, however, the final computation cutoffs are introduced at the
level of the effective particles, not at the level of the initial bare
particles.

The arrow 2 denotes the replacement of the whole effective hamiltonian
matrix by a limited matrix which is obtained from the full matrix by
introducing the model cutoffs $\tilde n$, $\tilde \delta$ and $\tilde
\Delta$.  The procedure of obtaining the small matrix will be discussed
below.  The spectrum of the small matrix may be very close to the
corresponding part of the spectrum of the full matrix because $\lambda$
is small (this will become clear later).  The accuracy must be verified
by relaxing the cutoffs $\tilde n$, $\tilde \delta$ and $\tilde \Delta$
and observing convergence of results as in the case of the arrow 1.
However, again, it is natural to expect that the cutoffs $\tilde n$,
$\tilde \delta$ and $\tilde \Delta$ may have to be varied only in a
small range which corresponds to $\lambda$.  Thus, a finite dynamical
problem to solve is defined.

The ``full'' renormalization group evolution can be calculated using the
approach of the present paper.  The ``limited'' evolution can be
calculated using the matrix element techniques introduced earlier by
G{\l}azek and Wilson \cite{GW1} \cite{GW2} who drew on the work of
Wilson \cite{W1} \cite{W2}.  The matrix elements techniques were
introduced for application to QCD.  \cite{W3} Alternatively, one can use
Wegner's flow equations for hamiltonian matrix elements in cases soluble
with the energy-independent width.  \cite{WEG} The present approach can
be viewed as a special case of the earlier procedures because the
hamiltonians which we consider transform by the same unitary
transformation as a single creation or annihilation operator.  However,
by having defined the renormalization group transformation for creation
and annihiltion operators, we remove the need to consider the model
space dependence of the renormalization procedure.

On the other hand, to make the transition to the bound state degrees of
freedom, such as mesons and barions built from quarks and gluons, the
matrix elements techniques are unavoidable.  The transition from
constituents to their bound states in the renormalization group flow,
although achievable in principle, is not discussed in the present paper.
In the present approach, quantum numbers of effective particles such as
spin, flavor or color, cannot change in the flow.

Both ways in the diagram which start from the initial hamiltonian
$H(\infty,n,\delta,\Delta)$ and go through (a) the arrow 1 and the
arrows ``limited'' and (b) the arrows ``full'' and the arrow 2, lead to
the numerically manageable $H(\lambda,\tilde n,\tilde \delta, \tilde
\Delta)$.  When calculations of some matrix elements of the effective
hamiltonian are done in perturbation theory by expanding in powers of a
coupling constant at a selected scale $\lambda$ up to a limited order in
the expansion, both ways of going through the diagram are, in principle,
equivalent, because only a finite range of particle numbers and momenta
can be reached in a limited number of steps of the size $\lambda$
starting from the finite values selected by the states used in
evaluation of the matrix elements in question.

Differences arise only when one solves for the spectrum of an effective
hamiltonian and when one attempts to vary the model space parameters
$\tilde n$, $\tilde \delta$ and $\tilde \Delta$.  In the ``full''
calculation, one obtains a single effective hamiltonian which one can
solve in successively enlargeable model spaces.  In the ``limited''
calculation, the model space restrictions are imposed at the beginning
and they lead to an effective hamiltonian whose action cannot be
considered in a larger model space without repeating the renormalization
group calculation in the larger space.

A priori, it is not clear how important different effective interaction
terms are.  Proportionality to different powers of the coupling constant
is certainly helpful in qualitative estimates.  But it is not known
precisely how to choose the basis states for evaluation of the effective
hamiltonian matrix elements.  Finding the basis which can span a good
approximation to the full solution will certainly
require extensive trial and error
studies.  This feature can be illustrated in an elementary example of
the following $2 \times 2$ matrix.
\begin{center}
$$ \left[
\begin{array}{ccc} a + b g^2 & g v \\ g v & c + d g^2 \end{array}
\right] $$
\end{center}
\noindent This matrix is a model of a full
effective hamiltonian calculated to second order in $g$ including all
couplings between all effective Fock sectors as given by the ``full''
calculation.  Thus, we have the hamiltonian terms order 1, order $g$ and
order $g^2$.  In a perturbative calculation using matrix elements, which
is focused on the upper sectors, one would calculate only the terms $a$
and $b g^2$.

Assume that $a$ and $c$ are of the same order, and $b$ and $d$ are of
the same order.  For arbitrarily small $g$, the eigenvalues are given by
$a$ and terms quadratic in $g$.  No terms linear in $g$ arise.  The
quadratic corrections include contributions due to the term $g v$ which
couples different sectors.  The r\^ole of this coupling needs to be
estimated.  The presence of $d g^2 $ seems to be irrelevant because it
couples to the upper sector through the off-diagonal terms order $g$.
Hence, it seems to contribute only in order $g^4$ to the eigenvalues.

It is well known that the above analysis is wrong in the case with
degenerate diagonal matrix elements.  It does not matter how the
degeneracy arises.  For example, consider the case of a finite $g$ such
that $ a + b g^2 = c + d g^2 $. The eigenvalues are equal $ a + b g^2
\pm gv$.  They are linear in $g$ instead of being quadratic, for
arbitrary $v$.  The lowest eigenvalue eigenstate is a superposition of
the upper and the lower sector instead of being dominated by the upper
one.  In this example, the degeneracy is not visible until the term $d
g^2$ is included in the calculation.  As a second example consider the
case with degenerate matrix elements $ c = a + b g^2 $ and $d$ not
included.  The simple nondegenerate perturbative expansion is again not
applicable.  But the addition of the term $d g^2$ can lift the
degeneracy and make the simple perturbative expansion work.

Corrections originating from the interaction terms such as $g v$ may be
suppressed in the power expansion in the coupling constant for small
$\lambda$ despite the degeneracy.  The argument is following.  The range
of $v$ in momentum variables is given by $\lambda$.  If $\lambda$ is
reduced in the renormalization group flow down to a number which implies
restriction on the energy transfers on the order of some positive power
of $g$ then, the resulting interaction can contribute to the eigenvalues
only in the order implied by $g$ and $\lambda$ together which is higher
than $g^2$.  In addition, the effective interaction $v$ may contain
small factors.  For example, in the effective $e^+e^-$ sector of
positronium, the emission of photons is proportional to the velocity of
electrons which is order $\alpha$, on average.  The interaction term $g
v$ which couples states with an additional photon, plays no role in the
eigenvalue in order $\alpha^2$ if the width $\lambda$ restricts energy
changes to order $\alpha^2 \, m_{electron}$ (c.f.  \cite{JPG}).

Terms such as $d g^2$ have been discussed in the light-front approach
to QCD.  \cite{Perry} In heavy quarkonia which are dominated by the $Q
\bar Q$ state, terms such as $d g^2$ in other Fock sectors than the $Q
\bar Q$, may be expected to lift up energies of effective gluons to a
sufficiently high value so that the effective hamiltonian part $a + b
g^2 $ in the $Q\bar Q$ sector alone may have eigenstates which are a
good approximation of the full solution for the heavy mesons.

But it is not known yet what happens in the eigenvalue equations of
hamiltonians calculated to higher order than second.  Unfortunately, the
space of states with fast moving fermions and bosons is highly
degenerate.  A large number of matrix elements must be calculated and
examined.  One needs to choose the states required for evaluation of the
hamiltonian matrix elements before it is known which of those states or
matrix elements are important.  Since the states are built using
creation and annihilation operators, the operator calculus for
hamiltonians provides a relatively small number of operators which
universally produce the very many matrix elements in question, such as
$g v$, $b g^2$ and $d g^2$ and similar higher order terms.  In
particular, the operator calculus becomes essential when the number of
states involved in the renormalization group flow is large.

In general, when one attempts to perform a multistep renormalization
group evolution which is indicated by the small vertical arrows, at each
step using an expansion in powers of the corresponding coupling constants, 
the degree of dressing of bare particles grows with the number of steps.
Eventually, the number of steps can approach infinity when the ratio of
a single step in the cutoff reduction to the initial cutoff tends to
zero.  In this case, it is not clear that both branches (a) and (b) in
the diagram above lead to the same hamiltonian $H(\lambda,\tilde
n,\tilde \delta, \tilde \Delta)$ for effective particles.

The reason is that the number of bare particles in the initial model
space may have to be enormously enlarged to support the effective
particle picture in the small final model space.  The same concerns the
momentum cutoffs.  The required enlargement of the initial space of
states may grow with the number of steps in the renormalization group
flow and even become infinite in the vicinity of a fixed point.  Then,
proceeding along the branch (a) by introducing the model limitations
$\tilde n$, $\tilde \delta$ and $\tilde \Delta$ at the beginning of the
renormalization group evolution (arrow 1) and working with matrix
elements of the hamiltonians in the limited space (arrows ``limited''),
may give different results from those that come out from the calculation
along the branch (b).  In the latter one, the model space cutoffs
$\tilde n$, $\tilde \delta$ and $\tilde \Delta$ are introduced only in
the final small energy transfer hamiltonian (arrow 2) when solving for
its spectrum, with a numerically verifiable accuracy.

An explicit example of a difference between proceeding along the two
branches of the diagram above is provided in Ref.  \cite{TD} which
discusses an approach analogous to the branch 1 and ``limited'' (c.f.
Ref.  \cite{HO}).  In the Tamm-Dancoff (TD) approach, there are
restrictions on the particle number which naturally lead to the
sector-dependent counterterms as described in Ref.  \cite{TD}, for
example, for masses.  On the other hand, in the procedure of the branch
``full'' and 2 no such sector dependent counterterms arise.  The present
paper will provide examples of sector-independent mass counterterms.

The above diagram and the $2 \times 2$ matrix model illustrate the
general structure of our light-front hamiltonian approach to quantum
field theory.  We summarize the steps here.

The first step is the calculation of the effective hamiltonian,

$$   H(\lambda) \quad = \quad S^\dagger_{\lambda,n,\delta,\Delta}
     \,\, H(\infty,n,\delta,\Delta) \,\, S_{\lambda,n,\delta,\Delta}
     \,\, . \eqno(1.1)$$

\noindent $S$ denotes the similarity transformation.  Eq.  (1.1)
corresponds to the arrows marked full in the diagram.  This equation
represents the ideal situation for massive particles where the
ultraviolet renormalization removes dependence of the effective
hamiltonian matrix elements on $n$, $\delta$ and $\Delta$ when the
regularization is being removed.  The similarity renormalization scheme
is presented in the next Sections.

The second step is the solution of the effective Schr\"odinger equation

$$   H(\lambda) \,\, |\psi\rangle \quad = \quad E \,\, |\psi\rangle
     \,\, . \eqno(1.2) $$

\noindent $H(\lambda)$ has the same dynamical content and eigenvalues as
$H(\infty,n,\delta,\Delta)$ and $E$ is independent of $\lambda$.  But
Eq.  (1.2) greately differs from the eigenvalue equation for
$H(\infty,n,\delta,\Delta)$.  The major difference is that the dynamics
of $H(\lambda)$ has a limited range on the energy scale.  Therefore, one
can hope it is possible to solve the eigenvalue problem scale by scale.
Scattering processes are described by the same hamiltonian.  Next
Sections will give 2nd order examples.

Solutions to Eq.  (1.2) provide renormalization conditions for the
finite parts of the counterterms.  When one has to go beyond
perturbation theory in the solution, as it is the case in the bound
state dynamics, a general method is necessary for reducing the whole
problem to a manageable one.  This step is marked by the arrow 2 in the
diagram.  In the case of the matrix model, this step corresponds to the
calculation of the model space hamiltonian in the upper-left corner of
the matrix.  The similarity renormalization scheme guarantees that this
step is free from ultraviolet divergences because the width $\lambda$ is
finite.

In the general case, one can apply the well known Bloch \cite{Bloch}
technique of calculating the model space hamiltonians (see Ref.
\cite{W2} for the application of this technique in the usual
renormalization group approach).  Suppose we want to evaluate the model
two-body hamiltonian using $H(\lambda)$ with $\lambda < m$, where $m$ is
the one-body mass.  We can introduce the projection operator $P$ on the
effective two-particle sector with a limited center-of-mass energy.  
We also introduce the operator $R$ which generates 
the multi-particle and high energy components of eigenstates from their 
limited mass two-body part.  By assumption, $R$ satisfies the
conditions $(1-P)R = RP = R$ and $PR = R(1-P) = 0$ and the equation $P +
R - 1) \, H(\lambda) \, (P + R) = 0$.  Then, the model two-body dynamics
is described by the hamiltonian

$$  H_2 \quad = \quad (P + R^\dagger R)^{-1/2}
\,\, (P+R^\dagger) \,\, H(\lambda) \,\, (P + R) \,\,
     (P + R^\dagger R)^{-1/2} \,\, . \eqno(1.3) $$

\noindent The same approach can be used for larger model spaces.  The
model space is characterized by the parameters $\tilde n$, $\tilde
\delta$ and $\tilde \Delta$ in the diagram.  So, the operation $R$
depends on these parameters.  But the resulting spectrum in the range of
interest should not depend on the model space boundary once it encloses
the dominant basis states in the selected range of scales.

The heuristic Eq.  (1.3) can be applied in perturbation theory even for
sizable coupling constants since the interaction strength is
considerably reduced by the similarity factors.  These factors limit the
interaction range in the relative momentum space of the interacting
particles to be of the order of $\lambda$.  Thus, the perturbative
expansion relies on the form factor suppression of the interaction
vertices which change the particle number.  In principle, no
$\lambda$-dependence is generated in the model dynamics because the form
factors are generated through the transformation $S$.  This feature will
be illustarted in the lowest order examples in Section 3. Note, that
the operation $R$ is ultraviolet finite.

The general scheme outlined above is still prone to the infrared
regularization dependence which may appear in Eq.  (1.3), in particular,
in gauge theories. Extensive studies are required to find out if
the model dynamics for gauge invariant quantities is infrared
convergent.  One may discover new efective interactions coming from 
the infrared region. \cite{W3} \cite{SUS} \cite{GSR}

Construction of a general hamiltonian approach to particle dynamics by
necessity touches upon fundamental issues.  Therefore, these issues are
discussed in this paper as they appear.  Some issues, however,
especially those related to the ground state formation, are not
discussed extensively because such discussion would have to include
description of states containing very large numbers of bare particles.
Dynamics of such states is not understood.  For example,
it is not known if going through the branch ``full'' of the
diagram above one can obtain interactions which couple few-particle
states with practically infinitely many wee-particle states.  Similarly,
issues concerning gauge invariance and rotational symmetry are not
extensively dicussed because they require better understanding.
For example, current conservation leads to important
cancelations in the on-energy-shell hamiltonian matrix elements.  Such
cancelations are absent in the off-energy-shell hamiltonian matrix
elements.  The on-energy-shell hamiltonian matrix elements exhibit
rotational invariance which is absent in the off-energy-shell
hamiltonian matrix elements.

Granted all the reservations, the calculus we develop in this paper is,
in a way, similar to the transformation discussed by Melosh
\cite{Melosh}.  The important differences are that we provide a
renormalized dynamical theory applicable to particles of different
kinds.  If one restricts attention to QCD, the calculus is
expected to help in establishing a connection between the constituent
quark model, Feynman parton model, and perturbative quantum
chromodynamics.

This paper is organized as follows.  Section 2 describes the general
formalism.  Section 3 contains examples of lowest order applications.
Section 4 concludes the paper.  The list of references is focused on
the similarity renormalization group approach to light-front hamiltonian 
dynamics in the Fock space. The reader should be aware of this limtation.

\vskip.3in
{\bf 2. EFFECTIVE HAMILTONIANS}
\vskip.1in

This Section is divided into three parts.  The first part describes a
general method of calculating effective hamiltonians in the Fock space.
The second part presents our regularization scheme for bare
hamiltonians.  The last part discusses renormalization conditions and
the effective eigenvalue problem.

\vskip.3in
{\bf 2.a Similarity transformation}
\vskip.1in

We construct a family of effective hamiltonians in the light-front Fock
space.  The family is parametrized by a scale parameter $\lambda$ which
ranges from infinity to a finite value.  $\lambda$ limits energy
transfers in the interaction terms.

The hamiltonians are written in terms of sums of ordered products of
creation and annihilation operators.  The hamiltonian labeled by
$\lambda$ is expressed in terms of creation and annihilation operators
which correspond to $\lambda$.  We commonly denote these operators by
$q_\lambda$.  In addition, the creation and annihilation operators carry
labels of quantum numbers such as momentum, spin, flavor or color.  We
will not indicate those numbers in the initial presentation, unless it
is necessary.

All hamiltonians in the family are assumed to be equal. Thus,

$$        H_{\lambda_1} (q_{\lambda_1}) = H_{\lambda_2} (q_{\lambda_2}).
\eqno (2.1) $$

For $\lambda = \infty$, the hamiltonians $H_\infty$ are expressed in
terms of operators creating and annihilating bare particles, $q_\infty$.
Hamiltonians $H_\infty$ can be constructed from the canonical
field theoretic expressions for the energy-momentum density tensors.

Unfortunately, expressions for $H_\infty$ in local field theories are
divergent.  They need to be regularized by introducing a bare
ultraviolet cutoff which we shall denote by $\epsilon$.  The ultraviolet
cutoff $\Delta$ from the previous Section corresponds to
$\Lambda^2/\epsilon$ where $\Lambda$ is an arbitrary finite constant
which carries the necessary dimension of a mass.  The limit of removing
the bare ultraviolet cutoff will correspond to $\epsilon \rightarrow 0$.

$H_{\lambda= \infty} = H_\epsilon$ for all values of $\epsilon$.  For
the limit $\epsilon \rightarrow 0$ to exist the hamiltonians $H_\infty$
must include a number of additional terms (called counterterms) whose
structure will be determined later.

$H_\infty$ may have to include an infrared regulator, generically
denoted by $\delta$.  For example, this is required in QED with massless
photons and in QCD with massless gluons.  $\delta \rightarrow 0$ when
the infrared regularization is removed.  The parameter $\delta$ is
indicated explicitly only if needed.

Our key assumption is that the particle degrees of freedom for all
different scales $\lambda$ are unitarily equivalent to the bare particle
degrees of freedom:

$$   q_{\lambda} = U_\lambda q_\infty U^\dagger_\lambda \, .\eqno (2.2) $$

\noindent This assumption says that the quantum numbers of bare and
effective particles are the same for all values of $\lambda$.  A few
examples will illustrate the physical origin of this assumption.  (1)
Constituent quarks have the same quantum numbers as current quarks.  (2)
We use the same quantum numbers for photons and electrons independently
of the kind of processes we consider in QED or in related effective
theories such as the nonrelativistic Schr\"odinger equation with Coulomb
potentials between charges.  (3) Pions and nucleons in nuclear physics
have the same quantum numbers quite independently of what kind of
models, pion-nucleon vertex form factors or dynamical assumptions one
uses.

It follows from Eq.  (2.2) that creation and annihilation operators for
$\lambda_1$ and $\lambda_2$ are unitarily equivalent and connected by
transformations of the form $U_{\lambda_1} U^\dagger_{\lambda_2}$.  The
transformations $U_{\lambda_1}$ or $ U_{\lambda_2}$ will depend on the
bare cutoffs but the transformation $U_{\lambda_1}
U^\dagger_{\lambda_2}$ for finite $\lambda_1$ and $\lambda_2$ will have
to be finite in the limit $\epsilon \rightarrow 0$.  On the other hand,
these transformations may depend on the infrared cutoff in the case of
massless particles, unless there is a mechanism removing this
dependence.

We shall define transformations $U_{\lambda}$ using equations of the
similarity renormalization scheme of G{\l}azek and Wilson \cite{GW1}
\cite{GW2}.  That scheme was originally developed for application to QCD
\cite{W3}.

The transformation $U_\lambda$ is defined indirectly through a
differential equation for effective hamiltonians.  Hamiltonians with
labels $\lambda_1$ and $\lambda_2$ are connected by integration of
the differential equation from $\lambda_1$ to $\lambda_2$.  Our guiding
principle in writing the differential equation is that {\it the
transitions between states of effective particles with considerably
different relative momenta should be suppressed}.

A few examples of effective theories illustrate why we adopt this
principle.  (1) Emission or absorption of short wavelength photons are
not essential in the formation of atoms.  (2) Emission of hard pions
from a nucleon is not important in nonrelativistic nuclear physics.  (3)
Constituent quarks have moderate momenta and their effective dynamics
seems to be independent of the very hard gluon emissions.  Similarly,
(4) the high momentum transfer phenomena are independent of the small
momentum transfer effects such as binding.  A standard way of achieving
this kind of picture in theoretical models is to include form factors in
the interaction vertices.  The form factors quickly tend to zero when
momenta change by more then the size of a specific cutoff parameter.

An alternative mechanism for decoupling of different momentum scales is
the quick rise of the free energy as a function of the momentum and,
effectively, appearance of the energy gap between different momentum
scales.  For example, this second mechanism works in atomic physics
where the relativistic expression for a free electron energy is replaced
by the nonrelativistic one.  The linear dependence of the energy on a
large momentum is replaced by a quadratic one.  The quadratic rise of
the energy suppresses contributions of high energy intermediate states.
This suppression stabilizes the nonrelativistic bound state theory with
a small coupling constant in the small relative momentum range.  In a
relativistic case, we have to include form factors to suppress
transitions between different momentum scales.

The cutoff parameter in the form factors sets the scale for allowed
changes of momenta.  It determines the range or width of the interaction
in momentum space.  Precisely this notion of the interaction width in
momentum space is the origin of our scale $\lambda$ which labels
renormalized effective hamiltonians.  The effective hamiltonians will
contain factors that are analogous to the vertex form factors which are
commonly used in nonlocal models (see also Ref.  \cite {BGP}).

The transformation $U_\lambda$ transforms the regularized hamiltonian
$H_\infty$ of a local theory into a nonlocal effective hamiltonian
$H_\lambda$ in which the large momentum transfer dynamics is
integrated out.

The momenta of individual particles are not restricted.  This is
required to obtain a boost invariant spectrum of solutions.  The reason
is that boosts change the momenta unlimitedly.  Restricting the
individual particle momenta would exclude boost invariance beyond the
width scale. This would be a problem because the width needs to be 
small for the hamiltonian eigenvalue equation to be soluble in practice.

Also, the larger is a relative momentum the larger change is generated
by a boost.  Therefore, when the free energy of interacting particles in
their center-of-mass frame (i.e. the free light-front invariant mass) is
larger than $\lambda$ then, the immediate change of energy due to the
interactions will be limited by the large energy itself instead of
$\lambda$.  This construction also reflects the property of quantum
systems that strong interference occurs between waves of similar
wavelengths within a range of wavelengths on the order of 
the wavelengths themselves. \cite{W1}

The infinitesimal transformations are constructed in such a way that
small energy denominators cannot appear in perturbative calculations of
the effective hamiltonians.\cite{GW1} \cite{GW2} Only large energy
changes are integrated out.  Therefore, the dynamical interference
effects for states of similar free energies are not included in the
derivation of an effective hamiltonian.  The calculation of the
strong coherence effects is deferred to the later step of solving for the
eigenstates of the effective hamiltonian.  This second step may be
nonperturbative.  For example, the Coulomb potential of QED is formally
of the first order in $\alpha$ but it leads to an overwhelmingly rich atomic
structures beyond perturbation theory.

Our differential equations will require a separation of the changes in
the creation and annihilation operators from the changes in the
coefficients in front of their products.  It is also initially assumed
that terms with large numbers of operators in a product will not
dominate or mediate the dynamics of interest. If the latter assumption
turns out to be invalid then, the present formalism may only provide a
way to approach the resulting problems, but that case is not in the
focus of the present paper.  The only comment due here is that, if the
ground state dynamics leads to spontaneous symmetry breaking, or
condensates, in a well defined renormalized hamiltonian theory then, we
will have a tool to study those phenomena theoretically in great detail,
c.f. Refs. \cite{W3}, \cite{SUS} and \cite{GSR}.
Nothing more can be said about it at this point.  All hamiltonians we
consider can, by assumption, be very well approximated by a finite sum
of finite products of creation and annihilation operators.

The unitary equivalence of the creation and annihilation operators for
the scale $\lambda$ and the creation and annihilation operators at the
infinite scale, i.e. those appearing in $H_\infty = H_\epsilon$, and the
equality of the hamiltonians at all scales, imply that

$$ H_\lambda (q_\lambda) = U_\lambda H_\lambda(q_\infty) U^\dagger_\lambda
   = H_\infty (q_\infty). \eqno (2.3) $$

\noindent We denote $H_\lambda(q_\infty) = {\cal H} _\lambda$
and obtain

$$  {\cal H} _\lambda = U^\dagger_\lambda H_\infty U_\lambda  . \eqno (2.4) $$

\noindent Thus, the effective hamiltonian $H_\lambda$ for effective
particles at the scale $\lambda$ is obtained from the hamiltonian ${\cal
H} _\lambda$ by replacing the creation and annihilation operators for
the bare particles by the creation and annihilation operators for the
effective particles with the same quantum numbers.  The bare creation
and annihilation operators are independent of $\lambda$.  We only need
to calculate the $\lambda$-dependent coefficients in front
of the products of $q_\infty$ in ${\cal H}_\lambda$.

The differential equation for ${\cal H}_\lambda$ is proposed in the
following form \cite {GW2}

$$ {d \over d\lambda} {\cal H}_\lambda = [ {\cal H}_\lambda, {\cal
T}_\lambda] \, ,\eqno (2.5) $$

\noindent where

$$ {\cal T}_\lambda = U^\dagger_\lambda {d \over d\lambda} U_\lambda .
\eqno (2.6) $$

Instead of calculating $U_\lambda$ we shall be using ${\cal T}_\lambda$
in our calculations.  With the help of Eq.  (2.5), all calculations can
be carried out using the bare creation and annihilation operators.
Subsequently, at the end of the calculations, one replaces the bare
operators by the dressed, effective ones and one obtains the desired
effective hamiltonian $H_\lambda(q_\lambda)$ for the scale $\lambda$.

Description of the key elements of the calculation of ${\cal H}_\lambda$
starts here.  ${\cal H}_\lambda$ has the following structure

$$ {\cal H}_\lambda = F_\lambda[{\cal G}_\lambda].  \eqno (2.7) $$

\noindent $F_\lambda[ {\cal G}_\lambda ]$ denotes an operation on the
operator ${\cal G}_\lambda$ which inserts special numerical factors and
makes ${\cal H}_\lambda$ have the properties we want ${\cal H}_\lambda$
to have to represent dynamics of effective particles at the scale
$\lambda$ after $q_\infty$ in ${\cal G}_\lambda$ is replaced by
$q_\lambda$.  Using the unitary equivalence, we also have

$$ H_\lambda(q_\lambda)= F_\lambda[G_\lambda(q_\lambda)], \eqno (2.8) $$

\noindent where

$$ G_\lambda(q_\lambda)= U_\lambda {\cal G}_\lambda U^\dagger_\lambda .
\eqno (2.9) $$

\noindent A similar relation holds for $T_\lambda(q_\lambda)$ and
${\cal T}_\lambda$ since the latter is expressed in terms of $q_\infty$.

The operation $F_\lambda$ will be defined now in a number of steps.  We
begin with a definition of operators which $F_\lambda$ acts on.  The
operator ${\cal G}_\lambda$, or any other operator we will consider, is
a superposition of terms each of which is an ordered product of creation
and annihilation operators.  The ordering is arbitrary but needs to be
defined and the sign in front of each term needs to be determined.  For
example, we will adopt this order:  creators of fermions, creators of
bosons, creators of antifermions, annihilators of antifermions,
annihilators of bosons, annihilators of fermions.  So, each term
contains a product of creation operators standing to the left of a
product of annihilation operators.  At least two operators must appear
in a product and at least one creation and one annihilation operators
must appear, i.e. no product contains only creation or only annihilation
operators, by definition.  This is a universal property of light-front
hamiltonians.  Other forms of hamiltonian dynamics do not have this
property and immediately lead to the problem of understanding the ground
state formation before any of its excitations can be considered beyond
perturbation theory. We shall not further elaborate on this issue
here (see e.g. \cite{SUS} and \cite{GSR}).

The operator ${\cal G}_\lambda$ is divided into two parts, ${\cal
G}_{1\lambda}$ and ${\cal G}_{2\lambda}$.  ${\cal G}_{1\lambda}$ is the
part of ${\cal G}_\lambda$ which is a superposition of the terms of the
form $a^\dagger a$ for $a$ equal $q_\infty$ of any kind.  In principle,
one could also include in ${\cal G}_{1\lambda}$ some terms with a larger
number of creation and annihilation operators, e.g. terms containing two
creation and two annihilation operators.  This option may be useful in
cases where the effective particles interact strongly and the formation
of resonances or bound states needs to be taken into account already
when one evaluates the effective hamiltonians.  However, in general, the
plane-wave Fock space basis states are not eigenstates of such operators
and one is forced to consider more complicated basis states.  Such cases
are beyond the scope of the present article.  Here, we limit ${\cal
G}_{1\lambda}$ to terms $a^\dagger a$.

${\cal G}_{1\lambda}$ becomes the effective free part of $G_\lambda$,
denoted $G_{1\lambda}$, after $q_\infty$ is replaced by $q_\lambda$.
The effective free hamiltonian part will be equal to $G_{1\lambda}$
because $G_{1\lambda}$ will not be changed by the operation $F$.
Eigenvalues of $G_{1\lambda}$ will be also called {\it free energies}.

The remaining part of ${\cal G}_\lambda$, i.e.  ${\cal G}_{2\lambda} =
{\cal G}_\lambda - {\cal G}_{1\lambda}$, contains all terms besides
terms of the form $a^\dagger a$.  To obtain the corresponding part of
the effective hamiltonian $H_\lambda$, one first replaces $q_\infty$ by
$q_\lambda$ and obtains $G_{2\lambda}$.  Then, one applies the operation
$F_\lambda$ to $G_{2\lambda}$.  $F_\lambda$ inserts the vertex form
factors.

Let the momentum labels of all creation operators in a single product in
an interaction term be $k_1, k_2, ..., k_I$ and the momentum labels of
all annihilation operators be $k'_1, k'_2, ..., k'_J$.  Each momentum
has three components, $k^+ $ ranging from 0 to $\infty$ and two
transverse components $k^\perp = (k^1,k^2)$, both ranging from $ -
\infty$ two $+ \infty$.  The $z$-axis is distinguished by our choice of
the light-front.  The three light-front momenta are conserved.  The sum
of momentum labels of all creation operators, $\sum_{i=1}^I k_i$, equals
the sum of momentum labels of all annihilation operators in this term,
$\sum_{j=1}^J k'_j$.  We denote these sums by $P = (P^+, P^\perp)$,
($P^+$ is positive).  Each momentum label has three components, $p=
(p^+, p^\perp)$.  Each $k^+$ or ${k'}^+$ is a positive fraction of
$P^+$; $x_i= k^+_i /P^+ $, $1 > x_i > 0$ and $x'_j = k'^+_j/P^+ $, $1 >
x'_j > 0$.  We have $\sum_{i=1}^I x_i = \sum_{j=1}^J x'_j = 1$.  We also
define

$$ \kappa^\perp_n = k^\perp_n - x_n P^\perp \eqno (2.10) $$

\noindent for all momenta in the hamiltonian term.  $\sum_{i=1}^I
\kappa^\perp_i = \sum_{j=1}^J {\kappa'}^\perp_j = 0$.

The above definitions of momentum variables are standard for wave
functions in the light-front Fock space.  But the way they are used here
is not standard.  We introduce this notation for all translationally
invariant operators.  The definition of $P$ is not connected with any
particular state.  $P$ is not changed as a whole by the action of the
hamiltonian term.  It is only redistributed from the set of momenta of
the annihilated particles to the set of momenta of the created
particles.

Thus, each term in the hamiltonian is characterized by $P$ and two sets
of variables, $X_I=\{(x_i, \kappa^\perp_i)\}^{i=I}_{i=1}$ for creation
operators and $X'_J=\{(x'_j,{\kappa'}^\perp_j)\}^{j=J}_{j=1}$ for
annihilation operators, in that term.  For example, if a product
contains two creation operators and one annihilation operator, we have
$x_1 = x$, $x_2 = 1-x$ and $x'_1 = 1$.  Also, $\kappa^\perp_1 = -
\kappa^\perp_2 = \kappa^\perp$ and ${\kappa'}^\perp_1 = 0$.  But $P$ can
be arbitrary and only a part of the total momentum of some state in
which action of the product in question replaces one particle of
momentum $P$ by two particles of momenta $x P + \kappa$ and $(1-x) P -
\kappa$ (the latter expressions are valid for $+$ and $\perp$
components, $\kappa^+=0$).  It is convenient to speak of $P$ as a {\it
parent} momentum and about the individual particle momenta as {\it
daughter} momenta.  The parent momentum in a hamiltonian term equals one
half of the sum of momenta labeling all creation and annihilation
operators in the term.  Each daughter particle carries a fraction of the
parent momentum.  The parent momentum may be carried by one or more
particles.

The operation $F_\lambda$ on a product of creation and annihilation
operators produces 

$$  F_\lambda \left[
\prod_{i=1}^I a^\dagger_{k_i} \prod_{j=1}^J a_{k'_j} \right] =
    f_\lambda(X_I,X'_J)
\prod_{i=1}^I a^\dagger_{k_i} \prod_{j=1}^J a_{k'_j}
      . \eqno (2.11) $$

The function $f_\lambda (X_I,X'_J)$ can be any suitable function of the
daughter momentum variables which represents our physical intuition and
suits our calculations.  The arguments of $f_\lambda$ are explicitly
invariant with respect to seven kinematical Poincar\'e transformations
of the light-front frame.  This feature leads to explicit symmetry of
our theory with respect to these transformations.  There are three
additional conditions that the function $f_\lambda$ will have to
satisfy.

The first condition is that $f_\lambda$ should be expressible through
the eigenvalues of $G_{1\lambda}$ corresponding to the sets $X_I$ and
$X'_J$ so that $f_\lambda$ equals 1 for small differences between the
eigenvaules and it quickly goes to zero when the differences become
large.  This is an application of the basic condition of the similarity
renormalization scheme for hamiltonians.\cite{GW1} \cite{GW2} The width
of $f_\lambda$ is set by $\lambda$.  One can consider functions
$f_\lambda$ which depend on $X_I$ and $X'_J$ in a more general way than
through the eigenvalues of the free hamiltonian but that option will not
be discussed here.

The first condition defines the effective nature of the hamiltonian
labeled by $\lambda$.  Namely, the effective particle states which are
separated by the free energy gap which is much larger than $\lambda$,
are not directly coupled by the interactions.  In other words, $\lambda$
limits the free energy changes induced by the effective interaction.
Moreover, as a consequence of $f_\lambda \sim 1$ for similar energies,
$1-f_\lambda$ is close to zero for the similar energies and it vanishes
proportionally to a power of the energy difference.  The higher is the
power the smaller is the role of states of similar energies in the
calculation of the effective hamiltonian.  This will be explained later.
Consequently, the higher is the power the smaller is the role of
nonperturbative phenomena due to energy changes below the scale
$\lambda$, in the calculation of the effective hamiltonian.  Thus, there
exists a chance for that the full hamiltonian diagonalization process
can be divided into two parts:  a perturbative calculation of the
effective renormalized hamiltonian and a nonperturbative diagonalization
of that effective hamiltonian. This is our factorization hypothesis in
the hamiltonian approach.

The second condition is that both, $1-f_\lambda$ and $d f_\lambda / d
\lambda $, must vanish faster than linearly in the free energy
difference.  This condition is required to exclude small energy
denominators in perturbation theory.  This will become clear below.  The
second condition implies that $1-f_\lambda$ vanishes as at least second
power of the energy difference near zero.

The third condition for $f_\lambda$ is most difficult to satisfy and a
trial and error approach will be required to verify if this condition is
satisfied.  The third condition is defined by saying that multi-particle
interactions (especially interactions that change the number of
effective particles by many) should not be important in the effective
hamiltonian dynamics which is characterized by changes of energies below
the scale $\lambda$.  This may be possible if {\it $f_\lambda$ as a
function of the daughter variables approximates the shape of one particle
irreducible vertices which is characteristic to the theory under
consideration}.  Structure of $G_\lambda$ depends on the choice of the
function $f_\lambda$.  Some choices will lead to more complicated
effective interactions than others.  The best choices for the most
efficient description of physical phenomena at some scale $\lambda$ are
such that the effective particles interact in a way that is most easy to
understand and which can be parametrized with the least possible number of
parameters over the range of scales of physical interest.  Therefore,
one can conceive variational estimates for the best choices of
$f_\lambda$ that minimize complexity of the effective hamiltonians.
Nothing more can be said in this work about the accuracy one can achieve
in satisfying the third condition in general.  But it is clear that
creation of effective particles will be suppressed when the width
$\lambda$ becomes comparable to the effective masses of those particles.

To satisfy the first condition above we define a boost invariant gap
between free energy eigenvalues for effective particles which is to be
compared with the running cutoff parameter $\lambda$.  The reason to
avoid dependence of $f_\lambda$ on the parent momentum $P$ is that
otherwise the effective dynamics could depend on the frame of reference,
even within the class of frames reachable by kinematical symmetries of
light-front dynamics.  The difficulties may emerge despite the fact that
no physical result depends on the effective cutoff scale $\lambda$, by
construction, and no dependence on the frame of reference can emerge
through dependence of $f_\lambda$ on $\lambda$.  However, when
approximations are made, the symmetry is no longer exact.  The cutoff
independence may appear only in a window corresponding to the
approximations made.  The problem with boost invariance is that boosts
can change momentum arbitrarily and in order to obtain fully boost
invariant results the window would have to be infinite.  Otherwise, one
would obtain frame-dependent results for, a priori, frame-independent
quantities.

If the explicit boost invariance is not preserved by the procedure of
calculating the effective hamiltonian then, the boost invariance
condition for bound state masses provides a good check on the theory.
But no clue is provided within the perturbative calculation of the
effective hamiltonian for what to do if the nonperturbative
diagonalization of the effective hamiltonian does not produce boost
invariant eigenmasses.  Therefore, we choose to preserve kinematical
symmetries of light-front dynamics explicitly.  The price we have to pay
for this great simplification is the length of definition of our
procedure.

As a side remark, we should mention that rotational symmetry of the
spectrum, if obtained, depends on interactions.  Our procedure does not
explicitly preserve rotational symmetry.  Nevertheless, it has been
shown that if counterterms provide enough freedom through their finite
parts and multi-particle effects are suppressed then, there exist
reasons to believe that the rotationally invariant results may be
obtained.  \cite{GP} \cite{JUNKO} \cite{JPG}

The free energy eigenvalues relevant to a particular hamiltonian
term with daughter variables $X_I$ and $X'_J$ are

$$ \sum_{i=1}^I { k^{\perp 2}_i + m^2_i(\lambda) \over k^+_i}
= { P^{\perp 2} + {\cal M}^2_I  \over P^+ } \eqno (2.12) $$

\noindent and

$$ \sum_{j=1}^J { {k'}^{\perp 2}_j + m^2_j(\lambda) \over {k'}^+_j}
= { P^{\perp 2} + {\cal M}^2_J  \over P^+ }  , \eqno (2.13) $$

\noindent where

$$ {\cal M}^2_I = \sum_{i=1}^I { \kappa^{\perp 2}_i + m^2_i(\lambda)
\over
   x_i } \eqno (2.14) $$

\noindent and

$$ {{\cal M}'}^2_J = \sum_{j=1}^J { {\kappa'}^{\perp 2}_j +
m^2_j(\lambda)
    \over x'_j } . \eqno (2.15) $$

\noindent
The individual effective particle masses are allowed to
depend on the effective hamiltonian width parameter $\lambda$.
We define {\it the mass difference} for a hamiltonian term to be

$$  \Delta {\cal M}^2 = {{\cal M}'}^2_J - {\cal M}^2_I , \eqno (2.16) $$

\noindent and {\it the mass sum} to be

$$  \Sigma {\cal M}^2 = {{\cal M}'}^2_J + {\cal M}^2_I. \eqno (2.17)  $$

Now we will describe details of the function $f_\lambda(X_I,X'_J)$.  We
introduce a parameter $z_\lambda$.  Different ways of defining this
parameter are useful for different purposes.  It is important to keep in
mind the physical requirement of the renormalization group approach and
quantum mechanics that large free energy states will be allowed to
interact with considerably larger free energy changes than small free
energy states are.  Consequently, the mass differences for masses
smaller than or on the order of $\lambda$ will be limited by $\lambda$
and the mass differences for masses much larger than $\lambda$ will be
limited by the size of the large masses.  Following the similarity
renormalization scheme \cite{GW1} \cite{GW2}, $z_\lambda$ can be chosen
in such a way that $z_\lambda$ is close to zero for $\Delta {\cal M}^2$
small in comparison to $\lambda^2$ or $\Sigma {\cal M}^2$ and
$|z_\lambda |$ is close to 1 for $\Delta {\cal M}^2$ large in comparison
to $\lambda^2$ or comparable to $\Sigma {\cal M}^2$.  For example,

$$ z_\lambda = {\Delta {\cal M}^2 \over \Sigma {\cal M}^2 + \lambda^2 }.
\eqno (2.18) $$

\noindent This type of definition (including $\Sigma {\cal M}^2$) is
useful for estimates that are necessary in high order perturbation
theory, especially in the analysis of overlapping divergences.
\cite{GW1} The new feature here is that the introduction of $\Sigma
{\cal M}^2$ does not violate the light-front boost invariance and basic
cluster decomposition properties.  Other options exist too.  For
example, one can consider functions of the ratio $\Delta {\cal M}^2 /
\Sigma {\cal M}^2$.  Such functions would depend on $\lambda$ as a
separate parameter so that the ratio would have to be closer to 0 for
smaller $\lambda$'s.  There exists also an option of replacing Eq.
(2.18) by a more complicated function of $\Delta {\cal M}^2$, $\Sigma
{\cal M}^2$ and $\lambda$ which could accelerate or slow down the
decrease of the effective hamiltonian width with the decrease of
$\lambda$ or the masses themselves.  In principle, one can also use
variables $x$ and $\kappa$ in other combinations than ${\cal M}^2$.

The function $f_\lambda(X_I,X'_J)$ is defined for the purpose of this
article to be a function of $z_\lambda^{2^n}$, $ n \geq 1$, which is
analytic in the vicinity of the interval $[0,1]$ on the real axis,
equals 1 for $z_\lambda=0$ and quickly approaches 0 for $z_\lambda \sim
1$;

$$ f_\lambda(X_I,X'_J) = f(z_\lambda^{2^n}) \, . \eqno (2.19) $$

\noindent For example,

$$ f(u) = \left[ 1 + \left({u (1-u_0) \over u_0 (1-u)}\right)^{2^m}
          \right]^{-1}, \eqno (2.20) $$

\noindent where $1 > u_0 > 0$ and $m \geq 1$.  The larger the exponent
$m$ the closer $f(u)$ approaches $\theta(u_0 - u)$ for $0 \leq u \leq
1$.  Eq.  (2.20) concludes our definition of the operation $F_\lambda$.

The smallest possible value of $\Sigma {\cal M}^2$ in Eq.  (2.18) is
$\left[ \sum_{i=1}^I m_i(\lambda) \right]^2 + \left[ \sum_{j=1}^J
m_j(\lambda) \right]^2 $. Thus, $z_\lambda$ is small for small positive
$\lambda^2$ when $\Delta {\cal M}^2$ is small in comparison to particle
masses.  Therefore, $u_0$ must be much smaller than 1 to force $\Delta
{\cal M}^2$ to be small in comparison to $\Sigma {\cal M}^2$ when
$\lambda^2$ is small.

One can also force $\Delta {\cal M}^2$ to be small in comparison to the
particle masses by making $\lambda^2$ negative so that it subtracts from
$\Sigma {\cal M}^2$ its minimal value.  Then, the mass difference is
compared to the sum of kinetic energies due to the relative motion only.

It is also useful to limit the small mass differences by choosing an
infinitesimally small $u_0$ and introducing $\lambda^2 = u_0^{-1/2^n}
\tilde \lambda^2$.  Then, $|\Delta {\cal M}^2| \leq \tilde \lambda^2$ in
the $\theta$-function limit.  In this case, the band-diagonal
hamiltonian width becomes independent of the mass sum for as long as the
latter is small in comparison to $\lambda^2$.

Next, we define the infinitesimal transformation ${\cal T}_\lambda$ in
Eq.  (2.5).  Eq.  (2.5) can be rewritten using Eqs.  (2.8) and (2.9),
into the form

$$  {\cal H}' = f'{\cal G} + f {\cal G}' = [f{\cal G}_1, {\cal T}]
+ [f{\cal G}_2, {\cal T}] \, . \eqno (2.21) $$

\noindent The prime denotes differentiation with respect to $\lambda$.
We have simplified the notation of $F_\lambda[{\cal G}_\lambda]$ to
$f{\cal G}$.

In the further development, three universally valid relations will be
often used without saying.  Namely, $f{\cal G}_1 = {\cal G}_1$ and
$(1-f){\cal G}_1 = f'{\cal G}_1 = 0$.

Equation (2.21) involves two unknowns, $\cal G$ and $\cal T$.
Additional conditions are required to define $\cal T$.  One recalls that
if the interaction is absent, i.e. when ${\cal G}_2 = 0$, then no
evolution with $\lambda$ may appear.  Therefore, in the limit of
negligible interactions, both ${\cal G}' = 0$ and ${\cal T} =0$ . Thus,
we expect that ${\cal G}'$ should differ from zero if and only if the
interactions are important.  The first term on the right-hand side is
order ${\cal T}$ since ${\cal G}_1$ is order zero in powers of the
interactions.  The second term on the right-hand side is at least of
second order in the interactions.  The first term can be used as a seed
for defining ${\cal T}$ through a series of powers of the interaction.

We associate the derivative of $\cal G$ with the second term on the
right-hand side.  The first term on the right-hand side and a part of
the second term which is left after the derivative of $\cal G$ is
defined, together determine $\cal T$ . Thus, we come to the definition
of $\cal T$ through the commutator $[{\cal G}_1, {\cal T}]$.  We write

$$ A = \{ B \}_{{\cal G}_1} \eqno (2.22) $$

\noindent when

$$ [A, {\cal G}_1] = B   . \eqno (2.23) $$

\noindent The subscript of the curly bracket in Eq.  (2.22) will be
often omitted in later discussion unless this omission might lead to a
confusion with the usual curly brackets.  The bracket operation of Eqs.
(2.22) and (2.23) can be written in the explicit form since the most
general structure of the operators $A$ and $B$ we shall encounter is a
superposition of products of creation and annihilation operators.
Suppose B contains a term which involves a product

$$ \prod_{i=1}^I a^\dagger_{k_i} \prod_{j=1}^J a_{k'_j} . \eqno
(2.24) $$

\noindent Then, $\{B\}_{{\cal G}_{1\lambda}}$ contains the same product
(as a part of the same expression) with an additional factor equal

$$ \left[ \sum_{j=1}^J { {k'}^{\perp 2}_j + m^2_j(\lambda) \over
{k'}^+_j} -
\sum_{i=1}^I { k^{\perp 2}_i + m^2_i(\lambda) \over k^+_i}
\right]^{-1}    .  \eqno (2.25) $$

\noindent For translationally invariant operators, the sums of
individual energies satisfy Eqs.  (2.12) and (2.13), respectively, and
the factor (2.25) equals

$$ \left[ \Delta {\cal M}^2 \over P^+ \right]^{-1} ,  \eqno (2.26) $$

\noindent where $P^+$ is the parent momentum for the product under
consideration and the mass difference is defined in Eq.  (2.16).  When
all terms in the operator $B$ are multiplied by the corresponding
factors (or, equivalently, divided by the energy denominators) one
obtains the operator $A$ defined in Eq.  (2.22).

The factor (2.25) explodes to infinity when the denominator approaches
zero.  Hence, for the operator $A$ to be well defined, the coefficients
of products of the form (2.24) in the operator $B$ must vanish at least
as fast as the energy denominator itself when the denominator approaches
zero.  Therefore, our definition of $[{\cal G}_1, {\cal T}]$ must be
given in terms of an operator which has such property.  In order to
satisfy this condition, Eq.  (2.21) is split into two equations as
follows.

$$  f{\cal G}' = f[f{\cal G}_2, {\cal T}] \,  , \eqno (2.27) $$

$$ [{\cal T}, {\cal G}_1] = (1-f) [f{\cal G}_2, {\cal T}] - f' {\cal G}
 \,  .  \eqno (2.28) $$

\noindent These equations imply that $\cal T$ does not contain small
energy denominators provided the functions $f_\lambda$ and
$f'_\lambda$ satisfy our initial assumptions.  Our second condition
introduced below Eq.  (2.11) on the functions $1 - f_\lambda(X_I, X'_J)$
and $f'_\lambda(X_I, X'_J)$ guarantees that ${\cal T}$ is well defined
and tends to zero in the region of vanishing energy denominators because
the right-hand side of Eq.  (2.28) vanishes at least as fast as the
first power of the energy differences.

Equation (2.27) is a first order differential equation.  One has to
provide initial conditions to define the hamiltonian theory.  The
initial conditions will be set in examples discussed in this paper by
canonical light-front hamiltonians and the corresponding counterterms.
The difficulty of the renormalization theory is to determine the
counterterms so that the effective hamiltonians have well defined limits
when the bare cutoff is removed.  In other words, one has to find
the class of initial conditions at $\lambda = \infty$ which imply
$\epsilon$-independent $H_\lambda$'s for all finite $\lambda$'s when
$\epsilon \rightarrow 0$.

A general iterative procedure for calculating the effective hamiltonians
will be described now in complete analogy to Refs.  \cite{GW1} and
\cite{GW2}.  However, instead of the iteration with two coupled
equations for $H_\lambda$ and $T_\lambda$, we describe here an iteration
based on a single equation for $H_\lambda$ with an explicit solution for
$T_\lambda$ already built in.  Simple algebra and substitution of Eq.
(2.28) into Eq.  (2.27), lead to

$$     {d \over d\lambda} {\cal G}_\lambda =
\left[ f_\lambda {\cal G}_{2\lambda}, \left\{ {d \over d\lambda}
(1-f_\lambda) {\cal G}_{2\lambda} \right\}_{{\cal G}_{1\lambda}} \right] .
\eqno (2.29) $$

\noindent Equation (2.29) is the main element of the renormalization
group formalism in this paper.  Note that the right-hand side is given
in terms of a commutator.  Therefore, {\it the effective renormalized
hamiltonians contain only connected interactions}.  This is essential
for cluster decomposition properties of the effective hamiltonians.
\cite{WE1}

Equation (2.29) is of the form

$$     {d \over d\lambda} {\cal G}_\lambda =
   T_\lambda[{\cal G}_\lambda] . \eqno (2.30) $$

\noindent The right hand side contains terms which are bilinear in the
effective interaction strength.  The initial condition for Eq.  (2.29),
or (2.30), is given at $\lambda = \infty$:  ${\cal G}_{\lambda=\infty}
= {\cal G}_\epsilon$.

With accuracy to the first order in powers of the interaction strength,
${\cal G}_\lambda$ is independent of $\lambda$ and ${\cal G}_\epsilon$
is equal to the initial regularized hamiltonian expression one intends
to study, denoted by $H^{(0)}_\epsilon$.  In this initial approximation,
${\cal H}^{(0)}_\lambda = f^{(0)}_\lambda {\cal G}^{(0)}_{\lambda}$,
where ${\cal G}^{(0)}_\lambda = H^{(0)}_\epsilon$ and $f^{(0)}_\lambda$
is the similarity factor calculated using eigenvalues of ${\cal
G}^{(0)}_{1 \lambda}$.  ${\cal H}^{(0)}_\lambda$ forms our first
approximation to the similarity renormalization group trajectory of
operators ${\cal G}_\lambda$ parametrized by $\lambda$.

Eq.  (2.30) can now be written in the iterative form for successive
approximations to the trajectory ${\cal G}_\lambda$.  Namely,

$$ {d \over d\lambda} {\cal G}^{(n+1)}_\lambda =
   T^{(n)}_\lambda[{\cal G}^{(n)}_\lambda]
 .  \eqno (2.31)  $$

\noindent This is a conveniently abbreviated version of

$$ {d \over d\lambda} {\cal G}^{(n+1)}_\lambda = \left[ f^{(n)}_\lambda
{\cal G}^{(n)} _{2\lambda}, \left\{ {d \over d\lambda} (1-f^{(n)}_\lambda)
{\cal G}^{(n)} _{2\lambda} \right\}_{{\cal G}^{(n)}_{1\lambda}} \right]
. \eqno (2.32) $$

\noindent $f^{(n)}_\lambda$ denotes a function of $z^{(n)}_\lambda$
expressed through eigenvalues of ${\cal G}^{(n)}_{1\lambda}$, such as in
Eqs.  (2.19) and (2.20).  The initial condition is set by ${\cal
G}^{(n+1)}_\infty = {\cal G}^{(n+1)}_\epsilon$.  Thus, the solution is

$$ {\cal G}^{(n+1)}_\lambda = {\cal G}^{(n+1)}_\epsilon -
\int_\lambda^\infty T^{(n)}_s[{\cal G}^{(n)}_s] . \eqno (2.33) $$

By definition, ${\cal G}_\infty$ contains the $\epsilon$-regulated
canonical hamiltonian terms and counterterms.  The counterterms remove
the part of the integral in Eq.  (2.33) which diverges for finite
$\lambda$ when $\epsilon \rightarrow 0$.  This definition is based on
the requirement that matrix elements of the hamiltonian of the effective
theory have a limit when $\epsilon$ is made very small.  The condition
that the necessary ${\cal G}_\infty$ exists is the hamiltonian version
of renormalizability.  It does not require the number of counterterms to
be finite, although a finite number has the clear advantage of
simplicity.

Denote the part of the integrand in Eq.  (2.33) which leads to the
divergence by $\left[ T^{(n)}_s[{\cal G}^{(n)}_s] \right]_{div}$, and
the remaining part by $\left[ T^{(n)}_s[{\cal G}^{(n)}_s]
\right]_{conv}$.  ${\cal G} ^{(n+1)}_\epsilon$ contains the initial
regulated hamiltonian terms and counterterms.  We define the
counterterms in ${\cal G}^{(n+1)}_\epsilon$ from the condition that
matrix elements of the corresponding effective hamiltonians with finite
$\lambda$, i.e. of $F^{(n+1)}_\lambda [G^{(n+1)}_{\lambda}] $, are
independent of $\epsilon$ when $\epsilon \rightarrow 0$.  This condition
amounts to the requirement that the coefficients of products of creation
and annihilation operators are finite for finite free energies
corresponding to the momentum labels of the operators in the product.

Note that $F^{(n+1)}_\lambda [G^{(n+1)}_{1 \lambda}] = G^{(n+1)}_{1
\lambda} $ and it is not necessary to know $F^{(n+1)}_\lambda$ to
calculate $G^{(n+1)}_{1 \lambda}$.  One calculates $F^{(n+1)}_\lambda$
after $G^{(n+1)}_{1 \lambda}$ is made independent of $\epsilon$ when
$\epsilon \rightarrow 0$.

The way of deriving ${\cal G}^{(n+1)}_\epsilon$ from divergences in the
integral on the right-hand side of Eq.  (2.33) is following.

The diverging dependence of the integral on $\epsilon$ when $\epsilon
\rightarrow 0$, is typically of the form $\epsilon^{-1}$ or
$\log{\epsilon}$ times operator coefficients.  The operator coefficients
can be found by integrating the diverging part of the integrand from
some arbitrary finite value of $\lambda$, say $\lambda_0$, to infinity.
The divergence originates from the upper limit of the integration and it
is independent of $\lambda_0$.  The remaining finite part of the
integral is sensitive to the lower limit of integration and depends on
$\lambda_0$.  The counterterm does not depend on $\lambda_0$ but it
contains an arbitrary finite part which emerges in the following way.

The counterterm subtracts the diverging part of the integral.  But
subtracting terms with diverging functions of $\epsilon$ times known
operators does not tell us what finite parts times the same operators to
keep.  Thus, one needs to add arbitrary finite parts to the numbers
$1/\epsilon$ and $\log{\epsilon}$ in the counterterms.  These finite
parts are unknown theoretically and have to be fitted to data.  In
particular, observed symmetries may impose powerful constraints on the
finite parts.

The diverging part of the integrand may be such that the lower limit of
integration produces the same operators as the upper limit but the
diverging numbers such as $\epsilon^{-1}$ or $\log{\epsilon}$ from the
upper limit are replaced by finite numbers at the lower limit.  Those
finite numbers depend on $\lambda_0$ but, once they are replaced by the
required unknown finite parts, one obtains a valid expression for the
counterterm.  The replacement can be achieved by adding to the integral
the same operators multiplied by the numbers which are equal to the
differences between the unknown numbers and the numbers resulting from
the lower limit of the integration.  Thus, the unknown numbers we need
to add to the integral of the diverging part of the integrand from
$\lambda_0$ to infinity, depend on $\lambda_0$.  We can write ${\cal
G}^{(n+1)}_{\epsilon}$ which is independent of $\lambda_0$ as ${\cal
G}^{(n+1)}_{\epsilon\,{finite}}(\lambda_0) + \int_{\lambda_0}^\infty
\left[ T^{(n)}_s[{\cal G}^{(n)}_s] \right]_{div}$.  The free finite
parts of the counterterms are contained in ${\cal
G}^{(n+1)}_{\epsilon\,{finite}}(\lambda_0)$ and one can fit them to data
using effective hamiltonians at some convenient scales $\lambda$.

More than one scale $\lambda$ may become necessary for accurate
determination of the free parameters when their values have to be of
considerably different orders of magnitude and require knowledge of
physical phenomena at different scales.  For simplicity, let us assume
here that a single scale $\lambda = \lambda_0$ is sufficient for
practical calculations.  In this case, the renormalization conditions
are set through the description of selected physical phenomena using
$H_{\lambda_0}$.  One may also consider renormalization conditions for
parameters in $H_{\lambda_0}$ which are set at some nearby scale
$\lambda_1 \neq \lambda_0$.  This will be illustrated in the next
Section.

The complete recursion including construction of counterterms in Eq.
(2.33) is given by

$$ {\cal G}^{(n+1)}_\lambda = {\cal G}^{(n+1)}_{\epsilon\,
finite}(\lambda_0) + \int_{\lambda_0}^\lambda ds
\left[ T^{(n)}_s[{\cal G}^{(n)}_s] \right]_{div}
- \int_\lambda^\infty ds
\left[ T^{(n)}_s[{\cal G}^{(n)}_s] \right]_{conv}
 . \eqno (2.34) $$

\noindent In the limit $n \rightarrow \infty$, if the limit exists,
one obtains

$$ {\cal G}_\lambda = {\cal G}_{\epsilon\,
finite}(\lambda_0) + \int_{\lambda_0}^\lambda ds
\left[ T_s[{\cal G}_s] \right]_{div}
- \int_\lambda^\infty ds
\left[ T_s[{\cal G}_s] \right]_{conv}
   . \eqno (2.35) $$

\noindent $H_\lambda$ is obtained from Eq.  (2.35) through the
replacement of $q_\infty$ by $q_\lambda$ (to obtain $G_\lambda$) and
action of $F_\lambda$ on $ G_\lambda$.  These steps complete our general
definition of the renormalized effective hamiltonians in the light-front
Fock spaces.  The structure of the formalism is general enough for
extension to other applications but those are not discussed in this
paper.

The renormalized effective hamiltonians may still depend on the infrared
regulators in the case of theories with massless particles.  The limit
of the infrared regulators being removed ($\delta \rightarrow 0$ ) may
lead to infrared singularities which cannot be avoided without solving
for eigenstates of the effective hamiltonians.  It depends on how the
effective hamiltonians couple Fock sectors with many massless particles
in the eigenvalue problem.  For example, gauge theories contain massless
spin-1 bosons with singular dependence of the polarization vectors on
the small longitudinal momentum.  This region contributes infrared
cutoff dependent interaction terms.  To build a bridge between the
renormalized effective hamiltonians in the light-front Fock space and
the formal field theory with a highly symmetric local lagrangian, such
as for gauge theories, one has to understand what conditions justify
calculations of the effective hamiltonians in separation from a
nonperturbative solution of the eigenvalue problem when the infrared
cutoff is important.  For example, a neutral state of localized charged
fermions will not radiate bosons with wavelength much larger than the
distance between the fermions.  A precise general answer to the question
how the infrared cutoff disappears from the hamiltonian spectrum for
gauge theories is not known to the author.  Such infrared factorization
property in the hamiltonian approach requires extensive studies in gauge
theories.  The infrared regularization and renormalization conditions in
second order perturbation theory will be discussed in the next Sections.

Our description of the perturbative calculus for renormalized effective
hamiltonians starts here.  The key observation is that one can calculate
the rate of change of ${\cal G}_\lambda$ by expanding it
into a power series in the effective interaction ${\cal G}_{2\lambda}$
at the same running scale $\lambda$.  This is obtained by repeated
application of Eq.  (2.29).  One rewrites Eq.  (2.29) as

$$ {d {\cal G}_\lambda \over d\lambda} = \left[ f_\lambda {\cal
G}_{2\lambda}, \left\{ -f'_\lambda {\cal G}_{2\lambda} \right\}_{{\cal
G}_{1\lambda}} \right] + \left[ f_\lambda {\cal G}_{2\lambda}, \left\{
(1-f_\lambda) {d {\cal G}_{\lambda} \over d\lambda}  \right\}_{{\cal
G}_{1\lambda}} \right] . \eqno (2.36) $$

\noindent Then, one can substitute ${\cal G}'_\lambda $ in
the last term on the right-hand side of Eq.  (2.36) by the preceding
terms on the right-hand side.  Two successive substitutions produce an
expression for $ {\cal G}'_\lambda $ with four
explicit powers of the effective interactions and the remaining terms
are of higher order [note that $ (1 - f_\lambda) {\cal G}'_{1 \lambda} =
0 $].

$$     {d {\cal G}_\lambda \over d\lambda} =
\left[ f{\cal G}, \left\{ -f'{\cal G}\right\} \right]
+ \left[ f{\cal G}, \left\{ (1-f)
\left[ f{\cal G}, \left\{ -f'{\cal G}\right\} \right]
\right\} \right] $$
$$+ \left[ f{\cal G}, \left\{ (1-f)  \left[ f{\cal G}, \left\{ (1-f)
\left[ f{\cal G}, \left\{ -f'{\cal G}\right\} \right] \right\}
\right] \right\} \right] + o({\cal G}^5)  . \eqno (2.37) $$

\noindent We have omitted subscripts 2, $\lambda$ and ${\cal
G}_{1\lambda}$ on the right-hand side.  All the subscripts should appear
in the same pattern as in Eq.  (2.36).  Correspondingly, the infinite
chain of substitutions produces an expression ordered by explicit powers
of the effective interactions, to infinity.

$$ {d \over d\lambda} {\cal G}_\lambda =
   \sum_{n=0}^\infty \left[ f{\cal G},
   ( \left\{(1-f)\left[ f{\cal G}, )^{(n)}
   \{-f'{\cal G}\} ( \right] \right\} )^{(n)} \right] .
   \eqno (2.38) $$

\noindent The round bracket raised to the $n$-th power means
$n$ consequtive repetitions of the symbols from within the round
bracket. The subscripts are omitted for clarity as in Eq. (2.37).

The above expansion in powers of the effective interactions provides a
systematic order by order algorithm for building an expression for the
effective hamiltonian.  The energy denominators and functions
$f_\lambda$ are calculated using eigenvalues of ${\cal G}_{1\lambda}$.
Therefore, in Eqs.  (2.37) and (2.38), the derivatives of the function
$f_\lambda$ contain two kinds of contributions:  those resulting from
differentiating the explicit $\lambda$ dependence in the arguments
$z_\lambda$ (for example, $\lambda^2$ in Eq.  (2.18) ), and those
resulting from differentiating the free energy eigenvalues (for example,
$\Delta {\cal M}^2$ in Eq.  (2.18)).  Since the free energy terms
include effective masses which depend on the width $\lambda$, the
derivatives of the effective masses appear in the equations on the
right-hand side.  Moving them to the left-hand side leads to coupled
nonlinear differential equations for the effective hamiltonians.

The general iterative approach in Eq.  (2.34) or the expansion in Eq.
(2.38), can be analysed using expansion in the running coupling
constants.  One can divide ${\cal G}_{1\lambda}$ into two parts:  one
which is independent of the coupling constants and another one which
vanishes when the coupling constants are put equal to zero.  The parts
depending on the coupling constants are moved over to ${\cal
G}_{2\lambda}$ and treated as an interaction.  After ${\cal
G}_{1\lambda}$ is reduced to the part which is independent of the
interactions, the derivatives of $f_\lambda$ in Eqs.  (2.36) to (2.38)
do not introduce additional powers of the interaction strength and the
series is strictly ordered in powers of the interactions according to
their explicit appearance in the formula (2.38).  This series then provides
the perturbative expansion in terms of the running couplings.

The simplest case of the perturbative expansions is the case where one
expands in powers of a single coupling at a single scale.  Firstly, one
expands the renormalization group equations into a series of terms
ordered by powers of the bare coupling $g_0$.  Secondly, one evaluates
the effective coupling $g_1$ at the chosen scale $\lambda_1$ as a power
series in the bare coupling.  Thirdly, the latter series is inverted and
the bare coupling is expressed as a series in the effective coupling
$g_1$.  Then, one can pursue perturbative calculations in terms of the
effective coupling.  In particular, one can reduce the hamiltonian width
to $\lambda_2 < \lambda_1$ and calculate $g_2$ as a series in $g_1$.
Such steps can be repeated.  For example, one can reduce the width in
each step by a factor 2. \cite{W1} \cite{W2} $N$ steps will reduce the
width by the factor $2^{-N}$.  This way one can build the
renormalization group flow indicated by the chains of small arrows in
the diagram discussed in Section 1.

\vskip.3in
{\bf 2.b Regularization}
\vskip.1in

A canonical bare hamiltonian obtained from a local field theory is
divergent.  This Section describes how the ultraviolet singularities in
the canonical hamiltonian are regularized with the bare cutoff
$\epsilon$.  We also discuss the infrared regularization.  Our
presentation is ordered as follows.  First, we briefly explain the
relation between the ultraviolet and infrared regularizations.  Then, we
proceed with definitions of the canonical hamiltonian terms.  For that
purpose, we have to discuss the fundamental set of scales in the
hamiltonian approach and explain the role of classical field theoretic
lagrangians in the construction.  Then, we describe details of the
ultraviolet and infrared regularizations.

The ultraviolet and infrared regularizations in the light-front
hamiltonians are closely related.  The infrared structure is influenced
by the masses in the initial hamiltonian $H^{(0)}_\epsilon$.
$H^{(0)}_\epsilon$ carries the superscript 0 to indicate that it is the
initial hamiltonian from the previous Section - it does not include the
counterterms which need to be calculated.

An initial mass value is generically denoted by $m^{(0)}_\epsilon$.  A
light-front energy, $p^-_m$, of a free particle with a four-momentum
$p_m = \left(p^+, p^\perp, p^-_m = (p^{\perp 2} + m^2 )/p^+ \right)$,
tends to infinity when $m^2 > 0$ and $p^+$ tends to zero.  But $p^-_m$
may be finite or even approach zero in this limit if $m^2 = 0$ and
$p^\perp$ approaches zero too.  The limit of small momentum $p^+$ is
always a high-energy limit when $m^2 > 0$.  But it ceases to be the
high-energy limit for the small transverse momenta when $m^2 \rightarrow
0$.  Thus, $m^{(0)}_\epsilon$ in the initial hamiltonian is capable of
switching from the high-energy regime in the longitudinal direction to
the low-energy one when we take the limit $m^{(0)}_\epsilon \rightarrow
0$.  Conversely, introducing masses turns the infrared low-energy regime
into the high-energy regime.

We begin the construction of $H^{(0)}_\epsilon$ with an enumeration of
momentum scales which are involved in it.  We distinguish a set of
fundamentally different scales related to the boundary conditions for
fields at spatial infinity, small momentum cutoffs, phenomenological
parameters and large momentum cutoffs.

The bare hamiltonian $H^{(0)}_\epsilon$ is defined in terms of the
operators $q_\infty$.  We consider quantum fields to be built from these
creation and annihilation operators.  We consider quantization rules for
classical field theory as secondary to the quantum theory of particles
(see also Ref.  \cite{WE1}).

The initial basis in the Fock space is built from the vacuum state
$|0\rangle$ by successive actions of the creation operators $q_\infty$.
We will need more detailed notation.  Fermion, antifermion and boson
creation and annihilation operators are conventionally denoted by
$b^\dagger$, $d^\dagger$, $a^\dagger$, $b$, $d$ and $a$, respectively.
For example, $|k \sigma> = b^\dagger_{k\sigma} |0\rangle$ denotes a
state of one bare fermion of momentum $k=(k^+, k^\perp)$.  Spin $z$-axis
projection, flavor, color or other quantum numbers, are denoted by a
common symbol $\sigma$.  The momentum variables in the subscripts of
creation and annihilation operators are distinguished in order to set
the scales involved in the definition of the hamiltonian.  The momentum
scales will be ordered from the smallest to the largest.  The order of
scales in momentum space will be reverse to the order of scales in
position space.

The largest scale to be considered in the position space is the
quantization volume.  This volume will never be allowed to enter any
physical consideration.  The reason is that we want to describie
phenomena which we believe to be independent of any conditions at remote
distances.  If this belief is not justified our formulation will be
invalid.

We assume the quantization volume to be so large that the boundary
conditions are of no importance.  Since we define states of bare
particles by assigning them momenta, two different values of momenta are
separated by at least one unit of the inverse size of the quantization
volume.  By assumption, no physically interesting question will concern
wavelengths comparable with that size.

The Poincare symmetry of our theory will hold only approximately because
the space of momentum variables in our hamiltonian quantum mechanics
will not be continuous.  But we shall insure by our choices of scales
that the granulation in momentum space will never be noticeable and the
quantization volume will be effectively infinite for all our purposes.
Thus, we will universally adopt continuous notation for momentum
variables.  For example, invariance under boosts along $z$-axis will
only mean that there is sufficiently many different momentum states so
that a change of $k^+$ by a prescribed factor can be approximated using
a change by a rational factor with better accuracy than any gedanken
experiment could verify.  The same concerns rotations.

Our assumption that the quantization volume can be large enough for its
size to become irrelevant to physics, seems to contradict potential
relevance of topological features which can be related to the boundary
conditions in classical field theory.  This is not necessarily the case.
The place where a large volume boundary conditions can appear in the
hamiltonian theory is in the formation of dynamically stable infrared
components in physical states.  These might be called ``wee''.
Expectation values of field operators in states containing many wee
particles may be responsible for producing the approximate long distance
or large correlation length classical field dynamics.

Spontaneous symmetry breaking 
will be allowed to appear through the wee
components but it is not known in the quantum theory that such components do
emerge in physically interesting cases.  It is also not known how many
different long wavelength structures may form.  Most importantly, we do
not know the saturation mechanism which leads to the formation of
such stable components.  Therefore, their scale, even if present, escapes our
control at this time and requires more studies.

We cannot exclude that some topological features will eventually result
from our assumptions.  But we cannot address this issue before
calculating the effective hamiltonians and their eigenstates.  The issue
will have to be considered if it turns out that the large scale features
do matter in the formation of the eigenstates.

Some hypotheses about zero-modes and spontaneously broken symmetry in
light-front quantum field theories were recently discussed in Ref.
\cite{W3} which quotes important earlier literature on the subject.
Basically, one may expect that new terms emerge in the effective
hamiltonians and they account for the large scale dynamical effects.  
Susskind and co-workers have proposed a way to
think about the wee parton dynamics in a model.  \cite{SUS} 
Ref.  \cite{GSR} discusses the QCD sum rules techniques. However, the
original quantum dynamics of the vacuum formation and spontaneous
symmetry breaking in the present approach are not yet understood and
cannot be discussed further here.

The next smaller size in position space is the inverse of the infrared
regulator.  Three main types of the infrared regularization appear in
the light-front dynamics.  The first is a straightforward lower bound on
$k^+$, denoted by $\delta^+$.  The second is a lower bound, denoted by
$\delta$, on the parent $+$-momentum fraction that can be carried by a
particle in an interaction.  The third is a mass parameter $\mu_\delta$.
$\mu_\delta$ appears as the mass parameter $m^{(0)}_\epsilon$ in the
initial hamiltonian $H^{(0)}_\epsilon$.  It is introduced for massless
bare particles.  $\mu_\delta$ may effectively cut off the small
longitudinal momentum region at a small scale order $\mu_\delta/
\Delta$, where $\Delta$ is some ultraviolet momentum upper bound.  When
the infrared regularization is being removed, $\delta^+$, $\delta$ or
$\mu_\delta$ are sent to zero.

One should note that $\delta$ is a dimensionless number while $\delta^+$
and $\mu_\delta$ have dimensions.  In order to make a connection one
needs to introduce a number with the dimension of a mass.  It is assumed
that the finite detector sensitivity and experimental wave packet widths
are much larger than the infrared regularization scale and the wide gap
between the two scales lets us choose the dimensionfull number in such a
way that physical quantities will be independent of the small infrared
cutoff scale.

One should also note that $\delta^+$ and $\delta$ distinguish the
longitudinal direction from the transverse directions on the light-front
while the mass regulator combines them.  The mass regulator seems to
avoid the distinction.  However, introducing a mass term for photons or
gluons which have only two polarizations does explicitly distinguish
directions. In the gauge theories, understanding the mechanism by which the
infrared regularization mass scale  becomes
irrelevant to physics is one of the main theoretical challenges in the
hamiltonian theory.  Apparently, the lack of sensitivity to the infrared
cutoffs which distinguish directions is related to the rotational
symmetry and to the gauge symmetry.  The extraordinary interest in
understanding the mechanism of this insensitivity, if it exists, 
stems from the fact that it must combine the hamiltonian quantum 
mechanics with the Lorentz and gauge symmetries.

The next smaller scale in position space is set by the size of the
volume used for preparation of incoming and detection of outgoing
particles (including bound states) and the corresponding time scale.
Physics to be discussed in our approach will be contained within this
scale.  Wave packets mean values will be contained within distances on
that scale and the momentum space widths of the wave packets will be
limited from below by its inverse.  Physical observables are allowed to
depend on this scale since the preparation and detection of states is a
part of a physical process.

The order of magnitude of momenta larger than the experimental wave
packet widths will be characterized initially in terms of three
different scales.  Namely, (1) masses of particles, (2) the width of the
effective hamiltonian (i.e.  $\lambda$), and (3) the bare cutoff scale
$\epsilon^{-1}$.  Later, when solving for the hamiltonian spectrum, a
new scale may emerge dynamically, determined by the effective coupling
constants, masses and width of the effective hamiltonian.  Scale
invariance at large momenta may be violated through a dimensional
transmutation even if all mass scales are negligible in comparison to
the momenta and $\lambda$.

The masses of particles appear in the initial bare hamiltonian.
Renormalization effects lead to different renormalized masses in
the effective hamiltonians.  The effective masses will depend on the
width $\lambda$ and the strength of the interactions.

The width $\lambda$ will be ranging from $\infty$ to some convenient
finite values. Efficient description of physical phenomena which
involve energy-momentum transfers of the order of $k$ will require
$\lambda$ to be larger than $k$.  It will also be useful to use
$\lambda$ not too large in comparison to $k$ in order to avoid too much
detail in the effective hamiltonian dynamics. Useful values of $\lambda$
in nonrelativistic systems are smaller than the effective masses.
For example, in QED, the convenient
effective cutoff $\lambda$ on the electron momentum in a hydrogen atom
is much larger than the electron binding energy and much smaller
than the electron mass.

The bare cutoff scale $\epsilon^{-1}$ will be related to much larger
momentum scales than any scale of interest in physics.  It should be
stressed that the formal limit $\lambda \rightarrow \infty$ is used here
only to remove $\lambda$ dependence from the hamiltonian regulated by
$\epsilon$.  In other words, no $\lambda$ dependence appears in the
hamiltonians with $\lambda$ larger than the scale implied by
$\epsilon^{-1}$.  No physical quantity will be allowed to depend on
$\epsilon$ when $\epsilon \rightarrow 0$.  The similarity
renormalization scheme for hamiltonians is built to achieve this goal to
all orders in perturbation theory (cf.  \cite {GW1} \cite{GW2}).

One would have to make additional subtractions if dependence on
$\epsilon$ were recovered in the process of finding eigenstates of the
effective hamiltonians beyond perturbation theory.  Such unfavorable
result would indicate that perturbation theory in the running
interactions is not reliable in deriving the effective hamiltonians.
Although this possibility will be neglected in the present paper, we
mention it because we lack understanding of the mechanisms of the ground
state formation and the insensitivity to the infrared regularization. 
The point is that the order of removal of the ultraviolet and
infrared cutoffs may matter.  The infrared regularization by
$\mu_\delta$ turns the infrared divergences into the ultraviolet ones.
Then, the ultraviolet renormalization can remove the divergences by
removing the $\epsilon$-dependence from the effective hamiltonians.  But
the limit of $\mu_\delta \rightarrow 0$ destroys this correlation of the
ultraviolet and infrared divergences.  Therefore, it is not excluded
that the removal of the infrared regularization may lead to different
results depending on the order of limits one takes.  Such considerations
are beyond the scope of the present article.  This remark concludes our
description of scales involved in the construction of hamiltonians.

We proceed to the explicit construction of simplest terms in the
hamiltonian $H^{(0)}_\epsilon$.  The counterterms are not known exactly
from the outset.  Light-front power counting rules are helpful \cite{W3}
in determining the structure of the counterterms but more detailes are
required in practice.  The similarity renormalization group is to
provide the required insight.

All starting hamiltonians we consider will contain a free part which we
denote by ${\cal G}^{(0)}_1$.  The free part for fermions and bosons
has the form

$$ {\cal G}^{(0)}_1 = \sum_\sigma \int [k] \left[ {k^{\perp 2} +
m^{(0)\, 2}_\epsilon \over k^+} ( b^\dagger_{k\sigma} b_{k\sigma} +
d^\dagger_{k\sigma} d_{k\sigma}) + {k^{\perp 2} + \mu^{(0)\,2}_\epsilon
\over k^+} a^\dagger_{k\sigma} a_{k\sigma} \right] . \eqno (2.39) $$

\noindent We adopt the following conventions. Summation over $\sigma$
denotes a sum over all quantum numbers except the momentum.

$$ \int [k] = {1 \over 16 \pi^3} \int_0^\infty {dk^+ \over k^+}
\int d^2 k^\perp . \eqno (2.40) $$

\noindent The creation and annihilation operators in Eq.  (2.39) are the
bare ones commonly denoted in Section 2.a by $q_\infty$.  They
satisfy standard commutation or anticommutation relations

$$ \left[ a_{k\sigma}, a^\dagger_{k'\sigma'} \right] =
   \left\{ b_{k\sigma}, b^\dagger_{k'\sigma'} \right\} =
   \left\{ d_{k\sigma}, d^\dagger_{k'\sigma'} \right\} =
   16 \pi^3 k^+ \delta^3 (k-k') \delta_{\sigma \sigma'}   \eqno (2.41)$$

\noindent with all other commutators or anticommutators equal zero as
dictated by the spin and statistics assignments of Yukawa theory, QED or
QCD.

The initial mass parameters $m^{(0)}_\epsilon$ and $\mu^{(0)}_\epsilon$
do not include effects of any interactions and are independent of the
interaction strength.  We may have to consider limits where the mass
parameters are close to zero, in comparison to all other quantities of
relevance to physics.  For example, $\mu^{(0)}_\epsilon$ may be the
infrared regulator mass denoted by $\mu_\delta$.  Recall that the
subscript $\epsilon$ indicates that the mass parameters stand in the
hamiltonian with $\lambda = \infty$.

The initial hamiltonian contains an interaction part, ${\cal G}^{(0)}_2
= H^{(0)}_\epsilon - {\cal G}^{(0)}_1$.  For example, electrons may emit
photons.  One writes the corresponding interaction term in QED in the
form

$$ \sum_{\sigma_1 \sigma_2 \sigma'_1} \int [k_1][k_2][k'_1] 16\pi^3
\delta^3(k_1+k_2-k'_1) {\bar u}_{m^{(0)}_\epsilon k_1 \sigma_1} e
\not\!\varepsilon^*_{k_2 \sigma_2} u_{m^{(0)}_\epsilon k'_1 \sigma'_1}
b^\dagger_{k_1 \sigma_1} a^\dagger_{k_2 \sigma_2} b_{k'_1 \sigma'_1} .
\eqno (2.42) $$

\noindent We use here conventions to be specified shortly. The
hamiltonian term (2.42) is contained in the expression

$$ h = \int dx^- \,d^2x^\perp \,\, \left[e \bar \psi_{m ^{(0)}
_\epsilon}(x) \not\!\!  A (x) \psi_{m^{(0)}_\epsilon}(x) \right]_{x^+=0}
\eqno (2.43) $$

\noindent where the fields $\psi_{m ^{(0)} _\epsilon}(x)$ and $A^\nu
(x)$ for $x^+=0$ are defined by writing

$$ \psi_{m ^{(0)} _\epsilon}(x) = \sum_{\sigma} \int [k] \left[
u_{m^{(0)}_\epsilon k \sigma} b_{k \sigma} e^{-ikx} + v_{m ^{(0)}
_\epsilon k \sigma} d^\dagger_{k \sigma} e^{ikx} \right] \eqno (2.44) $$

\noindent and

$$ A^\nu (x) = \sum_{\sigma} \int [k] \left[ \varepsilon^\nu_{k \sigma}
a_{k \sigma} e^{-ikx} + \varepsilon^{\nu *}_{ k \sigma} a^\dagger_{k
\sigma} e^{ikx} \right] . \eqno (2.45) $$

\noindent Spinors $u_{m k \sigma}$ and $v_{m k \sigma}$ are defined by
boosting spinors for fermions at rest, $u_{m \sigma}$ and $v_{m
\sigma}$, to the momentum $k$, as if the fermion mass were $m$.  This is
done using the light-front kinematical boost representation for fermions

$$ S(m,k) = (mk^+)^{-1/2} [\Lambda_+ k^+ + \Lambda_- (m + \alpha^\perp
k^\perp)]. \eqno (2.46) $$

\noindent Namely, $u_{m k \sigma} = S(m,k) u_{m \sigma}$ and $v_{m k
\sigma} = S(m,k) v_{m \sigma}$.  Solving constraint equations for the
free fermion fields in canonical field theory amounts to using these
spinors.  The same boost operation defines the polarization vectors for
photons which are independent of the photon mass.  We have
$\varepsilon_{k \sigma} = \left( \varepsilon^+_{k \sigma} = 0,
\varepsilon^-_{k \sigma} = 2 k^\perp \varepsilon^\perp_\sigma/k^+ ,
\varepsilon^\perp_{k \sigma} = \varepsilon^\perp_\sigma \right)$.  The
spin label $\sigma$ denotes the spin projection on the z-axis.  We adopt
a number of conventions from Ref.  \cite{BL}.  It is useful to work with
the above spinors and polarization vectors because they provide insight
into the physical interpretation of the calculated matrix elements.  For
example, the spinors and polarization vectors help in tracing
cancelations which result from the current conservation (e.g. see Eq.
(3.103) etc. in the next Section).

Equation (2.43) includes 5 well known terms in addition to (2.42).  The
other 5 terms lead to emission of photons by positrons, absorption of
photons by electrons or positrons, or to transitions between
electron-positron pairs and photons.  There is no term leading to
creation of an electron-positron pair and a photon, or to annihilation
of such three particles.  This is the distinguished property of the
light-front hamiltonians:  conservation of momentum $k^+ > 0$ excludes a
possibility that the three momenta sum up to zero.

Strictly speaking, one has to limit each $k^+$ from below by a nonzero
positive lower bound in order to make sure that the three $+$-momentum
components cannot add up to zero.  This lower bound is provided by the
inverse of the quantization volume.  Our approach will ensure that this
largest of spatial scales in the theory does not need to be invoked in
the description of physical phenomena.  The regularization procedure
will cut off such small momenta long before they will have a chance to
become relevant. If high-order perturbation theory subsequently leads to
effective hamiltonians which describe universal low
momentum components in all physical states then, the
notion of a nontrivial vacuum will have to be taken
seriously into account for practical computational reasons.  A priori,
we cannot exclude this will happen.  But we postpone considerations of
such a situation until it becomes necessary in the future work.

The product $\bar \psi \not\!\!  A \psi$ denotes a sum of 6 basic
interactions.  The products of creation and annihilation operators are
ordered as indicated at the beginning of this Section.  However, Eq.
(2.43) requires additional steps before one can assign it a well defined
meaning because operators such as (2.42) can easily produce states of
infinite norm.  In the light-front hamiltonian approach, one needs to
define the individual terms such as (2.42) in order to provide meaning
to the whole combination of similar terms in Eq.  (2.43)

First of all, there are inverse powers of $k^+$ in Eq.  (2.42) and $k^+$
may be arbitrarily close to 0. For example, when $k_3^+$ and $k_1^+$ in
(2.42) are similar (and they are allowed to be arbitrarily close to each
other no matter what their own size is), the photon momentum
$k_2^+=k_3^+ - k_1^+$ is arbitrarily close to zero.  The problem is that
the photon momentum appears in the photon polarization vector in the
denominator:  $\epsilon^-_{k_2 \sigma_2} = 2 k_2^\perp
\epsilon^\perp_{\sigma_2}/k_2^+$.  Unless $k_2^\perp$ is close to zero,
the resulting emission strength will approach $\infty$ for $k_2^+
\rightarrow 0$.  Therefore, even for a very small coupling constant $e$,
the interaction can be arbitrarily strong.  There will be reasons for a
cancelation of this divergence in special circumstances.  For example,
in the tree diagrams for the S-matrix elements in QED, the cancelation
will be a consequence of the presence of more terms in the hamiltonian
and the energy conservation in physical processes.  However, for the
off-energy-shell matrix elements of the $T$-matrix, in loop diagrams, or
in bound state equations, such cancelations will not be ensured
automatically and could lead to untractable expressions.

In particular, one has to keep in mind that in the perturbative
calculation of the S-matrix it is possible to apply energy and momentum
conservation laws for incoming and outgoing particles on their
mass-shells.  In contrast, in the bound state calculations, the
individual particle momenta cannot simultaneously be on the individual
mass-shells and still sum up to the bound state momentum - the bound
state dynamics is always off-shell and perturbative mechanisms for
cancelations cease to be sufficient.

In Eq.  (2.42), the inverse powers of the longitudinal momentum also
appear in the fermion spinors.  These can be a source of divergences,
too.  However, the examples we describe in this article do not lead to
problems with infrared fermion divergences and we will not dwell on this
subject here.

Secondly, the spinor matrix elements depend on the transverse momenta of
the fermions and the boson polarization vector depends on the transverse
momentum of the boson.  The strength of the interaction grows when the
relative transverse momenta grow.  Again, even in the case of a very
small charge $e$, divergences arise and the question is if the
interaction is finite in any way.

For example, the finitness problem manifests itself clearly when one
attempts to evaluate the ratio of norms of the states $h
|k\sigma\rangle$ and $|k\sigma\rangle$.  This ratio is certainly not
finite and it is ill-defined.

Therfore, one might ask if it is useful to consider this ill-defined
hamiltonian term.  The answer is unambiguous:  yes.  The reason is that
the scattering amplitudes calculated using this term in combination with
two other terms in second order perturbation theory, agree very well
with observable scattering of electrons and photons.  No loop
integration appears in these calculations to indicate the divergence
problem.

It is well known that the terms one should put into the light-front
hamiltonian are provided by the formal lagrangian density for
electrodynamics ${\cal L} = -{1 \over 4} F^{\mu \nu} F_{\mu \nu} + \bar
\psi (i\not\!\! D - m )\psi$.  One can rewrite the lagrangian density
into a corresponding light-front hamiltonian density by using an
expression for the energy-momentum tensor density $T^{\mu\nu}$.
Integrating $T^{+-}$ over the light-front hyperplane gives the
expression one starts from in building the light-front hamiltonian for
QED.

The initial hamiltonian $H^{(0)}_\epsilon$ for QED results from formal
operations on fields $\psi_+$ and $A^\perp$.  \cite{Y} One uses the
gauge $A^+=0$ and solves the constraint equations, substitutes
expansions of the form (2.44) and (2.45) into the formal expression for
$T^{+-}$, integrates the density over the light-front hyperplane and
normal-orders all terms.  The normal-ordering produces terms that
involve numerically divergent momentum integrals.  The classical field
theory does not tell us what to do with the divergences resulting from
the ordering of operators.

To deal with all divergences one has to regularize the theory from
the outset.  The naive connection between the classical theory and the
quantum theory as given by the quantization rules is broken.  The
regularization turns out to introduce large terms into the initial
hamiltonian.  To gain control on the regularization effects one has to
construct a renormalization theory for hamiltonians.  The diverging
terms which result from normal ordering can be safely dropped in the
form they appear ill-defined in the canonical approach, because the
renormalization procedure introduces other terms of the same operator
structure to replace them.

Reasons to adopt the perturbatively and semi-classically motivated
strategy of building the initial hamiltonian from a classical
largrangian are less clear in the Yukawa theory and QCD than in QED.

Application of the pseudoscalar Yukawa interaction in nuclear physics
suggests that the coupling constant for nucleons emitting pions is order
10.  Perturbation theory is questionable.  However, it is important to
realize that the vertices for pions and nucleons which are used in the
relativistic nuclear physics involve pion-nucleon vertex form factors.
The vertex form factors are analogous to the similarity functions
$f_\lambda$.  Thus, the well-known pion-nucleon interaction could be
identified with an interaction term in an effective hamiltonian of an
unknown original theory rather than the initial bare hamiltonian term.

Following this observation, one can try to calculate an effective
hamiltonian in a pseudoscalar Yukawa theory for $\lambda$ on the order
of the cutoff parameters known in nuclear physics and ask what terms and
what sizes of the coupling constants would have to appear in that
hamiltonian to match experimental data, if it is possible.  \cite{BGP}
The narrow similarity vertex factors can, in principle, weaken the
effective strength of the interaction by reducing its width in momentum
space to a very small range so that the coupling constant itself being
order 10 may be too small to invalidate perturbation theory in the whole
interaction.  Then, one could study meson-barion physics using
perturbation theory with such narrow effective hamiltonians.  The goal
would be to correlate the vertex form factors parameters with the
coupling strength and masses of mesons and barions which are used in the
effective theory.

Thus, it becomes interesting to see what kind of effective hamiltonians
result from an abstract initial $H^{(0)}_\epsilon$ obtained from the
Yukawa lagrangian with a very small coupling constant or from similar
field-theoretic lagrangians, such as for $\sigma$-models.  For one could
see this way examples of the small width effective hamiltonian
structures implied by the local field theory and one could study
off-shell mechanisms which may lead to covariant results in the
renormalized hamiltonian approach.  The size of the coupling is of
secondary importance from that point of view.  Moreover, the range of
converegence of expansion in the coupling constant in the effective
theory is not known and it should be studied because it may be large in
the small momentum width effective hamiltonians.

One should also keep in mind that the effective hamiltonians may have a
number of universal chracteristics which are independent of details of
the initial theory.  For example, there exists some resemblance between
effective low-energy interactions in theories with pseudo-scalar direct
and derivative couplings between bosons and fermions.

In QCD, one proceeds by analogy with QED assuming that the constituent
quark picture of hadrons approximates the solution to QCD. \cite{W3}

On the one hand, confinement invalidates close analogy with QED.  On the
other hand, asymptotic freedom makes the limit of bare cutoffs being
removed more plausible than in QED.  One hopes that the strength of
effective interactions never grows too large when the scale $\lambda$ is
lowered and the effective hamiltonian for the constituent quark picture
for hadrons may be obtained.  Again, the important feature of the
similarity factors is that the strength of the interactions is not given
solely by the size of the coupling constant.  The width of the function
$f_\lambda$ is also important.  The smaller is the width the smaller is
the strength of the interactions.  Therefore, the running coupling
constant may still lie within the perturbative domain for the evaluation
of the effective hamiltonians of reasonably small widths despite the
coupling constant growth when the width gets smaller.

To explain the regularization for light-front hamiltonians which we apply to
expressions resulting from field theoretic lagrangian densities, we
first discuss the term (2.42).  In that term, the parent momentum $P$
equals $k'_1$.  The spinors and polarization vectors conveniently group
a number of terms with different momentum dependences into a combination
which is invariant under the light-front kinematical symmetry
transformations.  Among those terms, there are terms containing masses,
terms which in field theory result from derivatives 
$i\partial^\perp$ or $i\partial^+$,
or from inverting the operator $i \partial^+$.  All those derivatives
are replaced in the term (2.42) by momenta of particles created or
destroyed by that term.

We first introduce the daughter momentum variables for the created
electron and photon.  We have briefly introduced daughter momentum
variables in a similar configuration in Section 2.a when defining the
similarity functions $f_\lambda$.  Here, we use the daughter momentum
variables for the purpose of regularization.  The variables are

$$  x_1 = k_1^+/{k'}^+_1 = x , \eqno (2.47.a)  $$

$$  x_2 = k_2^+/{k'}^+_1 = 1-x , \eqno (2.47.b)  $$

$$  x'_1 = {k'}^+_1 / {k'}^+_1 = 1, \eqno (2.47.c) $$

$$ \kappa^\perp_1 = k_1^\perp - x_1 P^\perp = \kappa^\perp, \eqno
(2.47.d) $$

$$ \kappa^\perp_2 = k_2^\perp - x_2 P^\perp = - \kappa^\perp , \eqno
(2.47.e) $$

$$ {\kappa'}^{\perp}_1 = {k'}_1^\perp - x'_1 P^\perp = 0 . \eqno
(2.47.f) $$

\noindent For each of the creation or annihilation operators in
the interaction term (2.42), we define a {\it daughter energy}
variable. Namely,

$$ e_1 = { \kappa_1^{\perp 2} + m^{(0) \,2}_\epsilon \over x_1} =
         {\kappa^{\perp 2} + m^{(0)\,2}
         _\epsilon \over x}, \eqno (2.48.a) $$

$$ e_2 = { \kappa_2^{\perp 2} + \mu^{(0)\,2}_\epsilon \over x_2} =
         {\kappa^{\perp 2} + \mu^{(0)\,2}
         _\epsilon \over 1-x}, \eqno (2.48.b) $$

$$ e'_1 = { {\kappa'}_1^{\perp 2} + m^{(0)\,2}_\epsilon \over x'_1} =
          m^{(0)\,2}_\epsilon . \eqno (2.48.c) $$

\noindent For each creation and annihilation operator in the interaction
term (2.42), we introduce a factor which is a function, $r(y_i)$, of the
variable $y_i = \epsilon e_i/\Lambda^2$, where the subscript $i$ denotes
the operator in question.  In the no cutoff limit, $\epsilon \rightarrow
0$.  $\Lambda$ is an arbitrary constant with dimension of a mass ($\hbar
= c = 1$).  All masses and momenta are measured in units of $\Lambda$.
In this article, we choose $r(y) = (1+y)^{-1}$.  Thus, the term (2.42)
is regulated by introducing the factor

$$ (1+ \epsilon e_1/\Lambda^2)^{-1}  (1+ \epsilon e_2/\Lambda^2)^{-1}
(1+ \epsilon e_3/\Lambda^2)^{-1} \eqno (2.49) $$

\noindent under the integral.  The third factor in the above expression
can be replaced by 1, since $m^{(0)}_\epsilon$ is a finite constant and
it cannot compensate the smallness of $\epsilon$.  We shall make such
replacements wherever the parent momentum is carried by a single
creation or annihilation operator.

An interesting feature arises in this regularization scheme.  Namely, if
there is no infrared divergences due to the small values of $x$, one can
remove masses from the regularization factors.  It is the small $x$
divergences that invite keeping the masses in the daughter energy so
that the latter grows towards small $x$ even for the zero transverse
momenta.

In the case of terms which contain only 1 creation and 1 annihilation
operator, i.e. in ${\cal G}^{(0)}_1$, no regularization is introduced.
This is important because any restrictions on the particle momenta in
these terms would violate kinematical symmetries of the light-front
hamiltonian dynamics.  The reason is that the particle momenta in these
terms are equal to the parent momenta and limiting the parent momenta
violates the light-front boost invariance.

In the initial expressions for hamiltonian densities of Yukawa theory,
QED or QCD, only terms with products of up to four fields appear.
Therefore, we have only two more situations to consider in addition to
the cases such as ${\cal G}^{(0)}_1$ and terms of the type (2.42).  In
the first situation we have three creation operators and one
annihilation operator or vice versa, and in the second situation we
have two creation and two annihilation operators.

Independently of the number of creation and annihilation operators in a
product, the regularization is introduced by multiplying every creation
and annihilation operator in the product by a function $r(y)$ such as in
the factor (2.49), where $y=\epsilon e_d /\Lambda^2$ and $e_d$ is
the corresponding daughter energy variable.  Later, after counterterms
are calculated, the same regularization factors are introduced in the
counterterms.

An additional step is required in the case of hamiltonian terms which
originate from the products of four fields including inverse powers of
$i\partial^+$ acting on a product of two fields.  This additional step
will be described now.

In the terms containing the products of four fields and inverse powers
of $i\partial^+$, we introduce two kinds of a fifth daughter momentum
and two corresponding daughter energy variables, $e_{512}$ and
$e_{534}$.  The numbering originates from assigning numbers to the
fields in the product according to the schematic notation $\phi_1 \phi_2
(i\partial^+)^{-n} \phi_3 \phi_4$.  One of the fifth daughter energy
variables is associated with the operators coming from the fields number
1 and 2, and the other one is associated with the operators coming from
the fields number 3 and 4. The regularized terms will contain an
additional product of functions $r(y_{512})$ and $r(y_{534})$ with the
arguments $y_{512} = \epsilon e_{512}/\Lambda^2$ and $y_{534} = \epsilon
e_{534}/\Lambda^2$.

The auxiliary daughter energy variables $e_{512}$ and $e_{534}$ are
calculated as if they represented daughter energy variables for an
intermediate particle, a boson or a fermion, created and annihilated in
the vertices which contained the products $\phi_1 \phi_2$ and $\phi_3
\phi_4$, respectively.  Those vertices are treated as if each of them
contained three fields instead of two but the field of the intermediate
particle was contracted so that the corresponding creation operator and
the corresponding annihilation operator are absent in the resulting
term.  This particular definition of a gedanken intermediate particle
does not refer to any particular Fock states and remains valid when the
operators $q_\infty$ are replaced with $q_\lambda$ by the unitary
transformation $U_\lambda$.  The definition was inspired by Refs.
\cite{Y} and \cite{BRS} where the correspondence between the
intermediate states with backward moving particles with spin in the
infinite momentum frame and the light-front seagull interaction terms is
extensively described.

Mathematically, the definition of $e_{512}$ and $e_{534}$ is introduced
in the following way.  Every creation and annihilation operator in the
fields $\phi_1$, $\phi_2$, $\phi_3$ and $\phi_4$ is assigned a
corresponding number $s_i$, i=1,2,3,4.  $s_i$ equals $+1$ for a creation
operator and $s_i$ equals $-1$ for an annihilation operator.  We define
$k^+_5 = |s_3 k_3^+ + s_4 k_4^+|$ and $s_5 = (-s_3 k_3^+ - s_4
k_4^+)/k_5^+$.  The gedanken particle is thought to be created in the
product of fields including $\phi_3 \phi_4$ when $s_5 = 1$ and it is
thought to be annihilated in that product when $s_5 = -1$.  We define
the momentum $k_5= (k_5^+, k_5^\perp)$ by the relation $s_5 k_5 = - s_3
k_3 - s_4 k_4 = s_1 k_1 + s_2 k_2$.  We also introduce two auxiliary
parent momenta, $P_{34} = {1 \over 2} (k_5 + k_3 + k_4)$ and $P_{12} =
{1 \over 2} (k_5 + k_1 + k_2)$.  Then, we introduce the daughter
momentum and energy variables

$$ x_{512} = k^+_5 /P^+_{12}   , \eqno (2.50.a) $$

$$ \kappa^\perp_{512} = k^\perp_5 - x_{512} P^\perp_{12},
   \eqno (2.50.b) $$

$$ e_{512} = {\kappa^{\perp 2}_{512} + m^{(0)\,2}_{\epsilon 5} \over
x_{512} }, \eqno (2.50.c) $$

$$ x_{534} = k^+_5 /P^+_{34}    , \eqno (2.50.d) $$

$$ \kappa^\perp_{534} = k^\perp_5 - x_{534} P^\perp_{34} ,
   \eqno (2.50.e) $$

$$ e_{534} = {\kappa^{\perp 2}_{534} + m^{(0)\,2}_{\epsilon 5} \over
x_{534} } , \eqno (2.50.f) $$

\noindent where $m^{(0)}_{\epsilon 5}$ equals $m^{(0)}_\epsilon$ for
regularization of the terms involving $(i\partial^+)^{-1}$ and
$m^{(0)}_{\epsilon 5}$ equals $\mu^{(0)}_\epsilon$ for regularization of
the terms involving $(i\partial^+)^{-2}$.  This step completes our
definition of the ultraviolet regularization of initial hamiltonians.

We proceed to the definition of the infrared regularization.  Inverse
powers of $i\partial^+$ for massive particles are already regulated when
the ultraviolet regularization is imposed.  This was explained above.

For each creation and annihilation operator of an initially massless
particle, we introduce a factor which limits the daughter momentum
fraction $x$ for that operator to be greater than $\delta$.  An example
of such a factor is given by $(1 +\delta/x)^{-1}$.  Note that our
definition also implies that the same regularization factor is inserted
for the gedanken particles with $x_{512}$ defined in Eq.  (2.50.a) and
$x_{534}$ defined in Eq.  (2.50.d).

Our procedure of limiting the longitudinal momentum fractions in the
interaction terms preserves the light-front boost invariance explicitly
but may lead to problems in the renormalization procedure.  The problems
may arise when one expands various functions in powers of momentum using
typical light-front boost invariant combinations of relative momentum
variables and the expansion leads to new strongly
divergent terms, so that the corresponding chain of counterterms will be too
hard to calculate and the effective hamiltonians will not converge.
\cite{Wpriv} The argument is not iron-clad:  it is not proven that such
a situation has to arise.

We suggest that the way out of the potential problem can be looked for
in the expansions around nonzero values of $x$ so that terms such as
$\kappa^2/x$ will not combine transverse singularities at large $\kappa$
with longitudinal singularities at small $x$.  At this stage, we can
only assume that the expansions around the right values of the
longitudinal mometum fractions, such as $x = 1/2$ for the quarks in
heavy quarkonia or electrons in positronium or, such as $x = \alpha$ for
the gluons or photons, remove the danger of creating new divergences in
the high order perturbation theory.

On the other hand, when one limits $p^+$ from below by $\delta^+$,
instead of the fractions, to exclude appearance of the complicated
boost invariant combinations then, the boost invariance is not
explicitly preserved.  But we have to remember that the renormalization
procedure cannot potentially be complete without specifying
renormalization conditions.  Theoretical calculations of observables
which can provide the renormalization conditions require solving for the
eigenvalues of the hamiltonian.  This is an intrinsically complicated
nonperturbative step.  The approximations which are necessary in the
calculation will most probably violate boost invariance and this may
exclude predictions for the boost invariant quantities of sufficient
accuracy to define the renormalization conditions.

Fortunately, one can preserve the explicit light-front boost invariance
throughout the calculation of the effective interactions by using the
limits on the momentum fractions.  In this case, the renormalization
conditions play no role in obtaining the light-front symmetry because
the effective hamiltonians are explicitly symmetric.  One can also make
the necessary approximations in the diagonalization procedures without
violating the symmetries.  Thus, the exactly symmetric renormalization
conditions can be imposed.

Besides introducing the cutoff $\delta$ on the momentum fractions
carried by massless particles, we can also introduce for each initially
massless particle a finite regularization mass term which is denoted by
$m_\delta$.  In other words, in the case of the initially massless
particles, $m^{(0)}_\epsilon = m_\delta$.  Such finite masses
in the daughter energies lead to additional suppression of the 
infrared longitudinal momentum region. The reason for allowing 
the additional mass terms follows from anticipation of counterterms.

The relative momentum cutoffs lead to the counterterms which have unknown
finite parts.  In particular, the mass counterterms for the initially
massless particles contain unknown finite parts.  Since, at this point,
the finite parts are arbitrary and they are not known to be zero, we
introduce the mass terms and investigate their r\^ole.

However, the introduction of masses for photons or gluons in ${\cal
G}_1$ brings the masses into the denominators of perturbation theory for
effective hamiltonians and leads to additional singularities which do
not appear in the case of massless particles.  Therefore, additional
modifications may be required to remove such diverging effects of
the masses.  For example, the instantaneous vector boson exchange
interaction may be multiplied by the factor $\kappa^2/(\kappa^2 +
\mu^2_\delta)$ with a suitably defined $\kappa^2$.  The operational
definition for such a term involving four fields (two creation operators
and two annihilation operators or, three creation operators and one
annihilation operator or, one creation operator and three annihilation
operators) labeled by momenta $k_i$, $i = 1, 2, 3, 4$, is to calculate
$k_5$ using definitions introduced for Eqs.  (2.50.a) to (2.50.f) and
put $\kappa$ equal to $|k_5^\perp|$ with the parent transverse
momentum for the whole interaction term, $P^\perp = {1\over 2} (k_1
^\perp + k_2^\perp + k_3^\perp + k_4^\perp)$, set to 0.

If the mass parameters are treated as small perturbations, one does not
include them in ${\cal G}_1$.  Then, they do not modify energy
denominators in perturbation theory and can hopefully be taken into
account successively in a power expansion.  \cite{Wpriv}

In a close analogy to the mass parameter, one can also introduce a lower
bound on the transverse momenta.  This alternative has the advantage of
not modifying the free energy term and, therefore, does not introduce
additional divergences in lowest order calculations.  This option is not
further discussed in the present article.

The infrared regularization completes our definition of the initial
hamiltonians.

\vskip.3in
{\bf 2.c Renormalization conditions}
\vskip.1in

The free finite parts of the counterterms are determined by
renormalization conditions.  The renormalization conditions result from
the comparison of theoretical predictions with data.  However, the
calculations of observables require solutions to the bound state or
scattering problems for the renormalized hamiltonians.  In principle,
one could work with hamiltonians of any width $\lambda$.  In practice,
one is limited to consider some subspaces in the Fock space.  Therefore,
the issue of setting renormalization conditions is subtle.

Various approximations may influence the fit to data and it may be
necessary to scan a large number of options.  If a fixed point is
discovered and the bare cutoff can be made sufficiently large, one can
pursue a search for a full set of fundamental parameters that
characterize a theory.  But we will not achieve this kind of quality in
this paper.  Only second order perturbative examples are analysed here.

In theories without confinement, one has an option of defining
on-mass-shell renormalization conditions for single particles.  In the
present formalism, it means that one can determine free parameters in
the effective hamiltonians by demanding that single particle eigenstates
and the $S$-matrix have the required properties.  The key examples to be
discussed in detail here are nucleons and pions in Yukawa theory
(pseudoscalar coupling) and electrons and photons in QED.  The on-shell
scattering amplitudes can be used for direct determination of the
coupling constants, but in the second order calculations described in
this paper no coupling constant renormalization appears because it 
requires higher orders of the perturbation theory.

When one wants to include confinement, there is a trouble with deciding
how to choose the values of masses for colored particles.  Quarks and
gluons are not supposed to be observable.  Should or should not they
have diverging infrared cutoff dependent mass terms in the initial
hamiltonian?  We do not have any definite answer to this question today.

For the confined particles, one can adopt the strategy of fitting
observables for their bound states (masses, form factors, decay rates,
scattering cross sections).  This is a very demanding strategy which
must include solutions to a number of problems that are not solved yet.
For example, in order to obtain a rotationally invariant spectrum for
the bound states one has to explain how the apparent rotational symmetry
violations in the hamiltonian structure off the energy-shell combine to
produce a degenerate mass spectrum once a suitable choice of the finite
parts of the counterterms is made.  Unfortunately, this problem is not
solved even in theories without confinement, the chief example being
QED.  Another condition is the demand that a bound state current matrix
element has some definite structure which is required by the S-matrix
theory for electroweak external probes.  In fact, the present formalism
can be described as an attempt to provide a framework in which these
types of problems can be sufficiently well defined to seek their
solutions.

The key question we have to answer in practice is how many effective
particles have to be taken into account in the eigenvalue problem for an
effective hamiltonian and how many can be included in a real
calculation.  A good example of a theoretical problem one can think of
is how the momentum of a proton is shared by its constituents.
The phenomenology of deep inelastic scattering of leptons on
nucleons suggests that a considerable number 
even at moderate momentum transfers.  If the number of
effective constituents has to be large, one may
encounter ambiguities in the determination of
free parts of counterterms because the proton observables will be
calculable only through complicated procedures.  On the other hand, the
constituent quark model suggests that the large number of constituents
is not necessary for an approximate description of the proton
properties.  Clearly, we do not know yet how hard it is to find the
finite parts of the counterterms using bound state observables but we
cannot exclude that it is doable.

${\cal G}_\lambda$ in Eq.  (2.35) contains the free finite parts of the
counterterms in the term ${\cal G}_{\epsilon\,finite}(\lambda_0)$.  The
hamiltonian $H_{\lambda_0}$ can be used for calculating the scattering
amplitudes and bound state properties.  The most familiar example is
QED.  It is almost purely perturbative.  One can calculate the physical
electron energy as defined by the lowest eigenvalue of the effective
hamiltonian for eigenstates with the electron quantum numbers.  Thanks to
the symmetries of the light-front dynamics, the eigenvalue will have the
form $(p^{\perp 2} + m^2_e)/p^+$ and $m_e$ will have to be equal to the
physical electron mass.  Note also that the effective mass term for the
interacting photons will have to be different from zero (and growing
with $\lambda$) in order to obtain massless photon eigenstates.
Examples of the renormalization conditions for QED will be presented in
the next Section but no discussion of effective scattering theory beyond 
the second order perturbative results is offered in the present paper.

When calculating the unknown renormalization parameters, one can also
take advantage of another aspect of the present theory.  The spectrum of
physical masses will consist of multiplets, as implied by rotational
symmetry, if the hamiltonian belongs to a renormalized algebra of
Poincar\'e generators.  The Poincar{\'e} algebra commutation relations
should be studied order by order in perturbation theory to find out if
they impose constraints on the free parameters.  The initial algebra 
is spoiled only at the scale of the
regularization cutoff.  The effective algebra can be satisfied if the
dependence on the cutoff scale is removed in the renormalization
process.

The general structure of the similarity transformation which is defined
in terms of the creation and annihilation operators, provides a natural
way to extend the renormalization procedure to the whole Poincar\'e algebra.  
The renormalization group evolution of all generators is given by Eq.
(2.5).  The evolution preserves commutation relations.  Therefore, the
effective generators satisfy the same algebra as the initial ones.  

The renormalization conditions which are related to symmetries can be
imposed using commutation relations among the symmetry generators
without direct evaluation of physical quantities.  The
renormalization of the whole Poincar\'e algebra or other symmetry
generators is not further analysed in this article.

\vskip.3in
{\bf 3. EXAMPLES OF APPLICATION}
\vskip.1in

This Section describes a set of examples of lowest order calculations of
renormalized effective hamiltonians in the light-front Fock space using
the scheme from Section 2. We begin by the description of generic rules
for calculating the right-hand sides of the renormalization group
equations, i.e. Eq. (2.29).  The rules follow from the commutator
structure. Then, we discuss examples from Yukawa theory, QED and QCD.
\nopagebreak
\vskip.3in
{\bf 3.a Evaluation of commutators}
\vskip.1in
\nopagebreak

The right-hand sides of Eqs.  (2.29) and (2.38) are commutators.  This
implies that the interactions which appear in the effective renormalized
hamiltonians $H_\lambda(q_\lambda)$ and in the counterterms in
$H_\epsilon$, are connected.  This Section explains how this result
comes about.

The commutators can be evaluated in a number of equivalent ways but
some are more convenient than others. Suppose we are to evaluate

$$      {\hat H} = [ {\hat A}, \{ {\hat B} \}_{\hat C} ]   . \eqno (3.1)$$

\noindent
${\hat A} = A(X,Y) \prod_{i=1}^{I_A}
a^\dagger_{x_i}\prod_{j=1}^{J_A} a_{y_j}$,
${\hat B} = B(V, W) \prod_{k=1}^{I_B}
a^\dagger_{v_k}\prod_{l=1}^{J_B} a_{w_l}$
and ${\hat C} = \sum_z E(z) a^\dagger_z a_z$.
The right-hand side of Eq. (3.1) equals

$$ {\hat H} =
    A(X,Y)\prod_{i=1}^{I_A} a^\dagger_{x_i} \prod_{j=1}^{J_A} a_{y_j}
   {B(V,W)\over E_w - E_v}
          \prod_{k=1}^{I_B} a^\dagger_{v_k}\prod_{l=1}^{J_B} a_{w_l}$$
$$-{B(V,W)\over E_w - E_v}
          \prod_{k=1}^{I_B} a^\dagger_{v_k}\prod_{l=1}^{J_B} a_{w_l}
    A(X,Y)\prod_{i=1}^{I_A} a^\dagger_{x_i}\prod_{j=1}^{J_A} a_{y_j}
 , \eqno(3.2) $$

\noindent where $E_w = \sum_{l=1}^{J_B} E(w_l)$ and $E_v =
\sum_{k=1}^{I_B} E(v_k)$.  By commuting $\prod_{j=1}^{J_A} a_{y_j}$ in
the first term through $\prod_{k=1}^{I_B} a^\dagger_{v_k}$ one generates
the contracted terms with a number of contractions ranging from 1 to the
smaller of the numbers $J_A$ and $I_B$, and a term with
$\prod_{j=1}^{J_A} a_{y_j}$ standing to the right of $\prod_{k=1}^{I_B}
a^\dagger_{v_k}$.  Then, by commuting $\prod_{l=1}^{J_B} a_{w_l}$ in the
latter term through $\prod_{i=1}^{I_A} a^\dagger_{x_i}$, one obtains new
contracted terms with the number of contractions ranging from 1 to the
smaller of the numbers $I_A$ and $J_B$, and a term equal to the second
term in Eq.  (3.2) with an opposite sign which thereby is canceled out
leaving only connected terms in the result for $\hat H$.  The same
result holds for the boson and fermion creation and annihilation
operators independently of their commutation relations because each kind
of the operators is commuted an even number of times.

After the second term in Eq.  (3.2) is canceled one is left with a
number of partially contracted terms in which annihilation operators may
still stand to the left of creation operators. A number of ordering
transpositions need to be done before a generic ordering of operators
adopted in the previous Section is achieved.  In fact, the process of
commuting factors in $\hat A$ through factors in $\{ {\hat B} \}_{\hat
C}$ in the first term on the right-hand side of Eq.  (3.2) produced
above a number of terms with creation operators moved to the right of
annihilation operators unnecessarily. These transpositions have to be
undone to recover final asnwers with the adopted ordering.
Nevertheless, it is visible that disconnected terms cannot appear and
the following rule simplifies the calculations.

The right-hand side of Eq.  (3.2) equals the sum of the contracted terms
which result from ${\hat A} \{ {\hat B} \}_{\hat C}$ by moving
$\prod_{j=1}^{J_A} a_{y_j}$ through $\prod_{k=1}^{I_B} a^\dagger_{v_k}$
and, the contracted terms which result from $-\{ {\hat B} \}_{\hat C}
{\hat A}$ by moving $\prod_{l=1}^{J_B} a_{w_l}$ through
$\prod_{i=1}^{I_A} a^\dagger_{x_i}$.  All other terms cancel out.

\vskip.3in
{\bf 3.b Yukawa theory}
\vskip.1in
\nopagebreak
The standard procedure from Ref.  \cite{Y} leads from the lagrangian
density ${\cal L}_Y= {\bar \psi} (i\not\!\partial - m - g\phi)\psi +
{1\over 2}(\partial^\mu \phi \partial_\mu \phi - \mu^2 \phi^2)$ to the
light-front hamiltonian expression of the form

$$ H_Y = \int dx^- d^2 x^\perp \left[ \bar \psi_m \gamma^+
{-\partial^{\perp 2} +
m^2 \over 2 i\partial^+} \psi_m +
{1\over 2}\phi (-\partial^{\perp 2} + \mu^2 )\phi \right.$$
$$ \left. + g \bar \psi_m \psi_m \phi + g^2\bar \psi_m \phi {\gamma^+
\over 2i\partial^+}\phi \psi_m \right]_{x^+=0}. \eqno(3.3) $$

\noindent We replace fields $\psi_m(x)$ and $\phi(x)$ for $x^+=0$ by the
Fourier superpositions of creation and annihilation operators, order the
operators in all terms and drop the terms containing divergent integrals
which result from the contractions.  Then, we introduce the
regularization factors.

In the course of calculating effective hamiltonians we will also add new
terms to $H_Y$ which ought to be included due to the presence of the
regularization, in accord with the renormalization theory from the
previous Section.  For example, we will add a small term $\delta
m^2_\epsilon = m^2_\epsilon - m^2$ to $m^2$ in the first term and
$\delta \mu^2_\epsilon = \mu^2_\epsilon - \mu^2 $ to $\mu^2$ in the
second term.  We will calculate these additional terms below using the
renormalization theory to order $g^2$.  More additional terms can be
calculated when terms with higher powers of the coupling constant $g$
than second are included in the calculation of the renormalized
effective hamiltonians.  But the corresponding calculations go beyond
the scope of the present illustration of the theory and will not be
pursued here.

In order to obtain the theory for particles with the quantum numbers of
nucleons and pions one needs to include the isospin and replace the
scalar coupling by $i\gamma_5$.  \cite{BGP} However, for the purpose of
the illustration of the renormalization procedure to second order in the
coupling $g$, we do not have to introduce these factors explicitly.  The
additional factors merely lead to somewhat different algebra which can
be traced throughout the whole calculation and final results including
the isospin and $i\gamma_5$ can be read from the results in the Yukawa
theory.  In this Section we assume $m > \mu > 0 $.

The lagrangians with chiral symmetry would require massless fermions and
additional meson fields, such as $\sigma$.  When the spontaneous
symmetry breaking is considered in the light-front hamiltonian approach,
one has to define the hamiltonian including the symmetry violating
terms.  A discussion of scalar theories with symmetry breaking, in the
context of the hamiltonian renormalization theory, is given in Ref.
\cite{W3} (see also Ref.  \cite {SUS}).

\vskip.3in
\centerline {\bf Meson mass squared}
\vskip.1in

The simplest example of a second order expression for a term in an
effective hamiltonian in the Yukawa theory is provided by the meson mass
squared.  We first describe steps which produce this expression.  The
number of distinct steps in the procedure is 10:  defining the
regularized initial hamiltonian, calculation of the effective
hamiltonian, analysis of the cutoff dependence of finite matrix elements
of the calculated terms and extraction of the structure of the
divergence, evaluation of the counterterm, isolation of the finite part,
calculation of the effective hamiltonian knowing the structure of the
counterterm, solving a physical problem such as an eigenvalue problem or
a scattering problem, adjusting the finite part of the counterterm to
match data (including adjustements for the observed symmetries), and
computing the final expression for the effective hamiltonian with
the finite part determined from the fit to data.

The simplest example is described in full detail of all the steps.  Such
extensive presentation is not provided in later examples where more
complicated expressions would require too much space for such extended
presentation.  The first example is discussed in such detail despite the
fact that in this case one can also proceed without the explicit
discussion of all the steps and still obtain the correct result.
Correct means here in agreement with the similarity renormalization
group for hamiltonians.

For example, one might propose the structure of the counterterm using,
as is usually done, some scattering amplitude instead of the matrix
elements of an effective hamiltonian.  Note that one can also impose
renormalization condition using a scattering amplitude which results
from a calculation performed without use of the effective hamiltonian.

However, the systematic approach from Section 2 is the only tool we have
for dealing with more complicated cases of light-front hamiltonians and
their eigenvalue equations.  In other words, the simplest available case
is used to present all the steps in detail because it illustrates the
procedure in a familiar setting.  Then, we proceed to the expressions
for fermion masses and interactions between fermions.  Details are then
discussed only in cases where a new feature appears in the calculation.

Equation (2.36) implies to second order in ${\cal G}_{2\lambda}$ that

$$ {d \over d\lambda} {\cal G}_{1\lambda} =
\left[  {\cal G}_{12\lambda}
{{d \over d\lambda} f^2(z^2_\lambda) \over  {\cal
G}_{1\lambda}
- E_{1\lambda} } {\cal G}_{21\lambda} \right]_{11}
 + \sum_{p=3}^\infty \left[  {\cal G}_{1p\lambda}
{{d \over d\lambda} f^2(z^2_\lambda) \over  {\cal G}_{1\lambda}
- E_{1\lambda} } {\cal G}_{p1\lambda} \right]_{11} ,   \eqno(3.4) $$

\noindent where the double-digit subscripts refer to the number of
creation and annihilation operators (in that order) and the bracket
subscript denotes the part which contributes to the rate of change of
${\cal G}_{1\lambda}$ with $\lambda$.  $E_{1\lambda}$ is the eigenvalue
of ${\cal G}_{1\lambda}$ which corresponds to the creation and
annihilation operators indicated by the subscript 11.  The reason for
that only one free energy eigenvalue appears in the denominators is that
${\cal G}_{1\lambda}$ of Eq.  (3.4) is a one-body operator and quantum
numbers which label creation and annihilation operators in ${\cal
G}_{1\lambda}$ are the same, including momentum.  Therefore, the
free energy eigenvalues are also the same: both are equal to $E_{1\lambda}$.
Consequently, all commutators are written on the right-hand side of Eq.
(3.4) in the simplified form. The numerator similarity factors reduce to
the single derivatives of $f^2_\lambda$ (we have chosen $n=1$ in Eq.
(2.19)).  Terms with more than two intermediate particles ($p \geq 3$)
are of order $g^4$ or higher.

Assuming that $g$ in Eq.  (3.3) is extremely small, writing ${\cal
G}_{1\lambda}$ as a series in powers of $g$ and keeping only terms order
$g^2$, we obtain the following result from Eq. (3.4) for the meson free
energy term.

$$  {\cal G}_{1\,meson\,\lambda} =  \int[k] { k^{\perp 2} +
\mu^2_\lambda \over k^+ } a^\dagger_k a_k . \eqno (3.5) $$

\noindent A remarkable feature in this result is that no correction
arises to the term $k^{\perp 2} /k^+$ which is protected by the
kinematical symmetries of light-front frame.  Simply, the total
transverse momentum does not appear in a boost invariant expression.

The width dependence of $\mu_\lambda$ is determined by the equation

$$ {d \mu^2_\lambda \over d\lambda} = g^2 \int[x\kappa]
{d f^2(z^2_\lambda) \over d\lambda}
{ 8(x-{1\over 2})^2{\cal M}^2 \over {\cal M}^2 -\mu^2 }
r_\epsilon(x,\kappa) , \eqno (3.6) $$

\noindent where ${\cal M}^2 = (\kappa^2 + m^2) / x(1-x) $. $m^2$ and
$\mu^2$ are the original bare mass squared parameters from Eq.  (3.3).
They do not include terms order $g^2$ and higher because such terms
would lead to higher order corrections than $g^2$ for the whole
expression.  The terms order $g^2$ and higher are treated as
interactions in the perturbative calculation.

In terms of graphs for the effective hamiltonian calculus, Eq.  (3.6)
represents the contribution of a fermion loop on a meson line.  However,
the graphs are not provided in order to avoid confusion with other
diagrammatic techniques.

$$   \int [x\kappa] = (16\pi^3)^{-1} \int_0^1 {dx \over
x(1-x)} \int d^2\kappa^\perp . \eqno (3.7) $$

\noindent Using Eq. (2.20) in the limit of a $\theta$-function,
$f(u) = \theta(u_0 - u)$, one obtains

$$ f^2(z^2_\lambda) = \theta\left[ \lambda^2 + {1+\sqrt{u_0} \over
\sqrt{u_0}}\mu^2 - {1-\sqrt{u_0} \over \sqrt{u_0}} {\cal M}^2 \right].
\eqno (3.8.a) $$

\noindent For example, for $u_0 = {1 \over 4}$ one has $f^2(z^2_\lambda)
= \theta[\lambda^2 + 3\mu^2 - {\cal M}^2]$.  Therefore, the derivative
of $f_\lambda$ with respect to $\lambda$ forces the invariant mass of
the fermion-antifermion pair, ${\cal M}^2$, to be equal $\lambda^2 +
3\mu^2$.  In general, the derivative selects the range of energies in
the integral where the similarity function changes most rapidly.  The
regions where the function approaches a constant, i.e. 1 near the
diagonal and 0 beyond the hamiltonian width, are strongly suppressed.
The region that contributes is the edge of the diagonal proximum.
\cite{GW1} The derivative of $f_\lambda$ is large and positive in this
region and it approaches a $\delta$-function in the limit of Eq.  (3.8).

In the limit of an infinitesimally small $u_0$, as discussed below Eq.
(2.20), one would substitute $\lambda^2 = u_0^{-1/2} \tilde \lambda^2$.
Then,

$$ f^2(z^2_\lambda) = \theta\left[\tilde \lambda^2 + \mu^2 - {\cal M}^2
\right]\, . \eqno (3.8.b) $$

The numerator factor in the square bracket in Eq.  (3.6) originates from
spinors of the intermediate fermions, $Tr(\not\!p_m + m)(\not\!{\bar
p}_m - m)$ with $p_m^2 = {\bar p}_m^{\,2} = m^2$.  The subscript $m$
indicates that the $-$ component is calculated from the mass-shell
condition knowing $+$ and $\perp$ components.  $+$ and $\perp$
components of $p$ and $\bar p$ are constrained by the light-front
spatial momentum conservation law, $p + {\bar p} = k$, where $k$ is the
meson momentum.  The pseudoscalar interaction with $i\gamma_5$ gives the
same result with an additional term $+8m^2$ in the numerator.

According to Eq.  (2.49),

$$ r_\epsilon(x,\kappa) = \left[ 1 + \epsilon{{\cal M}^2 \over
\Lambda^2} + \left(\epsilon{{\cal M}^2 \over \Lambda^2}\right)^2
x(1-x)\right]^{-2} . \eqno (3.9) $$

\noindent No infrared regularization is required in Yukawa theory
with massive particles, $m > 0$ and $\mu > 0$.

We can see some useful flexibility in the regularization factors at this
point.  If the regularization factors in Eq.  (2.49) contain $e_i$
divided by $1-x_i$ in place of $e_i$ one obtains here

$$ r_\epsilon(x,\kappa) = \left[1+ \epsilon \left({{\cal M}^2 \over
\Lambda^2}\right) \right]^{-4} \eqno (3.10) $$

\noindent instead of Eq.  (3.9).  The spinor bracket on the right-hand
side of Eq.  (3.6) equals $8 (p^3)^2$ when one changes variables from
$x$ and $\kappa^\perp$ to $\vec p$ with $p^\perp = \kappa^\perp$ and
$p^3$ is determined from the relation ${\cal M} = 2 \sqrt{m^2 + {\vec
p}^{\,2}}$.  Thus, with the modified regularization factor of Eq.
(3.10) which is a reasonable option, the spinor bracket can be replaced
by ${1 \over 3} {\vec p}^{\,2} = {1\over 3}({1\over 4}{\cal M}^2 -m^2)$
and two separate integrations over $x$ and $\kappa^2$ can be reduced to
a single integral over ${\cal M}^2$.  Such simplifications are helpful
in a qualitative analysis of the cutoff dependences.

In the limit of Eq. (3.8.a) for $u_0={1\over 4}$ one obtains

$$ {d \mu^2_\lambda \over d\lambda^2} = {\alpha \over 16 \pi} \left(1 +
{\mu^2 \over \lambda^2 + 2\mu^2} \right) \theta(z^2_0)
\left({2\over a}\right)^4 \int_0^{z_0} dz{z^2 \over \left[
(1+ 2/a)^2 - z^2\right]^2} , \eqno (3.11) $$

\noindent where $a=\epsilon (\lambda^2 + 3\mu^2)/\Lambda^2$ and $z_0=
\sqrt{1-4m^2/(\lambda^2 + 3\mu^2)}$.  Note that for $\lambda^2 \leq 4m^2
-3\mu^2$ the derivative of the effective meson mass equals zero and the
mass stays at the width independent value $\mu^2_{4m^2 -3\mu^2}$.  If
one uses Eq.  (3.10) instead, the corresponding result is

$$ {d \mu^2_\lambda \over d\lambda^2} = {\alpha \over 48 \pi} \left(1 +
{\mu^2 \over \lambda^2 + 2\mu^2} \right) \theta(z^2_0) z^3_0 (1+a)^{-4}.
\eqno (3.12) $$

\noindent Eqs. (3.11) and (3.12) are the same for $a \ll 1$ which is the
limit of removing the regularization cutoff, $\epsilon \rightarrow 0$,
for a fixed hamiltonian width $\lambda$. In this limit one has

$$ {d \mu^2_\lambda \over d\lambda^2} = {\alpha \over 48 \pi} \left(1 +
{\mu^2 \over \lambda^2 + 2\mu^2} \right) \left(1-{4m^2\over\lambda^2
+3\mu^2}\right)^{3/2} \theta(\lambda^2 + 3\mu^2 -4m^2) \, .\eqno (3.13)
$$

If one assumes that the meson mass squared parameter in the effective
hamiltonian has some finite value, $\mu^2_0=\mu^2_{\lambda_0}$ at some
$\lambda_0$ such that $\lambda_0^2 \geq 4m^2 -3\mu^2$ then, the
integration of Eq.  (3.13) demonstrates that

$$ \mu^2_\lambda = \mu_0^2 + {\alpha \over 48 \pi} (\lambda^2 -
\lambda_0^2) + {\alpha \over 48 \pi} (\mu^2 - 6m^2)\log{\lambda^2 \over
\lambda_0^2} + \mu^2_{conv}(\lambda,\lambda_0) \, . \eqno (3.14) $$

\noindent $\mu^2_{conv}(\lambda,\lambda_0)$ denotes the result of
integrating the convergent part of the integrand,

$$ \mu^2_{conv} (\lambda,\lambda_0) = {\alpha \over 48 \pi}
\int_{\lambda_0^2}^{\lambda^2} ds\left[ \left(1 + {\mu^2 \over s +
\mu^2} \right) \left(1-{4m^2\over s +3\mu^2}\right)^{3/ 2} - 1 -
{\mu^2 -6m^2 \over s} \right] . \eqno (3.14.a)$$

\noindent $\mu^2_{conv} (\lambda,\lambda_0) $ has a finite (i.e.
$\lambda$-independent) limit for large $\lambda$.  It contains the terms
which vanish for large $\lambda$ as inverse powers of $\lambda^2$.  The
dependence of $\mu^2_{conv} (\lambda,\lambda_0)$ on $m$ and $\mu$ is not
indicated explicitly because we will not need it in the discussion of
counterterms.  However, one should keep in mind that the mass parameters
determine the value of $\lambda = \sqrt{4m^2-3\mu^2}$ where the
effective mass stabilizes.  We simplify our notation assuming that the
effective cutoffs are always above the point of stabilization.  Below
the stabilization point, the meson mass has the constant value which is
independent of $\lambda$.  We will show later that the constant value is
equal to the physical meson mass.

The mass squared term in the effective hamiltonian with a non-negligible
coupling $g$ grows linearly with $\lambda^2$.  A logarithmic correction
appears with an opposite sign due to the factor $z_0^3$ in Eq.  (3.13),
as indicated in Eq.  (3.14).  However, one cannot make contact in Eq.
(3.14) with the initial hamiltonian by letting $\lambda$ grow to
infinity because one would obtain an ill-defined result.  The factors
depending on $a$ in Eqs.  (3.11) or (3.12) remove the infinite growth of
$\mu_\lambda$ when $\lambda \rightarrow \infty$.  Eq.  (3.12) is simpler
than Eq.  (3.11) and illustrates the same idea so we start with Eq.
(3.12).

Equation (3.12) can be integrated over $\lambda$ from infinity to any
finite value because the factor $(1+a)^{-4}$ provides convergence for
$\lambda^2 > \Lambda^2/\epsilon$.  Nevertheless, the integral diverges
as a function of $\epsilon$ when $\epsilon \rightarrow 0$.  The
divergence appears as a single number.  Therefore, the counterterm is
also a number.  We add $\mu^2_\epsilon - \mu^2$ to $\mu^2$ in the
initial hamiltonian.  We also write $\mu^2_\epsilon$ as a series in
powers of $g$, $\mu^2_\epsilon = \mu^2 + \delta \mu^2_\epsilon + o(g^4)$
so that $\delta \mu^2_\epsilon \sim g^2$.  Thus,

$$ \mu_\lambda^2 = \mu^2 + \delta\mu^2_\epsilon -
\int_{\lambda^2}^\infty {\alpha \over 48 \pi} \left(1+{\mu^2\over s+
2\mu^2}\right) \left(1-{4m^2\over s +3\mu^2}\right)^{3/ 2}
\left(1+\epsilon {s+ 3\mu^2 \over \Lambda^2}\right)^{-4} +o(g^4).  \eqno
(3.15) $$

\noindent This is an example of Eq.  (2.33) in a perturbative
application to second order in powers of $g$ in Yukawa theory.  The
counterterm $\delta \mu^2_\epsilon$ will be calculated following the
steps described below Eq.  (2.33).

The diverging part of the integrand equals $\alpha/48\pi \,\,
[1+(\mu^2-6m^2)/s]$ and the remaining part is convergent.  The
convergent part of the integrand has the same structure as in
$\mu^2_{conv}(\lambda,\lambda_0)$ but the integral now extends from
$\lambda^2$ to infinity instead of from $\lambda_0^2$ to $\lambda^2$.
In the convergent part, one can replace the regulating factor by 1.
Simplifications occur in the limit $\epsilon \rightarrow 0$ and the
result of integration in Eq.  (3.15) is

$$ \mu_\lambda^2 = \mu^2 +\delta\mu^2_\epsilon + {\alpha \over 48
\pi}\left[ -{\Lambda^2 \over 3}{1 \over \epsilon} + (\lambda^2 +3\mu^2)
+(\mu^2-6m^2)\left( \log{\epsilon{\lambda^2\over\Lambda^2}} +{11\over
6}\right)\right]$$
$$ - \mu^2_{conv}(\infty,\lambda) + o(g^4), \eqno (3.16) $$

\noindent where the square bracket originates from the diverging
part.

Following the procedure described below Eq.  (2.33), we define the
counterterm $\delta \mu^2_\epsilon$ as the negative of the integral of
the diverging integrand for some arbitrarily chosen $\lambda=\lambda_0$
plus an unknown finite piece corresponding to $\lambda_0$ and denoted by
$\delta \mu^2_{\epsilon \, finite}(\lambda_0)$.  Namely,

$$ \delta \mu_\epsilon^2 = {\alpha \over 48 \pi}\left[{\Lambda^2 \over
3}{1 \over \epsilon}+ (\mu^2 - 6m^2) \log{1\over\epsilon} - (\lambda_0^2
+3\mu^2) - (\mu^2 - 6m^2) \left(\log{\lambda_0^2\over\Lambda^2}
+{11\over 6}\right)\right]$$

$$+\delta \mu^2_{\epsilon\,finite}(\lambda_0)+o(g^4),\eqno (3.17)$$

\noindent where

$$\delta \mu^2_{\epsilon\,finite}(\lambda_0) = {\alpha \over 48\pi}
\left[ {\Lambda^2 \over 3} c_1 + (\mu^2 - 6m^2)c_2 + (\lambda^2_0
+3\mu^2) + (\mu^2-6m^2) (\log{\lambda_0^2 \over \Lambda^2} + {11\over
6}) \right] + o(g^4) \eqno (3.17.a) $$

\noindent with the unknown finite numbers $c_1$ and $c_2$.  So, in fact,

$$\delta \mu^2_\epsilon = {\alpha \over 48\pi} \left[ \Lambda^2
({1\over\epsilon} + c_1) + (\mu^2-6m^2) (\log{1\over \epsilon} + c_2 )
\right] . \eqno(3.17.b)$$

Since the whole expression on the right-hand side of Eq.  (3.17) is
merely a number, it is not necessary to find $c_1$ and $c_2$ or any
other part of it separately at this point.  One can easily find the
number $\delta \mu^2_{\epsilon\,finite}(\lambda_0)$ from the knowledge
of $\mu^2_\lambda$ at some value of $\lambda$.  The resulting
counterterm will render well defined finite boson mass squared parameter
in the effective hamiltonians in the limit $\epsilon \rightarrow 0$.
Using Eqs.  (3.16) and (3.17) one obtains

$$ \mu^2_\lambda = \mu^2 +\delta\mu^2_{\epsilon\,finite}(\lambda_0) +
{\alpha \over 48\pi} \left[\lambda^2 -\lambda_0^2 +(\mu^2-6m^2)
\log{\lambda^2 \over \lambda_0^2}\right] - \mu^2_{conv}(\infty,\lambda)
+ o(g^4).  \eqno (3.18) $$

\noindent Equation (3.18) is an example of Eq. (2.35).

The unknown finite term, $\delta \mu^2_{\epsilon \, finite}(\lambda_0)$,
has to be found, in principle, by comparison with data.  We shall
discuss an example of a renormalization condition later in this Section.

Let us assume now that at some arbitrarily chosen value of $\lambda =
\lambda_1$ the effective meson mass squared required in Eq.  (3.5) by a
fit to data equals $\mu^2_1$, i.e.  $\mu^2_{\lambda_1} = \mu^2_1$.  We
can calculate $\delta \mu^2_{\epsilon \, finite}(\lambda_0)$ using Eq.
(3.18) with $\lambda= \lambda_1$ and $\mu^2_{\lambda_1}$ on the
left-hand side replaced by the number $\mu^2_1$ inferred from the
experimental data.  The result is

$$ \delta \mu^2_{\epsilon\,finite} (\lambda_0) = \mu_1^2 - \mu^2 -
{\alpha \over 48\pi} \left[\lambda_1^2 -\lambda_0^2 +(\mu^2-6m^2)
\log{\lambda_1^2 \over \lambda_0^2}\right] +
\mu^2_{conv}(\infty,\lambda_1) + o(g^4) . \eqno (3.19) $$

\noindent Note that  one has to include the contribution of the
convergent terms in the determination of the arbitrary constants
when using the value of $\mu_1^2$.  Knowing $\delta \mu^2_{\epsilon \,
finite}(\lambda_0)$ one can calculate $\mu^2_\lambda$.  Namely,

$$ \mu^2_\lambda = \mu_1^2 + {\alpha \over 48\pi} \left[\lambda^2
-\lambda_1^2 + (\mu^2 -6m^2) \log{\lambda^2 \over \lambda_1^2}\right] +
\mu^2_{conv}(\lambda,\lambda_1) + o(g^4) . \eqno (3.20) $$

\noindent As expected, Eq.  (3.20) is the same as Eq.  (3.14) when
$\lambda_1=\lambda_0$ and $\mu_1 = \mu_0$.  One can also trace the
origin of all the terms from Eq.  (3.14); the diverging and converging
terms and the counterterm in Eq.  (3.15).

Now, we can approach Eq.  (3.11) analogously to Eq.  (3.12) without
calculating all integrals explicitly.  Integrating both sides of Eq.
(3.11), we have

$$ \mu_\lambda^2 = \mu^2_\epsilon - \int_{\lambda^2}^\infty ds {d\mu^2_s
\over ds}.  \eqno (3.21) $$

\noindent By demanding that $\mu^2_\epsilon$ removes the diverging
(i.e. $\epsilon$-dependent) part of the integral, and leaving the finite
part of $\mu^2_\epsilon$ free so that at some $\lambda=\lambda_0$ the
effective boson mass squared parameter has a desired value $\mu^2_0$,
we obtain

$$ \mu_\lambda^2 = \mu^2_0 + \int_{\lambda_0^2}^{\lambda^2} ds {d\mu^2_s
\over ds} \, . \eqno (3.22) $$

\noindent The integrand is given by the right-hand side of Eq.  (3.11)
with $\lambda^2=s$.  Since $s$ ranges only from $\lambda^2_0$ to
$\lambda^2$ and both are finite we can take the limit $\epsilon
\rightarrow 0$ under the integral sign and the integrand becomes equal
to the right-hand side of Eq.  (3.13) with $\lambda^2$ replaced with
$s$.  Integration over $s$ produces exactly the same answer as given by
Eq.  (3.14).  Thus, we see that the result of Eq.  (3.14) is independent
of the regularization scheme.  It is determined by the initial
hamiltonian of Yukawa theory as given by Eq.  (3.3).  The only unknown
in Eq.  (3.14) is the value of $\mu^2_0$.  More precisely, we know that
$\mu^2_0 = \mu^2 + \delta \mu_0^2 + o(g^4)$ and the unkonwn term is
$\delta \mu^2_0 \sim g^2$.

Note that the above calculations can be carried out in a different way
using the following observation.  Equation (3.6) in the lowest order of
perturbation theory has a particulary simple structure.  Namely, the
only dependence on $\lambda$ comes from the similarity function
$f(z^2_\lambda)$ and both sides of the equation are equal to the derivatives
with respect to $\lambda$.  Therefore, one can directly integrate both
sides and obtain a compact integral expression for $\mu^2_\lambda$ for
arbitrary functions $f$.

One should remember that such simplifications do not occur in higher
order calculations or beyond simple perturbative expansion when the
coupling constant and mass parameters depend on $\lambda$ themselves.
Therefore, we stress that the dominant contribution to the rate of
change of $\mu^2_\lambda$ with $\lambda$ comes from the edge of the
diagonal proximum.  This fact remains generally valid and the procedure
applied above represents a generic situation despite simplicity of the
example.  This example does not involve a distinction between the bare
coupling $g$ and a renormalized coupling because to order $g^2$ there is
none.

In order to determine $\delta \mu_0^2$ we need to specify a
renormalization condition.  An example of a renormalization condition
which determines $\mu_0^2$ will be described now.

A natural requirement for $\mu^2_0$ is that the effective hamiltonian
has one boson eigenstates parametrized by momenta $p^+$ and $p^\perp$
with eigenvalues of the form $p^- = (p^{\perp 2} + {\tilde \mu}^2)/p^+$
where $\tilde \mu$ is equal to the physical boson mass.  Our approach
preserves kinematical symmetries of the light-front frame explicitly and
the eigenvalue is guaranteed to appear in that form.  Therefore, one can
calculate a whole spectrum of eigenvalues for eigenstates of different
momenta by calculating the single mass parameter $\tilde \mu$.

In order to write the effective eigenvalue equation and find out
$\mu^2_0$ which leads to the desired value of $\tilde \mu$ (if it is
possible), the following steps need to be taken.

One inserts Eq.  (3.14) into Eq.  (3.5).  Then, one replaces the bare
operators, $a^\dagger_k$ and $a_k$, in the expression for ${\cal
G}_{\lambda}$ by the effective ones, $a^\dagger_{\lambda k}$ and
$a_{\lambda k}$, to obtain $G_\lambda$.  Next, by applying the operation
$F_\lambda$, one obtains the effective hamiltonian with the form factors
in the vertices, $H_\lambda = F_\lambda[G_\lambda]$.

The effective eigenvalue equation for bosons is an example of Eq.
(1.2).  Here, it is sufficient to consider the eigenvalue equation for
$H_\lambda$ in the expansion to second order in $g$ since we are
interested in $\delta \mu^2_0$ which is proportional to $g^2$.  The only
terms that contribute are the free energy term including the effective
mass squared and the interactions which change the particle number by
one.  The latter equal the canonical interactions with the similarity
form factors.

To zeroth order in $g$, a physical meson state equals a single
effective meson state, and ${\tilde \mu}^2 = \mu^2$.

No terms order $g$ arise in ${\tilde \mu}^2$ and the next correction is
order $g^2$.  This correction can be calculated using the operation $R$
and the model hamiltonian defined in a perturbative expansion from Eq.
(1.3).  Using expansion into a series of powers of $g$ to second order,
one can restrict the model space to the single effective boson sector.
The effect of coupling to the fermion-antifermion pair states is
included in perturbation theory.

Another method is to arbitrarily limit the number of effective Fock
sectors and diagonalize the effective hamiltonian in that limited space.
Such procedure could be called the effective Tamm-Dancoff approach (ETD),
c.f.  Refs.  \cite{TD} and \cite{HO}.  The term of the second order in
$g$ in the eigenvalue will determine $\delta \mu^2_0$.  In this
calculation one can limit the space of states to the one effective boson
and the effective fermion-antifermion pair.  Note that in the ETD
approach no ultraviolet renormalization problem emerges.  The
interactions are tempered by the similarity factors of width $\lambda$
which is on the order of particle masses.  This way the ETD approach
avoids the old problem of renormalization in the TD approach.

The model calculation using the operation $R$ and the ETD calculation
have to agree with each other for small coupling constants.  The reason
is that corrections to the zeroth order approximation tend to zero when
$g \rightarrow 0$ in the presence of a finite energy gap between states
with different numbers of particles, which is the case here for $0 < \mu
< 2m$.

In this paper, we discuss only the second order corrections in $g$ and a
very limited number of the effective Fock sectors can count (i.e. 2).
Therefore, we can focus on a straight-forward perturbation theory
anyways.  Nevertheless, our simple calculations have two interesting
aspects.

The first one is that no coupling renormalization effects arise to order
$g^2$.  Therefore, the expansion in powers of $g$ up to $g^2$ is
equivalent to the expansion in powers of a well defined effective
coupling, no matter how the latter is defined.  Thus, it is important to
realize that the expansion in powers of $g$ is understood here to be the
expansion in powers of the effective coupling which appears in the
effective hamiltonian of width $\lambda$; $g_\lambda = g$ to order
$g^2$.  It is not meant to be the expansion in the initial field theory
coupling constant.

The second aspect is following.  The
perturbative expansions applied in the effective eigenvalue problem are
expansions in the interaction which is suppressed in strength by the
vertex form factor of width $\lambda$, provided by the similarity
function $f_\lambda$.  Therefore, the actual range in momentum space of
the effective interactions proportional to $g$ in the effective
hamiltonian is infinitely smaller than the momentum range of the
analogous interaction in the bare hamiltonian.  In other words, the
effective strength of the interactions is greatly reduced and much
smaller than the value of $g$ itself would imply if it stood in the
initial bare hamiltonian.  Our initial expansion in powers of $g$ can
now be understood as a shortcut to the expansion in powers of the
effective coupling.  The latter expansion may have a considerable range
of rapid convergence because the form factors reduce the size of
coefficients in the expansion.  The effective coefficients are much
smaller than in the case of the initial hamiltonian without form
factors.

Thus, the operation $R$ on $H_\lambda$ expanded in powers of the
effective interaction (the coupling constant itself can be sizable),
opens new options for studying the effective eigenvalue problem in the
whole Fock space using the basis built with the effective creation and
annihilation operators corresponding to the width $\lambda$.  One can
estimate contributions of various components by making different choices
of the model spaces and solving model dynamical problems numerically.
Wave functions are expected to fall off sharply with growing momenta and
particle number if $g$ is not too large for the vertices to contain
sufficiently narrow similarity factors so that the particle number
changing interactions are small perturbations.  In that case, the result
should still match the outcome of the first two approaches when all
quantities are defined recursively in powers of $g$.  The coupling needs
to be set equal to the right value.

The second order expression in perturbation theory implies

$$ {p^{\perp 2}+ {\tilde \mu}^2 \over p^+} \langle p'|p\rangle =
{p^{\perp 2}+ \mu^2_\lambda \over p^+}\langle p'|p \rangle -
\langle p'|F_{\lambda}[G_{12\lambda}]{1\over G_{1\lambda} -
{(p^{\perp 2}+ \mu^2 ) / p^+}}F_{\lambda}[G_{21\lambda}]|p\rangle \,
. \eqno(3.23) $$

\noindent $|p\rangle$ denotes a single effective meson state with
momentum $p^+$ and $p^\perp$, $\langle p'|p \rangle = 16\pi^3 p^+
\delta^3(p'-p)$.  The term order $g^2$ produces

$$ {\tilde \mu}^2 = \mu^2_\lambda - \int[x\kappa] gf(z^2_\lambda)
{8(x-{1\over 2})^2{\cal M}^2 \over {\cal M}^2 -\mu^2} gf(z^2_\lambda) +
o(g^4) , \eqno(3.24) $$

\noindent where the notation is the same as in Eq.  (3.6).  Using Eq.
(3.8) with $u_0= {1\over 4}$ at $\lambda=\lambda_0$ one obtains

$$ \mu^2_0 = {\tilde \mu}^2 + {\alpha \over 4\pi} \int_0^1dx
\int_0^\infty d\kappa^2 {8(x-{1\over 2})^2{\cal M}^2 \over x(1-x)({\cal
M}^2 -\mu^2)} \theta(\lambda^2_0 + 3\mu^2 -{\cal M}^2) + o(g^4) .
\eqno(3.25) $$

For $\lambda_0 \leq \sqrt{4m^2 - 3\mu^2}$ the effective meson mass
parameter equals the physical meson mass, as promised.  For fermions
with masses order 0.9 GeV, this implies no corrections to a light meson
mass, such as $\pi$, for cutoffs smaller than 1.8 GeV.  But one has to
remember that the actual expression for the correction is different for
the pseudoscalar $\pi N$ interaction than in Eq.  (3.25), i.e. the spin
factor has to be enlarged by $8m^2$ (see comments above Eq.  (3.9)).
The size of the coupling constant $\alpha$ may invalidate the
perturbative result for pions coupled to nucleons because the coupling
constant on the order of 10 may appear when one attempts to make
comparisons with data.

However, the actual measure of the off-shell effects is not given
directly by $\mu^2_0$ but by the sum of $\mu^2_0$ and the self-energy
resulting from the effective interactions.  According to Eqs.  (3.22)
and (3.24), the sum of both contributions in the physical pion mass
itself is actually equal zero to order $\alpha$ . It is not clear at
this stage of the development of the effective hamiltonian theory if the
off-shell corrections of physical interest must be large due to the
coupling constant order 10.

Using Eqs.  (3.21), (3.22) and (3.25), one can express the meson mass
squared term in the initial renormalized hamiltonian in terms of the
physical meson mass $\tilde \mu$ and the initial mass parameter $\mu$.
Namely, $\mu^2 = {\tilde \mu}^2 + o(g^2)$ and

$$ \mu^2_\epsilon = {\tilde \mu}^2 + g^2 \int[x\kappa] {8(x-{1\over
2})^2{\cal M}^2 \over {\cal M}^2 -\mu^2}r_\epsilon(x,\kappa) + o(g^4)
\, .\eqno(3.26) $$

\vskip.3in
\centerline {\bf Fermion mass squared}
\vskip.1in

In complete analogy to Eqs.  (3.4) to (3.6), one obtains the fermion
energy operator,

$$ {\cal G}_{1\,fermion\,\lambda} = \sum_\sigma \int[k] { k^{\perp 2} +
m^2_\lambda \over k^+ } (b^\dagger_{k\sigma} b_{k\sigma} +
d^\dagger_{k\sigma} d_{k\sigma})  .\eqno(3.27) $$

\noindent Results for fermions and antifermions are identical and we
will describe explicitly only fermions. We have

$$ {d m^2_\lambda \over d\lambda} = g^2 \int[x\kappa]
{df^2(z^2_\lambda) \over d\lambda}
{ {\bar u}_{m\sigma k}(\not\!p_m + m)u_{m\sigma k}
\over {\cal M}^2 - m^2}
\, r_\epsilon(x,\kappa) .\eqno (3.28) $$

\noindent ${\cal M}^2 = {(m^2 + \kappa^2)/ x} + {(\mu^2 + \kappa^2)
/ (1-x)}$. The regularization factor of Eq.  (2.49) for the
intermediate particles implies

$$ r_\epsilon(x,\kappa) =\left[ 1+{\epsilon\over\Lambda^2} {\cal M}^2 +
\left({\epsilon\over\Lambda^2}\right)^2{\kappa^2 +m^2 \over x}{\kappa^2
+\mu^2\over 1-x}\right]^{-2} . \eqno(3.29) $$

\noindent  The spin factor in Eq. (3.28) can be rewritten as

$$ {\bar u}_{m\sigma k}(\not\!p_m + m)u_{m\sigma k} = {\bar u}_{m\sigma
k}[ x\not\!k_m + m + {1\over 2} \gamma^+ (p^-_m - x k^-_m) ]u_{m\sigma
k}.  \eqno(3.30) $$

\noindent $\not\!k_m$ between spinors is equivalent to $m$.  The term
with $\gamma^+$ is typical in light-front calculations.  The term in it,
proportional to $k^{\perp 2} / k^+$, cancels out.  The term linear in
$k^\perp$ does not contribute because it is odd in $\kappa^\perp$ and
all other factors (especially the regularization factor) depend only on
the modulus of $\kappa^\perp$.  Regularizations which are not even in
$\kappa^\perp$ would produce a term linear in $k^\perp$.  In the present
case, the spin factor is reduced to

$$ {\bar u}_{m\sigma k}\left[ (x + 1) m + {\gamma^+ \over 2k^+}
{\kappa^2 +(1-x^2)m^2 \over x} \right] u_{m\sigma k}= {\kappa^2+(1+x)^2
m^2 \over x} . \eqno(3.31) $$

\noindent The analogous spin factor for a pseudoscalar interaction with
$i\gamma_5$ is the same as the right-hand side of Eq.  (3.31) except for
the opposite sign in front of $x$ in the numerator.  Inclusion of
the isospin introduces the number of bosons in the theory in front of
the integral in Eq.  (3.28).

We observe a similar structure in Eq.  (3.28) as in the meson mass
dependence on $\lambda^2$ in Eq.  (3.6).  Namely, there are terms
diverging linearly and logarithmically and there is a series of
convergent terms.  The integrals are slightly more complicated because
two different masses determine the integrand dependence on $x$.
Otherwise, the procedure is essentially the same.

Instead of evaluating the integrals explicitly we observe that the
divergences amount to a number which grows when $\epsilon \rightarrow
0$.  We integrate both sides of Eq.  (3.28) and obtain

$$ m^2_\lambda = m^2_\epsilon - g^2 \int[x\kappa] \left[1 -
f^2(z^2_\lambda)\right]
{ \kappa^2 + (1+x)^2 m^2 \over x ({\cal M}^2 - m^2) } \,
r_\epsilon(x,\kappa) ,\eqno (3.32) $$

\noindent where, according to Eq.  (2.18), $z_\lambda = ({\cal M}^2 -
m^2)/({\cal M}^2 + m^2 + \lambda^2)$.  The $\epsilon$-dependent terms
originate from the first term in the bracket.  It is independent of
$\lambda$.  The counterterm $\delta m^2_\epsilon \sim g^2$ in
$m^2_\epsilon = m^2 + \delta m^2_\epsilon + o(g^4)$ removes the
divergence.  The finite part of the counterterm is left to be determined
by data.  Thus, one can write

$$ m^2_\lambda = m^2 + \delta m^2_{\epsilon \, finite} + g^2
\int[x\kappa] f^2(z^2_\lambda) { \kappa^2+(1+x)^2 m^2 \over x
({\cal M}^2 - m^2) } , \eqno (3.33) $$

\noindent and one can determine the value of $\delta m^2_{\epsilon \,
finite}$ from the value of $m^2_\lambda$ which is required by some
physical condition to be satisfied by the effective hamiltonian at some
value of $\lambda$.  If we had defined the divergent part by an integral
from some value of $\lambda$, such as $\lambda_0$ in the meson case in
the previous example, then we would have to take into account that
$m^2_{\epsilon \, finite}$ depended on $\lambda_0$ to compensate for the
$\lambda_0$ dependence of the diverging integral.  When we define the
counterterm to be given by the whole $\lambda$-independent part of the
integral in Eq.  (3.32), plus a finite constant to be determined by
data, then $\delta m^2_{\epsilon \, finite}$ does not depend on
$\lambda_0$.  Nevertheless, it can be expressed in terms of
$m^2_{\lambda_0}$.  For example, if for some reason the effective
fermion mass squared at $\lambda = \lambda_0$ should be $m^2_{\lambda_0}
= m^2_0$ then, one obtains

$$ m^2_\lambda = m^2_0 + g^2 \int[x\kappa] \left[ f^2(z^2_\lambda) -
f^2(z^2_{\lambda_0}\right] {\kappa^2+(1+x)^2 m^2
\over x ({\cal M}^2 - m^2) } . \eqno (3.34) $$

\noindent $m^2_0 = m^2 + \delta m^2_0 + o(g^4)$ and $\delta m^2_0 \sim
g^2$ can be found from a renormalization condition for the physical
fermion mass.

A natural condition for fitting $m^2_0$ is that the effective
hamiltonian at the scale $\lambda_0$ has fermionic eigenstates with
eigenvalues of the form $p^- = (p^{\perp 2} + {\tilde m}^2) /p^+$, where
$\tilde m$ denotes the physical fermion mass.  In analogy to Eqs.
(3.23) and (3.24), one obtains

$$ {\tilde m}^2 = m^2_\lambda - \int[x\kappa] gf(z^2_\lambda)
{\kappa^2+(1+x)^2 m^2 \over x ({\cal M}^2 - m^2 ) } g f(z^2_\lambda)
+ o(g^4). \eqno (3.35) $$

\noindent So,

$$ m^2_0 = {\tilde m}^2 + g^2 \int[x\kappa] f^2(z^2_{\lambda_0})
 {\kappa^2+(1+x)^2 m^2 \over x({\cal M}^2 - m^2)} + o(g^4).
\eqno (3.36) $$

\noindent The initial $m^2_\epsilon$ can be calculated in terms
of $m^2$, $\tilde m^2$, $g$ and $\epsilon$ from Eq. (3.32). The
effective fermion mass parameter in the interacting hamiltonian
of width $\lambda$ is

$$ m^2_\lambda = {\tilde m}^2 + g^2 \int[x\kappa] f^2(z^2_{\lambda})
{\kappa^2+(1+x)^2 m^2 \over x ({\cal M}^2 - m^2) }  + o(g^4).
\eqno (3.37) $$

Analogous equation in the case of nucleons coupled to pions is the same
as Eq.  (3.37) except for $(1-x)^2$ instead of $(1+x)^2$ in the
numerator and the isospin factor 3 in front of the integral.  In the
limit of a $\theta$-function for the similarity factor $f$ one obtains
the result $m^2 = {\tilde m}^2 + o(g^2)$, ${\tilde m} = m_N$, and

$$ m^2_\lambda = m^2_N + 3 g^2 \int[x\kappa] \theta(\lambda^2 + 3m^2
-{\cal M}^2) {\kappa^2+(1-x)^2 m^2 \over x ({\cal M}^2 - m^2) } +
o(g^4) . \eqno (3.38) $$

\noindent For $\lambda^2 = (m + n \, \mu_\pi)^2 - 3m^2$, where $n$ is a
number order 1 or a few, one obtains ($\alpha = g^2/ 4\pi$)

$$  m^2_\lambda = m^2_N + m_N^2 {3 \alpha \over 4 \pi} c  \eqno (3.39)
$$

\noindent where $c= 4/3 (n \, \mu_\pi / m_N)^3 +o(\mu_\pi^4)$.  The
expansion formula for $c$ shows how small the correction is for small
meson masses.  Note that $\lambda^2$ must be negative for small $n$, as
explained below Eq.  (2.20).  The exact result for $c$ is that, for
example, $ n = 3 $ gives $c \sim 0.03 $ and $ n = 4 $ gives $c \sim
0.12$.  Thus, even for quite large couplings the effective mass
parameter in the hamiltonian deviates only a little from the physical
nucleon mass if the hamiltonian width allows momentum changes of the
order of a few meson masses only.  Moreover, the physically relevant
off-shell effects are not given directly by these numbers but by the
difference between these and the effects of the interactions present in
the effective hamiltonian.  The combined effect is zero for the nucleon
mass itself to order $g^2$.

Note that, if we used Eq.  (2.20) in the $\theta$-function limit with an
ifinitesimal $u_0$ and $\lambda^2 = \tilde \lambda ^2 /\sqrt{u_0}$ for
$n=1$ then, the $\theta$-function under the integral in Eq.  (3.38)
would be replaced by $\theta(\tilde \lambda^2 + m^2 - {\cal M}^2)$.

The present calculation does not say what is the size of the off-shell
effects in the theory of pions coupled to nucleons but Eq.  (3.39) does
not exclude that the effects can be calculable in perturbation theory.
This is encouraging for the program outlined in Ref.  \cite{BGP}.

\vskip.3in
\centerline {\bf Fermion-fermion interaction}
\vskip.1in

Our next example is the second order calculation of the effective
hamiltonian term which contains products of two creation and two
annihilation operators for fermions.  This term is particulary interesting
because it complements the one boson-exchange interactions which result
from emission and absorption of the effective bosons by the effective
fermions.

The differential equations we need to consider are given by Eq.  (2.29)
for the two-fermion terms.  We have

$$ {d \over d\lambda} {\cal G}_{22\lambda} = \left[ f_\lambda {\cal
G}_{12\lambda} \left\{ {d \over d\lambda} (1-f_\lambda) {\cal
G}_{21\lambda} \right\}_{{\cal G}_{1\lambda}} - \left\{ {d \over
d\lambda} (1-f_\lambda) {\cal G}_{12\lambda} \right\}_{{\cal
G}_{1\lambda}} f_\lambda {\cal G}_{21\lambda} \right]_{22} . \eqno
(3.40) $$

\noindent The subscript 22 denotes a term with two creation and two
annihilation operators for fermions. 21 denotes a term with one
annihilation operator and one creation operator for fermions and one
creation operator for bosons. 12 denotes a term which annihilates a
fermion and a boson and creates a fermion.  For a hermitean hamiltonian,
we have ${\cal G}_{12} = {\cal G}_{21}^\dagger$.  In higher order terms,
the subscript notation has to be generalized to represent the meaning of
the subscripts more precisely and, multiple subscripts are needed.  It
is not necessary to introduce the more general notation in second order
calculations in this paper.

The right-hand side of Eq.  (3.40) does not contain disconnected
interactions (it never does) and one can isolate the terms with two
creation and two annihilation operators for fermions by contracting one
creation operator and one annihilation operator for bosons.

The only term which contributes to the right-hand side of Eq.  (3.40) in
Yukawa theory, is the ordered and regularized third term on the
right-hand side of Eq.  (3.3).  Thus, in Eq.  (3.40), we have

$$ {\cal G}_{21\lambda} = \sum_{\sigma \sigma_f} \int [k_1 k k_2]
16\pi^3 \delta(k_1 + k - k_2)\, g{\bar u}_{m k_1 \sigma_f}u_{m k_2
\sigma}\, r(\epsilon e_f/\Lambda^2)\, r(\epsilon e_b/\Lambda^2)
b^\dagger_{k_1 \sigma_f} a^\dagger_{k} b_{k_2 \sigma} $$ $$ = \int [P]
{1 \over P^+} \sum_{\sigma \sigma_f}\int [x \kappa] g{\bar u}_{m
xP+\kappa \sigma_f} u_{m P \sigma} \, r(\epsilon e_f/\Lambda^2)\,
r(\epsilon e_b/\Lambda^2) \, b^\dagger_{xP+\kappa
\sigma_f}a^\dagger_{(1-x)P-\kappa} b_{P \sigma} \, , \eqno(3.41)$$

\noindent where $ e_f = ( \kappa^2 + m^2 ) / x $ and $ e_b = ( \kappa^2
+ \mu^2 ) / (1-x) $. This representation illustrates appearance of the
parent and daughter momentum variables in the interaction term.  The
factor $g\, {\bar u}_{m k_1 \sigma_f}u_{m k_2 \sigma}\, r(\epsilon
e_f/\Lambda^2)\, r(\epsilon e_b/\Lambda^2) b^\dagger_{k_1 \sigma_f}$
will be denoted by $g_{21}$.  The analogous factor in ${\cal
G}_{12\lambda}$ will be denoted by $g_{12}$.

Similarly, in the case of the four-fermion interaction term, we have

$$ {\cal G}_{22\lambda}
= \int [P] {1 \over P^+} \sum_{\sigma_1 \sigma_2 \sigma_3 \sigma_4}
\int [x\kappa][y\rho] \, g_{22\lambda} \,
b^\dagger_{xP+\kappa \sigma_1}b^\dagger_{(1-x)P-\kappa \sigma_3}
b_{yP+\rho \sigma_2}b_{(1-y)P-\rho \sigma_4}
, \eqno(3.42)$$

\noindent where $g_{22\lambda}$ is a function of the daughter momentum
variables $x$, $\kappa^\perp$, $y$, $\rho^\perp$ and the fermion spin
projections on the z-axis $\sigma_1$, $\sigma_2$, $\sigma_3$ and $
\sigma_4$.  Details of the notation will become clear shortly.

To order $g^2$, only $f_\lambda$ depends on $\lambda$ on the right-hand
side of Eq.  (3.40).  Equation (3.40) can be rewritten in a simplified
fashion in terms of the coefficient functions such as in Eq.  (3.42).
Namely,

$$ {d \over d\lambda} g_{22} = \left[ f \{ -f'\} - \{
- f'\} f \right] \left[ g_{12} g_{21} \right]_{22} .
\eqno (3.43) $$

\noindent The subscript $\lambda$ and arguments of the functions are not
indicated, to simplify the formula.  Expression in the first bracket is
called {\it the inner similarity factor} for ${\cal G}_{22}$.  The word
``inner'' is used to distinguish this factor from the form factor
introduced by the operation $F_\lambda$ when this operation is applied
to $G_{22}$.  The latter form factor can be called the {\it outer}
similarity factor because it depends on the incoming and outgoing 
invariant masses only, independently of the internal structure of the 
interaction.

Momentum variables in Eq.  (3.43) can be expressed by the daughter
momentum variables from Eq.  (3.42).  One needs to express the parent
and daughter variables of ${\cal G}_{12}$ and ${\cal G}_{21}$, and the
energy denominators, in terms of $x$, $\kappa^\perp$, $y$ and
$\rho^\perp$ from Eq.  (3.42).  This is done as follows.

The momentum labels of the fermion annihilation operators in Eq.  (3.42) are
denoted by $k_2$ and $k_4$ and the momentum labels of the fermion creation
operators in Eq.  (3.42) are denoted by $k_1$ and $k_3$ . The numbers
assigned to the fermion momenta correspond to the numbers labeling their spin
projections on the z-axis in Eq.  (3.42).  In terms of the origin of the
annihilation and creation operators in Eq.  (3.42), ${\cal G}_{21}$
provides the fermion creation operator with momentum $k_1$ and the
fermion annihilation operator with momentum $k_2$. ${\cal G}_{12}$
provides the fermion creation operator with momentum $k_3$ and the
fermion annihilation operator with momentum $k_4$.  There is a change of
sign due to the reordering of the fermion operators.  The boson
operators from ${\cal G}_{12}$ and ${\cal G}_{21}$ are contracted and
provide factors similar to the factors associated with an intermediate
particle in the old-fashioned hamiltonian calculations for scattering
processes.

It is useful to think about the effective hamiltonian in terms of a
scattering amplitude with two vertices but the reader should remember
that the formula we are describing is not for an $S$-matrix matrix
element.  Therefore, the ``scattering'' language has a limited meaning.  
The fine point is that, after
evaluation of ${\cal G}_\lambda$, one has to replace the bare creation
and annihilation operators by the effective ones in order to obtain
$G_\lambda$.  The term in ${\cal G}_\lambda$ is not directly related to any
scattering process before the replacement is made.  The replacement
prevents confusion between the hamiltonian calculus which uses the bare
operators, and the $S$-matrix calculus which uses the effective
hamiltonian and the corresponding incoming, outgoing and intermediate
states of effective particles.  The scattering language becomes
particulary confusing in theories with gauge symmetry, spontaneous
symmetry breaking and confinement.  None of these features appear here
in the calculation to order $g^2$.  Therefore, the scattering language
is useful in the current example.

The intermediate boson momentum is defined for $+$ and $\perp$
components as $k_5 = k_3 - k_4$ and $k^-_{5\mu} = (k^{\perp 2}_5 +
\mu^2)/k^+_5$.  These four components form the four-momentum of the
exchanged boson. It is denoted by $k_{5\mu}$ to indicate the mass which
determines the minus component.  The same result for $k_{5\mu}$ is
obtained by subtracting $k_1$ from $k_2$ instead of $k_4$ from $k_3$.
It is so because the translational invariance of the hamiltonian on the
light-front implies momentum conservation for the $+$ and $\perp$
components.

Thus, the inner similarity factor in Eq. (3.43) is

$$ \left[ f \{ - f'\} - \{ - f'\} f \right]
 =  f(z^2_{12 \lambda}) {[-f(z^2_{21 \lambda})]'\over \Delta E_{21}}
 - {[-f(z^2_{12 \lambda})]' \over \Delta E_{12}} f(z^2_{21 \lambda})
 . \eqno (3.44) $$

\noindent The prime denotes differentiation with respect to $\lambda$.

In the case of ${\cal G}_1$ in Eq.  (3.4), the whole inner similarity
factor of the analogous structure was equal to the derivative of
$f^2(z^2_\lambda)$ divided by a single denominator.  For both functions
$f$ in Eq.  (3.4) had the same argument $z^2_\lambda$ and the two
corresponding energy denominators were the same.  In Eq.  (3.44), we
have two different arguments of the similarity functions $f$ and two
different energy changes.  Namely, $z_{12 \lambda}$ and $\Delta E_{12}$
for the vertex of ${\cal G}_{12}$ with momenta $k_{3m}$, $k_{4m}$ and
$k_{5\mu}$, and $z_{21 \lambda}$ and $\Delta E_{21}$ for the vertex of
${\cal G}_{21}$ with momenta $k_{1m}$, $k_{2m}$ and $k_{5\mu}$.

The parent momentum for the vertex of ${\cal G}_{12}$ is $P_{12} =
(k_{5\mu} + k_{3m} + k_{4m})/2$ so that for the $+$ and $\perp$
components we have $P_{12} = k_3$.  Similarly, the parent momentum for
the vertex of ${\cal G}_{21}$ is $P_{21} = (k_{5\mu} + k_{1m} +
k_{2m})/2$ so that for the $+$ and $\perp$ components we have $P_{21} =
k_2$.

Now, the rules provided by Eqs.  (2.12) to (2.20) imply the following
formulae for the arguments of the similarity functions $f$.

$$ \Delta {\cal M}^2_{12} =
(k_{5\mu} + k_{4m})^2 - k^2_{3m} =
2 (k_{5\mu} + k_{4m} - k_{3m}) P_{12}
 . \eqno(3.45) $$

$$ \Sigma {\cal M}^2_{12} = {\cal M}^2_{12} + 2m^2 . \eqno (3.46) $$

$$ z_{12 \lambda} = { \Delta {\cal M}^2_{12} \over
            \Sigma {\cal M}^2_{12} + \lambda^2 } . \eqno(3.47)$$

$$ \Delta {\cal M}^2_{21} =
 k^2_{2m} - (k_{5\mu} + k_{1m})^2 =
 - 2 (k_{5\mu} + k_{1m} - k_{2m}) P_{21}
 . \eqno(3.48) $$

$$ \Sigma {\cal M}^2_{21} = - {\cal M}^2_{21} + 2m^2 . \eqno (3.49) $$

$$ z_{21 \lambda} = { \Delta {\cal M}^2_{21} \over
                     \Sigma {\cal M}^2_{21} + \lambda^2 } .
\eqno(3.50)$$

Equations (2.12) to (2.16) imply

$$  \Delta E_{12} = {\Delta {\cal M}^2_{12} \over P^+_{12}} \eqno(3.51)$$

\noindent and

$$  \Delta E_{21} = {\Delta {\cal M}^2_{21} \over P^+_{21}}.
\eqno(3.52)$$

In terms of the familiar parameters $x$, $\kappa^\perp$, $y$ and
$\rho^\perp$ from Eq.  (3.42), i.e.

$$ P = k_1 + k_3 = k_2 +k_4,  \eqno (3.53) $$

$$ x = k^+_1/P^+, \eqno (3.54) $$

$$ \kappa^\perp = k_1^\perp - x P^\perp, \eqno (3.55) $$

\noindent and

$$ y = k^+_2/P^+, \eqno (3.56) $$

$$ \rho^\perp = k_2^\perp - y P^\perp \, , \eqno (3.57) $$

\noindent the mass differences which determine the arguments of the
similarity functions read as follows.

$$ \Delta {\cal M}^2_{12} = (1-x) \left[
{(\kappa^\perp-\rho^\perp)^2 + \mu^2 \over y-x} +
{\rho^{\perp 2} + m^2 \over 1-y} -
{\kappa^{\perp 2} + m^2 \over 1-x} \right] . \eqno(3.58)
$$

$$ \Delta {\cal M}^2_{21} = y \left[
{\rho^{\perp 2} + m^2 \over y}
- {\kappa^{\perp 2} + m^2 \over x} -
{(\kappa^\perp-\rho^\perp)^2 + \mu^2 \over y-x} \right] .\eqno(3.59) $$

\noindent The corresponding energy denominators are

$$ \Delta E_{12} = \left[ {(\kappa^\perp-\rho^\perp)^2 + \mu^2 \over
y-x} + {\rho^{\perp 2} + m^2 \over 1-y} - {\kappa^{\perp 2} + m^2 \over
1-x} \right]/P^+ \eqno(3.60)$$

\noindent and

$$ \Delta E_{21} = \left[ {\rho^{\perp 2} + m^2 \over y} -
{\kappa^{\perp 2} + m^2 \over x} - {(\kappa^\perp-\rho^\perp)^2 + \mu^2
\over y-x} \right]/P^+ . \eqno(3.61)$$

In Eqs.  (3.40) to (3.61), always $y > x$.  When evaluating matrix
elements of the effective interaction between states of
indistinguishable fermions, one obtains results in which the momentum
and spin variables are properly symmetrized (antisymmetrized) as
dictated by the statistics.

Evaluation of the second bracket in Eq. (3.43) gives

$$ [g_{12} g_{21}]_{22} \, = \, - \, g \, {\bar u}_{m (1-x)P-\kappa
\sigma_3} u_{m (1-y)P-\rho \sigma_4} \, r(\epsilon e_4/\Lambda^2)
r(\epsilon e_{12}/\Lambda^2)$$
$${1 \over (y-x)P^+} \, g \, {\bar u}_{m xP+\kappa \sigma_1} u_{m
yP+\rho \sigma_2} \, r(\epsilon e_1/\Lambda^2) r(\epsilon
e_{21}/\Lambda^2) \, . \eqno(3.62)$$

\noindent The arguments of the regularization factors appear in the mass
differences.  Namely,

$$ \Delta {\cal M}^2_{12}= e_4 + e_{12} - m^2 =
   {\kappa^{\perp 2}_{12} + m^2 \over x_{12} } +
   {\kappa^{\perp 2}_{12} + \mu^2 \over 1 - x_{12} }
   -m^2 ,  \eqno(3.63.a) $$

\noindent where

$$  x_{12} = {1-y \over 1-x} , \eqno(3.63.b) $$

$$  \kappa^\perp_{12} = -\rho^\perp + x_{12} \kappa^\perp,
\eqno(3.63.c)$$

\noindent and

$$ \Delta {\cal M}^2_{21}= m^2 - e_1 + e_{21} =
   m^2 - {\kappa^{\perp 2}_{21} + m^2 \over x_{21} } -
   {\kappa^{\perp 2}_{21} + \mu^2 \over 1 - x_{21} }
  ,  \eqno(3.64.a) $$

\noindent where

$$  x_{21} = {x \over y} , \eqno(3.64.b) $$

$$  \kappa^\perp_{21} = \kappa^\perp - x_{21} \rho^\perp .
\eqno(3.64.c)$$

The daughter energies defined by Eqs.  (2.48) and (2.49) appear in Eqs.
(3.62) to (3.64).  The energy subscripts will be used to distinguish the
corresponding regularization factors in an abbreviated notation below.
Equation (3.43) reads, in the abbreviated notation, as follows.

$$ {d \over d\lambda} g_{22} =
   \left[
    {y \over y-x}\, {f_{12} f'_{21} \over \Delta {\cal M}^2_{21}}
- {1-x \over y-x}\, {f'_{12} f_{21} \over \Delta {\cal M}^2_{12}}
  \right]\theta(y-x) \,
g{\bar u}_3 u_4 \, r_4 r_{12} \, g{\bar u}_1 u_2 \, r_1 r_{21}
 . \eqno (3.65) $$

\noindent In the familiar limit of Eq.  (3.8) where the similarity
function $f$ approaches the $\theta$-function, i.e.  $f_{12} =
\theta[\lambda^2 + 2m^2 - \Delta{\cal M}^2_{12}]$ and $f_{21} =
\theta[\lambda^2 + 2m^2 + \Delta{\cal M}^2_{21}]$, the derivatives of
the similarity functions become $\delta$-functions and one can integrate
Eq.  (3.65) using the relation $\int_a^\infty ds \theta(s-b) \delta(s-c)
= \theta(c-b) \theta(c-a)$.  The result is

$$ g_{22\lambda} = g_{22\epsilon} + \left[
{ y \theta_{21-12} \theta_{21-\lambda} \over
(y-x) |\Delta {\cal M}^2_{21}|} +
{(1-x) \theta_{12-\lambda} \theta_{12-21} \over
(y-x) |\Delta {\cal M}^2_{12}|} \right] \theta(y-x)
\, g{\bar u}_3 u_4 \, r_4 r_{12} \, g{\bar u}_1 u_2 \,
r_1 r_{21} . \eqno (3.66.a) $$

\noindent The symbols $\theta$ with various subscripts denote the
following functions:

$$ \theta_{12-21} = 1 - \theta_{21-12} = \theta(|\Delta {\cal M}^2_{12}|
- |\Delta {\cal M}^2_{21}|) , \eqno(3.66.b)$$

$$\theta_{12-\lambda} = \theta(
|\Delta {\cal M}^2_{12}| - \lambda^2 - 2m^2), \eqno(3.66.c)$$

\noindent and

$$ \theta_{21-\lambda} =
\theta( |\Delta {\cal M}^2_{21}| - \lambda^2 - 2m^2). \eqno(3.66.d)$$

\noindent The initial value term of $g_{22\epsilon}$ at $\lambda =
\infty$ is absent in the canonical hamiltonian.  It should be set to 0
unless it is found that matrix elements of $H_{22\lambda} = F_\lambda
\left[ G_{22\lambda} \right]$ between finite free energy states depend
on $\epsilon$ when $\epsilon \rightarrow 0$ and a counterterm containing
nonzero $g_{22\epsilon}$ is required to remove this dependence.

Note again that, if we used Eq.  (2.20) in the $\theta$-function limit
with an ifinitesimal $u_0$ and $\lambda^2 = \tilde \lambda ^2
/\sqrt{u_0}$ for $n=1$ then, $\lambda^2 + 2 m^2$ in the
$\theta$-function arguments above would be replaced by $\tilde
\lambda^2$.  This feature will be used later in the case of
nonrelativistic bound states.

The easiest momentum configuration to analyse is the one where the sum
of free energies for the momenta of creation operators equals the sum of
free energies for the momenta of annihilation operators.  In this case,
we have $(k_{1m} + k_{3m})^2 = (k_{2m} + k_{4m})^2 = {\cal M}^2$.  We
will refer to this configuration as {\it the energy-diagonal part of the
interaction}.  In the energy-diagonal part of the interaction, we have
$\Delta E_{12} = - \Delta E_{21}$, and

$$ {\rho^{\perp 2} + m^2 \over 1-y} - {\kappa^{\perp 2} + m^2 \over 1-x}
 = {\kappa^{\perp 2} + m^2 \over x} - {\rho^{\perp 2} + m^2 \over y} .
\eqno (3.67) $$

\noindent Thus,

$$ {|\Delta{\cal M}^2_{12}|\over 1-x} =
   {|\Delta{\cal M}^2_{21}|\over y}   =
   {\mu^2 + {\vec q}^{\, 2} \over y-x} , \eqno(3.68) $$

\noindent where ${\vec q}^{\, 2} = (\kappa^\perp - \rho^\perp)^2 +
(y-x)^2 {\cal M}^2$.  These relations imply $\theta_{12-21} = 1 -
\theta_{21-12} = \theta(1-x-y)$, $\theta_{12-\lambda}= \theta[\mu^2 +
{\vec q}^{\, 2} - (y-x)(\lambda^2 + 2m^2)/(1-x)]$ and
$\theta_{21-\lambda}= \theta[\mu^2 + {\vec q}^{\, 2} - (y-x)(\lambda^2 +
2m^2)/y]$.  Therefore, $\theta_{12-\lambda} = 1$ in the same momentum
range where $\theta_{21-\lambda} = 1$.  Thus, the energy-diagonal part
of the fermion-fermion effective interaction order $g^2$ is

$$ g_{22\lambda} = g_{22\epsilon} + { g{\bar u}_3 u_4 \, r_4 r_{12}
\, g{\bar u}_1 u_2 \, r_1 r_{21} \over \mu^2 + {\vec q}^{\, 2}}
\theta(y-x) \theta \left[ \mu^2 + {\vec q}^{\, 2} - {y-x\over
max(y,1-x)} (\lambda^2 + 2m^2)  \right] . \eqno (3.69) $$

Equation (3.69) is helpful because it provides insight into the more
complicated interaction from Eq.  (3.66).  When the momentum transfer is
sufficiently large and $(y-x)(\lambda^2 + 2m^2)/max(y,1-x)$ is
negligible so that the $\theta$-functions and the regularization
functions $r_4$, $r_{12}$, $r_1$ and $r_{21}$ in Eq.  (3.69) equal 1,
the second term on the right-hand side of Eq.  (3.69) is equal to
Feynman's expression for the one boson-exchange scattering amplitude for
two fermions.  Namely, the numerator factors are standard for the Yukawa
interaction and the denominator equals $ \mu^2 - (k_{3m} - k_{4m})^2 =
\mu^2 - (k_{2m} - k_{1m})^2$.  The necessary antisymmetrization for
identical fermions requires evaluation of the matrix elements of the
hamiltonian term.  However, the $\theta$-functions and the
regularization factors in Eq.  (3.69) produce a difference between the
energy diagonal part of the effective hamiltonian matrix elements and
the on-shell Feynman scattering amplitude.

The $\theta$-functions force the momentum transfer carried by the
intermediate boson, $|{\vec q}|$, to be larger than
$\left[(y-x)(\lambda^2 + 2m^2)/y - \mu^2\right]^{1/2}$.  The size of
this number depends on how large $\lambda^2$ and the ratios of the
longitudinal momenta are.  If $\lambda^2$ is negative and compensates
$2m^2$, the lower bound on the momentum transfer is absent.  For larger
$\lambda^2$, the ratio of $y-x$, i.e. the intermediate meson
longitudinal momentum fraction, to the parent fermion momentum fraction
$y$ has to be smaller than $\mu^2(\lambda^2 + 2m^2)^{-1}$ for the lower
bound on the momentum transfer to be absent.  Otherwise, the momentum
transfer is limited from below.  This means that the effective
interaction term does not include the long distance part of the Yukawa
potential.

The regularization factors in the limit $\epsilon \rightarrow 0$
converge pointwise to 1. We shall explain how it happens in the more
general case below.  No diverging cutoff dependence is obtained when
evaluating matrix elements of $G_{22\lambda}$ between states of finite
invariant masses ${\cal M}^2$ so that the matrix elements of
$H_{22\lambda}$ are also free from divergences.  Therefore,
$g_{22\epsilon} = 0$. We can replace the regularization factors in
the limit $\epsilon \rightarrow 0$ by 1.

We proceed to the analysis of Eq.  (3.66).  No divergences appear in the
finite matrix elements of the effective hamiltonian of width $\lambda$
when $\epsilon \rightarrow 0$.  One can see that this is the case using
Eqs.  (3.63.a) and (3.64.a).  Namely, the arguments of the
regularization factors are finite for finite $\Delta {\cal M}^2_{12}$
and $\Delta {\cal M}^2_{21}$ and they approach 0 when $\epsilon
\rightarrow 0$.  One demands that the free invariant masses of the
states of fermions used to calculate the matrix elements are finite.
Then, the only possibility for $\Delta {\cal M}^2_{12}$ or $\Delta {\cal
M}^2_{21}$ to diverge emerges when $x$ approaches $y$, i.e. when the
longitudinal momentum transfer between the fermions approaches zero.  In
such case, $e_{12}$ and $e_{21}$ approach infinity even for a vanishing
transverse momentum transfer because the meson mass squared is greater
than zero.

Now, the remaining factors of spinors and energy denominators, the
latter multiplied by the boson phase-space factor of $y-x$, are all
finite in the limit $x \rightarrow y$.  The regularization factors
$r_{12}$ and $r_{21}$ deviate from 1 only in the small region in the
momentum space where $|x-y| < \epsilon\Lambda^2/\mu^2$ (or in a still
smaller region for a nonzero meson transverse momentum).  All other
factors in the interaction are finite in this region.  Therefore, for
finite wave packets used in the evaluation of the matrix element, this
small region produces a contribution which is proportional to
$\epsilon$.  Thus, it vanishes in the limit $\epsilon \rightarrow 0$.
Consequently, the matrix elements of $g_{22\epsilon}$ are equal 0 and
the regularization factors can be replaced by 1.

The full result for the effective fermion-fermion interaction in the
limit $\epsilon \rightarrow 0$ can now be rewritten as

$$ H_{22\lambda} = F_\lambda\left[G_{22\lambda}\right] $$
$$ = \int [P] {1 \over
P^+} \sum_{\sigma_1 \sigma_2 \sigma_3 \sigma_4} \int [x\kappa][y\rho]\,
g_{22\lambda}\, f(z^2_{22\lambda})\, b^\dagger_{\lambda xP+\kappa
\sigma_1}b^\dagger_{\lambda (1-x)P-\kappa \sigma_3} b_{\lambda
yP+\rho \sigma_2}b_{\lambda (1-y)P-\rho \sigma_4} \, ,
\eqno(3.70.a)$$

\noindent where

$$ g_{22\lambda} = \left[ { \theta_{12-21} \theta_{21-\lambda} \over
\mu^2 - q^2_{21} } + { \theta_{12-\lambda} \theta_{21-12}\over \mu^2 -
q^2_{12} } \right] \theta(y-x) g{\bar u}_3 u_4 g{\bar u}_1 u_2 \, , \eqno
(3.70.b)  $$

$$ q_{12} = k_{3m} - k_{4m} , \eqno(3.70.c)$$

$$ q_{21} = k_{2m} - k_{1m} , \eqno(3.70.d)$$

$$ \theta_{12-21} = 1 - \theta_{21-12} =
   \theta\left[(1-x)(\mu^2-q^2_{12})  - y(\mu^2-q^2_{21})\right]
   , \eqno(3.70.e) $$

$$ \theta_{12-\lambda} = \theta\left[(1-x)(\mu^2-q^2_{12}) -
(y-x)(\lambda^2 + 2m^2)\right] , \eqno(3.70.f) $$

$$ \theta_{21-\lambda} = \theta\left[y(\mu^2-q^2_{21}) -
(y-x)(\lambda^2 + 2m^2)\right] . \eqno(3.70.g) $$

\noindent The argument of the outer similarity factor $
f(z^2_{22\lambda}) $ which limits the width of the effective interaction
in momentum space, is given by

$$ z_{22\lambda} = {\Delta{\cal M}^2_{22}\over
\Sigma{\cal M}^2_{22} + \lambda^2 }. \eqno(3.70.h) $$

\noindent The mass difference, $\Delta{\cal M}^2_{22} = {\cal M}^2_{24}
- {\cal M}^2_{13}$, and the mass sum, $\Sigma{\cal M}^2_{22} = {\cal
M}^2_{24} + {\cal M}^2_{13}$, can be expressed by the fermion momenta
using relations ${\cal M}^2_{24} = (k_{2m} + k_{4m})^2$ and ${\cal
M}^2_{13} = (k_{1m} + k_{3m})^2$.

Equations (3.70.a) to (3.70.h) explain the structure of the
fermion-fermion effective interaction order $g^2$ in terms of the two
four-momentum transfers, $q_{12}$ and $q_{21}$.  The transfer $q_{12}$
appears in the vertex where the intermediate boson is annihilated and
the transfer $q_{21}$ appears in the vertex where the boson is created.
In the energy-diagonal part the two four-momentum transfers are equal.
In the general case they are different.  The $\theta$-functions
exclusively select which momentum transfer appears in the denominator.
The lower bounds on the momentum transfers depend on the ratio of
$|x-y|$ to $y$ and $1-x$ and on the masses and $\lambda^2$.

We can now evaluate matrix elements of the $T$-matrix between the
effective two-fermion states using the effective hamiltonian of width
$\lambda$ to second order in $g$. We use the formula

$$  T(E) = H_{I\lambda} + H_{I\lambda} {1\over E -
H_{0\lambda} + i\varepsilon} H_{I\lambda} \, . \eqno(3.71) $$

\noindent We have $H_{0\lambda} = G_{1\lambda}$ and $H_{I\lambda} =
F_\lambda\left[ G_{22\lambda} + G_{12\lambda} + G_{21\lambda}\right]$.
The first term on the right-hand side of Eq.  (3.71) contributes solely
through $H_{22\lambda}$.  In the second term, in $H_{I\lambda}$, only
$H_{12\lambda} + H_{21\lambda}$ contributes.

The first term in Eq.  (3.71) has its matrix element given by the
antisymmetrization of the right-hand side of Eq.  (3.70.b).  The
multiplication by $f(z^2_{22\lambda})$ does not matter because
$f(z^2_{22\lambda}) = 1$ in the energy-diagonal matrix elements.  Only
the energy-diagonal part contributes to the cross section.  The
energy-diagonal part of $g_{22\lambda}$ is given by

$$ g_{22\lambda} = { \, g{\bar u}_3 u_4 \, g{\bar u}_1 u_2 \, \over
\mu^2 - q^2} \, \theta(y-x) \, \theta_\lambda \, , \eqno (3.72) $$

\noindent where $ \theta_\lambda = \theta \left[ max(y,1-x)(\mu^2 - q^2)
- (y-x)(\lambda^2 + 2m^2) \right]$ and $q = q_{12} = q_{21}$. The
antisymmetrization of the right-hand side of Eq. (3.72) produces the
contribution of the first term in Eq. (3.71) to the scattering
amplitude.

The second term in Eq.  (3.71) provides the one-boson exchange amplitude
with form factors in the fermion-boson vertices.  The form factors are
the similarity functions $f_\lambda$.  The resulting amplitude is given
by the antisymmetrization of Eq.  (3.72) with $\theta_\lambda$ replaced
by the product of the form factors.  In the $\theta$-function limit, the
vertex form factors equal $1 - \theta_{21-\lambda}$ and $1 -
\theta_{12-\lambda}$.  Their product equals $1- \theta_\lambda$.  Thus,
the second term provides the same contribution as the first term but the
factor $\theta_\lambda$ is replaced by $1 - \theta_\lambda$.

The sum of both terms in Eq. (3.71) produces the matrix element of the
scattering matrix on-energy-shell which is independent of
$\lambda$. Our complete on-shell result is equal to the
well known one-boson-exchange scattering amplitude.

Higher order calculations require definitions of the incoming and
outgoing physical fermion states which differ from the single effective
fermions due to the interactions which dress effective fermions with
effective bosons.  Discussion of the corresponding scattering theory
goes beyond the scope of this paper.

There is an important property of the second order calculation above
which is worth a separate note.  When the hamiltonian width in the mass
difference, as determined by $\lambda^2 + 2m^2$ or $\tilde \lambda^2$,
becomes small then, the effective meson emission can no longer occur.
Thus, the effective theory describes fermions interacting by potential
forces.  The potentials are given by factors 
$f(z^2_{22\lambda})g_{22\lambda}$ 
in $H_{22\lambda}$.  Note the necessary presence of
the similarity form factor off-energy shell.  $g_{22\lambda}$ contains
also the inner similarity factors which force the intermediate boson to
form a sufficiently high invariant mass state but if the width is small
enough they become identically equal 1.

In particular, $f(z^2_{22\lambda}) g_{22\lambda}$ in the Yukawa theory
becomes equal to the relativistic potential which approaches the Yukawa
potential in the nonrelativistic limit.  The nonrelativistic theory is
defined by the limiting case where the width $\lambda$ is such that the
allowed energy transfers are much smaller than the effective fermion
masses.  This condition limits only the relative motion of the effective
constituents.  It does not limit their total momentum.  Thus, we have
accomplished a derivation of the potential order $g^2$ in the
relativistic quantum mechanics of fermions in the Yukawa theory.  The
relativistic one-boson exchange potential is a term in the effective
hamiltonian, including its off-shell behavior. Note that in the energy
diagonal part of the effective potential as well as in the on-energy shell
scattering amplitude the outer similarity factor equals 1 independently 
of the size of the momentum transfer. In other words, one cannot see
the outer similarity factor in the physical scattering amplitude 
order $g^2$ and the only trace of the effective nature of the potential 
are the form factors in the interaction vertices.

If we use the interaction $g \bar \psi i\gamma_5 \vec \tau \psi \vec
\phi$ instead of $g \bar \psi \psi \phi$ in writing the initial
hamiltonian of Eq.  (3.3), the resulting effective potential corresponds
to the one-pion exchange between nucleons.  Since the formalism is not
limited to the nonrelativistic domain of the fermion momenta or to the
lowest order perturbation theory, one can investigate this type of
potentials in a wide range of applications in meson-barion and
quark-pion physics.  However, the second order calculation cannot tell
us how the coupling constant depends on the width $\lambda$.  To
correlate the coupling strength with the hamiltonian width one needs to
calculate the effective hamiltonians including terms order $g^4$.

The remaining question is how can one explain the size of the coupling
constant if it is of order 10.  We cannot offer an explanation in the second
order calculations.  However, we can suggest that the similarity flow
may lead to such large couplings on the basis of the following
reasoning.

The outer similarity factor and the vertex similarity factors continue
to reduce the strength of the effective interactions when $\lambda$
continues to decrease.  The only way the effective hamiltonian of the
very small width may have the same eigenvalues as the unitarily
equivalent large width hamiltonians, is by the rise of the coupling
constant.  Therefore, we forsee that a complete calculation including
the coupling constant dependence on $\lambda$ will lead to the
nucleon-nucleon one-pion-exchange potentials with narrow form factors
and large coupling constants.  One may hope to derive similar structures
for other meson exchanges.  The general trend may be that the smaller is
the meson mass the larger is the effective coupling rise because heavier
effective meson emissions and absorptions are eliminated earlier.

\vskip.3in
\centerline {\bf Fermion-antifermion interaction}
\vskip.1in

The fermion-antifermion interaction order $g^2$ satisfies a differential
equation which is analogous to Eq.  (3.40) but more terms need to be
considered explicitly.  The operator subscripts must distinguish
fermions and antifermions and one has to include terms which result from
the annihilation channels.

The fermion-antifermion
effective interaction term is written as

$$ {\cal G}_{1\bar 1 \bar 1 1 \lambda}
= \int [P] {1 \over P^+} \sum_{\sigma_1 \sigma_2 \sigma_3 \sigma_4}
\int [x\kappa][y\rho] \, g_{1\bar 1 \bar 1 1 \lambda} \,
b^\dagger_{xP+\kappa \sigma_1} \, d^\dagger_{(1-x)P-\kappa \sigma_3}
\, d_{(1-y)P-\rho \sigma_4} \, b_{yP+\rho \sigma_2}
. \eqno(3.73)$$

\noindent Note the change of order of the spin numbering and momentum
assignments in comparison to Eq.  (3.42). The new order results
from the operator ordering defined in Section 2.a.
Momenta $k_1$ and $k_2$ are used for
fermion and $k_3$ and $k_4$ for antifermion operators with even
subscripts for annihilation operators and odd subscripts for creation
operators.

There are three terms contributing to the derivative of $g_{1\bar 1 \bar
1 1 \lambda}$ with respect to $\lambda$:  one due to the annihilation
channel and two due to the exchange of the boson. One of the latter two
contributions results from the emission of the boson by the fermion
and absorption by the antifermion and the other one is for the emission
by the antifermion and absorption by the fermion.  In each of the
terms, there are two similarity functions with different arguments.
Signs in front of each of the three terms are determined by
ordering of creation and annihilation operators.
The result is

$$ { d g_{1\bar 1 \bar 1 1 \lambda} \over d \lambda} =
S_1 g\bar u_1 v_3 \, g \bar v_4 u_2 \, r_{11} r_{13} r_{14} r_{12}
{1 \over P^+} $$
$$ -\left\{
 S_2 r_{21} r_{2512} r_{24} r_{2534} {\theta(y-x) \over (y-x)P^+}
+S_3 r_{32} r_{3512} r_{33} r_{3534} {\theta(x-y) \over (x-y)P^+}
    \right\} g\bar u_1 u_2 \, \, g \bar v_4 v_3 \, .
\eqno(3.74) $$

\noindent The inner similarity factors are

$$ S_i = f(z^2_{i2}) {[-f(z^2_{i2})]'\over \Delta E_{i2}} -
{[-f(z^2_{i1})]' \over \Delta E_{i1}} f(z^2_{i2}) . \eqno (3.75) $$

\noindent Equation (3.75) is similar to Eq.  (3.44) (the subscript
$\lambda$ is skipped for clarity).  The second subscript of the
arguments of the similarity function $f$ denotes the vertex, i.e. 1
stands for the vertex where the boson was annihilated and 2 stands for
the vertex where the boson was created.  In Eq.  (3.74), the fermion
regularization factors first subscript is the same as the corresponding
inner similarity factor subscript (i.e. the subscript of $S$) and the
second subscript is the same as the corresponding fermion momentum
subscript.  The boson regularization factors are distinguished by the
subscript $5$ following the convention from Eqs.  (2.50).  Their first
subscript is also the same as the corresponding inner similarity factor
subscript.  Last two subscripts of the boson regularization factors
equal subscripts of the fermion momenta from the vertex where the boson
regularization factor originated.  Arguments of the regularization
factors have the same subscripts as the regularization factors
themselves, i.e.  $r_i = r(\epsilon e_i /\Lambda^2)$.  The daughter
energies in the arguments are calculated according to the rules given in
Eqs.  (2.47) to (2.50).  We give the results below for completeness.
The same arguments will appear in all theories of physical interest.

$$e_{11} = {\kappa^{\perp \, 2} + m^2 \over x} \, . \eqno(3.76.a) $$

$$e_{13} = {\kappa^{\perp \, 2} + m^2 \over 1-x} \, . \eqno(3.76.b) $$

$$e_{14} = {\rho^{\perp \, 2} + m^2 \over y} \, . \eqno(3.76.c) $$

$$e_{12} = {\rho^{\perp \, 2} + m^2 \over 1-y} \, . \eqno(3.76.d) $$

$$ e_{21} = {\kappa^{\perp \, 2}_{212} + m^2 \over x_{212}} \, .
\eqno(3.77.a)$$

$$ e_{2512} = {\kappa^{\perp \, 2}_{212} + \mu^2 \over 1-x_{212}} \, .
\eqno(3.77.b)$$

$$ x_{212} = {x \over y}        \, . \eqno(3.77.c)$$

$$ \kappa^{\perp}_{212} = \kappa^\perp - x_{212} \rho^\perp \, .
\eqno(3.77.d)$$

$$ e_{24} = {\kappa^{\perp \, 2}_{234} + m^2 \over x_{234}} \, .
\eqno(3.77.e)$$

$$ e_{2534} = {\kappa^{\perp \, 2}_{234} + \mu^2 \over 1-x_{234}} \, .
\eqno(3.77.f)$$

$$ x_{234} = {1-y \over 1-x}        \, . \eqno(3.77.g)$$

$$ \kappa^{\perp}_{234} = -\rho^\perp + x_{234} \kappa^\perp \, .
\eqno(3.77.h)$$

$$ e_{32} = {\kappa^{\perp \, 2}_{312} + m^2 \over x_{312}} \, .
\eqno(3.78.a)$$

$$ e_{3512} = {\kappa^{\perp \, 2}_{312} + \mu^2 \over 1-x_{312}} \, .
\eqno(3.78.b)$$

$$ x_{312} = {y \over x}        \, . \eqno(3.78.c)$$

$$ \kappa^{\perp}_{312} = \rho^\perp - x_{312} \kappa^\perp \, .
\eqno(3.78.d)$$

$$ e_{33} = {\kappa^{\perp \, 2}_{334} + m^2 \over x_{334}} \, .
\eqno(3.78.e)$$

$$ e_{3534} = {\kappa^{\perp \, 2}_{334} + \mu^2 \over 1-x_{334}} \, .
\eqno(3.78.f)$$

$$ x_{334} = {1-x \over 1-y}        \, . \eqno(3.78.g)$$

$$ \kappa^{\perp}_{334} = -\kappa^\perp + x_{334} \rho^\perp \, .
\eqno(3.78.h)$$

\noindent Arguments of the similarity functions and energy denominators
which appear in Eq.  (3.75) are calculated according to the rules given
in Eqs.  (2.12) to (2.19) and (2.24) to (2.26).  The results are
universal for one-particle-exchange two-particle interactions and are
given below for completeness.

$$ \Delta {\cal M}^2_{11} = (k_1 + k_3)^2_\mu - (k_{1m} + k_{3m})^2$$
$$   = \mu^2 - e_{11} - e_{13} \, . \eqno(3.79.a) $$

$$ \Sigma {\cal M}^2_{11} = - \Delta {\cal M}^2_{11} + 2\mu^2
\, . \eqno(3.79.b) $$

$$ \Delta E_{11} = \Delta {\cal M}^2_{11}/P^+
\, . \eqno(3.79.c) $$

$$ \Delta {\cal M}^2_{12} = (k_{2m} + k_{4m})^2 - (k_2 + k_4)^2_\mu$$
$$   = e_{14} + e_{12} - \mu^2  \, . \eqno(3.79.d) $$

$$ \Sigma {\cal M}^2_{12} = \Delta {\cal M}^2_{12} + 2\mu^2
\, . \eqno(3.79.e) $$

$$ \Delta E_{12} = \Delta {\cal M}^2_{12}/P^+
\, . \eqno(3.79.f) $$

$$ \Delta {\cal M}^2_{21} = (k_{2534\mu}+ k_{4m})^2 - k_{3m}^2 $$
$$   =  e_{2534} + e_{24} - m^2  \, . \eqno(3.80.a) $$

$$ \Sigma {\cal M}^2_{21} = \Delta {\cal M}^2_{21} + 2m^2
\, . \eqno(3.80.b) $$

$$ \Delta E_{21} = \Delta {\cal M}^2_{21}/(1-x)P^+
\, . \eqno(3.80.c) $$

$$ \Delta {\cal M}^2_{22} = k_{2m}^2 - (k_{2512\mu} + k_{1m})^2 $$
$$   = m^2 - e_{2512} - e_{21} \, . \eqno(3.80.d) $$

$$ \Sigma {\cal M}^2_{22} = - \Delta {\cal M}^2_{22} + 2\mu^2
\, . \eqno(3.80.e) $$

$$ \Delta E_{22} = \Delta {\cal M}^2_{22}/yP^+
\, . \eqno(3.80.f) $$

$$ \Delta {\cal M}^2_{31} = (k_{3512\mu}+ k_{2m})^2 - k_{1m}^2 $$
$$   =  e_{3512} + e_{32} - m^2 \, . \eqno(3.81.a) $$

$$ \Sigma {\cal M}^2_{31} = \Delta {\cal M}^2_{31} + 2m^2
\, . \eqno(3.81.b) $$

$$ \Delta E_{31} = \Delta {\cal M}^2_{31}/xP^+
\, . \eqno(3.81.c) $$

$$ \Delta {\cal M}^2_{32} = k_{4m}^2 - (k_{3534\mu} + k_{3m})^2 $$
$$   = m^2 - e_{3534} - e_{33} \, . \eqno(3.81.d) $$

$$ \Sigma {\cal M}^2_{32} = - \Delta {\cal M}^2_{32} + 2\mu^2
\, . \eqno(3.81.e) $$

$$ \Delta E_{32} = \Delta {\cal M}^2_{32}/(1-y)P^+
\, . \eqno(3.81.f) $$

\noindent In all cases, the arguments of the similarity functions are
given by Eq.  (2.18), i.e.  $ z_i = {\Delta {\cal M}^2_i /( \Sigma {\cal
M}^2_i + \lambda^2) }$ for all subscripts appearing in Eq.  (3.75).

The same reasoning is used to integrate Eq.  (3.74) as in the case of
Eq.  (3.65) for the fermion-fermion interaction.  For the similarity
function $f$ approaching the $\theta$-function with $u_0 = 1/4$ in Eq.
(2.20), we have $f(z^2_i) = \theta(\lambda^2 + 2m_i^2 - |\Delta {\cal
M}^2_i|)$ with $m_i^2 = \mu^2$ in the first, and $m_i^2 = m^2$ in the
second and third inner similarity factors in Eq.  (3.74).

Integration of Eq. (3.74) gives

$$ g_{1\bar 1 \bar 1 1 \lambda} = g_{1\bar 1 \bar 1 1 \epsilon}
+ c_1 \, g\bar u_1 v_3 \, g \bar v_4 u_2 \, r_{11} r_{13} r_{14}
r_{12}$$
$$ +\left[
 c_2 \, r_{21} r_{2512} r_{24} r_{2534} \, \theta(y-x)
+c_3 \, r_{32} r_{3512} r_{33} r_{3534} \, \theta(x-y)
   \right]\, g\bar u_1 u_2 \, \, g \bar v_4 v_3 \, , \eqno(3.82.a) $$

\noindent where the coefficients are,

$$ c_1 =
  {\theta_{12-11} \theta_{12-\lambda} \over |\Delta {\cal M}^2_{12}|}
+ {\theta_{11-\lambda} \theta_{11-12} \over |\Delta {\cal M}^2_{11}|}
\, , \eqno(3.82.b) $$

$$ c_2 =
  {y \theta_{22-21} \theta_{22-\lambda} \over (y-x)|\Delta {\cal
M}^2_{22}|}
+ {(1-x)\theta_{21-\lambda} \theta_{21-22} \over (y-x)|\Delta {\cal
M}^2_{21}|}
\, , \eqno(3.82.c) $$

$$ c_3 =
  {(1-y) \theta_{32-31} \theta_{32-\lambda} \over (x-y)|\Delta {\cal
M}^2_{32}|}
+ {x \theta_{31-\lambda} \theta_{31-32} \over (x-y)|\Delta {\cal
M}^2_{31}|}
\, . \eqno(3.82.d) $$

\noindent The symbols for $\theta$-functions have the following meaning.
$\theta_{i-j} = \theta (|\Delta {\cal M}^2_i| - |\Delta {\cal M}^2_j|)$
and $\theta_{i-\lambda} = \theta (|\Delta {\cal M}^2_i| - 2m^2_i -
\lambda^2)$ with $m^2_i$ equal $\mu^2$ in $c_1$ and $m^2$ in $c_2$ and
$c_3$.

The next step is the construction of the interaction $F_\lambda\left[
G_{1\bar 1 \bar 1 1 \lambda}\right]$ from ${\cal G}_{1\bar 1 \bar 1 1
\lambda}$ of Eq.  (3.73) using Eqs.  (2.8) and (2.9).

Then, one has to find out if matrix elements of $F_\lambda\left[
G_{1\bar 1 \bar 1 1 \lambda}\right]$ between finite free invariant mass
states have a limit when $\epsilon \rightarrow 0$.  Stated differently,
one checks if the existence of the limit requires the initial value of
$g_{1\bar 1 \bar 1 1 \epsilon}$ to differ from zero to cancel potential
divergences in the limit.  Following the same steps as in the case of
Eqs.  (3.66) and (3.70), one can check that no divergences arise.
Therefore, $g_{1\bar 1 \bar 1 1 \epsilon} = 0$.

The final answer for the effective fermion-antifermion interaction is

$$ H_{1\bar 1 \bar 1 1 \lambda} = F_\lambda\left[ G_{1\bar 1 \bar 1 1
\lambda}\right] $$
$$ = \int [P] {1 \over P^+} \sum_{\sigma_1 \sigma_2 \sigma_3 \sigma_4}
\int [x\kappa][y\rho] \, g_{1\bar 1 \bar 1 1 \lambda}
f(z^2_{1\bar 1 \bar 1 1 \lambda}) \,
b^\dagger_{\lambda xP+\kappa \sigma_1} \,
d^\dagger_{\lambda(1-x)P-\kappa \sigma_3}
\, d_{\lambda (1-y)P-\rho \sigma_4} \, b_{\lambda yP+\rho \sigma_2}
, \eqno(3.83.a)$$

\noindent where

$$ g_{1\bar 1 \bar 1 1 \lambda} = c_1\, g\bar u_1 v_3 \, g \bar v_4 u_2
\,
   + \left[ c_2 \, \theta(y-x) + c_3 \, \theta(x-y) \right] \,
   g\bar u_1 u_2 \, \, g \bar v_4 v_3 \, . \eqno(3.83.b) $$

\noindent In terms of the fermion momenta,

$$ c_1 = {\theta ( s - 3\mu^2 - \lambda^2) \over s - \mu^2 } \, ,
\eqno(3.83.c) $$

\noindent with $ s = max( {\cal M}^2_{13}, {\cal M}^2_{24} )$, and

$$ c_2 \theta(y-x) + c_3 \theta(x-y) =
{\theta_{a-b} \theta_{a-\lambda} \over a} +
{\theta_{b-a} \theta_{b-\lambda} \over b} \, , \eqno(3.83.d)$$

\noindent with $ a = \mu^2 - q_{12}^2 $, $ b = \mu^2 - q_{34}^2 $,
$\theta_{a-b} = 1- \theta_{b-a} = \theta (m_{xy} a - m_{1-x1-y} b) $,
$\theta_{a-\lambda} = \theta \left[ m_{xy} a - |x-y|(2m^2 + \lambda^2)
\right] $, $\theta_{b-\lambda} = \theta \left[ m_{1-x1-y} b - |x-y|(2m^2
+ \lambda^2)\right] $, $m_{xy} = max(x,y)$ and $m_{1-x1-y} =
max(1-x,1-y)$.  The argument of the outer similarity factor in Eq.
(3.83.a), $ z_{1\bar 1 \bar 1 1 \lambda}$, is equal to $z_{22\lambda}$
from Eq.  (3.70.h).  Note that Eqs.  (3.83.a-d) provide the
generalization of Eqs.  (3.70.a-h) to the case of effective interactions
of distinguishable fermions.

When $\lambda^2$ is reduced below $4m^2 - 3\mu^2$, the internal
similarity factor in the annihilation term stays equal 1 independently
of the value of $\lambda$.  The effective interaction term provides the
full contribution of the annihilation channel to the fermion-antifermion
scattering amplitude of order $g^2$.  The fermion-antifermion-boson term
in the effective hamiltonian which could contribute to the scattering
acting twice, is zero for such low values of $\lambda^2$ because the
mass gap between the boson and the fermion pair is larger than $\lambda$
allows.

The internal similarity factor in the exchange term becomes equal 1
independently of $\lambda$ only when $\lambda^2$ becomes smaller than
$-2m^2 + 2 m\mu + \mu^2$.  The lower bound on $\lambda^2$ is $ - m^2 - (
m + \mu )^2$ (see the discussion of Eqs.  (2.18) and (2.20)).  In the
lower bound region, the effective boson emission and absorption vanish
and the exchange interaction term provides the full scattering amplitude
due to the one-boson exchange.  The amplitude is equal to the standard
result on-shell where the outer similarity factor equals 1.

If the boson mass is much smaller than the fermion mass then, for small
momenta, the fermion energies are quadratic functions of momentum while
the boson energy is a linear function of momentum.  Therefore, for
sufficiently small momenta, the boson energy is large in comparison to
the fermion kinetic energies and their changes.  Thus, the
one-boson-exchange interaction is mediated by a relatively high energy
intermediate state.  Consequently, it is represented by a term in the
effective hamiltonian.

For small $\lambda$, the effective hamiltonian contains potentials which
are equal to standard scattering amplitudes in the Born approximation.
The potentials differ from the Born amplitudes off-shell in a unique way
which is dictated by principles of the hamiltonian quantum mechanics and
the similarity renormalization group:  the outer similarity factor
reduces the strength of the interaction off-energy-shell.  In the
light-front dynamics this is equivalent to the off-shellness in the free
invariant mass.

For $\lambda$ outside the lower bound region, the scattering amplitudes
obtain also contributions from the effective interactions which change
the number of bosons by one in the transition through the intermediate
states.  Analysis of Eq.  (3.71) in application to the
fermion-antifermion scattering follows the same steps as for the
fermion-fermion scattering in the previous Section.  The resulting
on-shell scattering amplitude is independent of $\lambda$.  The
amplitude is equal to the well known perturbative result in Yukawa
theory to order $g^2$.

\vskip.3in
{\bf 3.c QED}
\vskip.1in
\nopagebreak

This Section describes calculations of the effective mass squared term
for photons, the effective mass squared term for electrons and the
effective interaction between electrons and positrons in QED.  The
calculated terms are order $e^2$.

The initial expression which we use to calculate the renormalized
hamiltonian of QED is obtained from the lagrangian ${\cal L} = -{1 \over
4} F^{\mu \nu} F_{\mu \nu} + \bar \psi (i\not\!\!D - m)\psi$ by the
procedure of evaluating the energy-momentum tensor $T^{\mu \nu}$
and integrating $T^{+-}$ over the light-front. \cite{Y} The formula we
use reads

$$ H_{QED} = \int dx^- d^2 x^\perp \left[ \bar \psi_m \gamma^+
{-\partial^{\perp 2} + m^2 \over 2 i\partial^+} \psi_m -
{1\over 2} A^\nu_0 \partial^{\perp 2} A_{0\nu} \right.$$
$$ \left.  + e \bar \psi_m \not\!\!A_0 \psi_m + e^2 \bar \psi_m \not\!\!A_0
{\gamma^+ \over 2i\partial^+} \not\!\!A_0 \psi_m + {e^2 \over 2} \bar
\psi_m \gamma^+ \psi_m { 1 \over (i\partial^+)^2} \bar \psi_m \gamma^+
\psi_m \right]_{x^+=0} \, ,  \eqno(3.84) $$

\noindent where $\psi_m$ is a free fermion field with mass $m$ and
$A^\nu_0$ is a free massless photon field with $A^+_0 = 0$.

We replace fields $\psi_m(x)$ and $A^\nu_0(x)$ for $x^+=0$ by the
Fourier superpositions of creation and annihilation operators, we order
the operators in all terms and we drop terms containing divergent
integrals which result from contractions.  This is done in the same way
as in the Yukawa theory but more terms need to be considered.  Then, we
introduce the regularization factors.

The ultraviolet regularization factors are already familiar and the same
as in the Yukawa theory.  The additional regularization is required due
to the infrared singularities.  Photons have diverging polarization
vectors when their $+$-momentum approaches 0. The corresponding seagull
term, i.e. the 5th term in Eq.  (3.84), is diverging too.  We have to
introduce the infrared regularization factors, e.g.  $(1 +
\delta/x)^{-1}$ described in Section 2.b.  We also introduce a photon
mass term $\mu^2_\epsilon = \mu^2_\delta$ by adding it to $ -
\partial^{\perp \, 2}$ in the second term in Eq.  (3.84).  The mass term
requires a few comments.

Consider $\mu^2_\delta$ which is finite and let $e \rightarrow 0$.  The
leading correction order $e^2$ is small in comparison to $\mu^2_\delta$.
The situation changes when $\mu^2_\delta \rightarrow 0$ while $e$ is
kept fixed.  At some point, the correction order $e^2$ becomes larger
than $\mu^2_\delta$ itself and one cannot consider the correction to be
a small perturbation on the chargeless theory.

Unfortunately, the limit $\mu^2_\delta \rightarrow 0$ is not completely
understood in the present approach.  The second order calculations
described in this paper are not sufficient.  Higher order calculations
need to be done to gain more understanding.  In particular, it is not
known what conditions correspond to gauge invariance identities used in
the lagrangian approach to make the massless theory exist.  It is also
important to include the running coupling effects.

The width dependence of the running coupling constant can be expected to
matter in the analysis of the initial value problem for the 
photon mass term.  The
perturbative mass correction for photons diverges when the ultraviolet
cutoff is being removed.  The finite part of the photon mass counterterm
is not known.  Therefore, we have to fit it to data.  The divergent term
is order $e^2$.  The finite part must be of the same order.  In QED, the
running coupling is expected to grow with the hamiltonian width.  Thus,
one may expect that a very small mass term order $e^2$ in low energy
effective QED may correspond to a much larger photon mass term in the
initial hamiltonian with $\lambda = \infty$ due to the much larger
coupling constant at the scale $1/\epsilon$. However, the small size
of the effective fine structure constant makes the issue rather academic. 

Introduction of the photon mass is relevant to the effective infrared
dynamics.  The cutoff $\delta$ allows photons 
with $x \sim \delta$ to appear in the
effective interactions of a finite width $\lambda$.  Such a photon may
contribute only a little to an intermediate state energy if it has
sufficiently small transverse momentum.  Therefore, emissions and
absorptions of such photons are allowed even by the narrow similarity
factors.  The massless photons with small $x$ introduce dependence on
the cutoff $\delta$ into the effective hamiltonian.  Moreover, large
numbers of such photons can appear in the eigenstates.  But large
occupation numbers for bosons enhance the strength of the emission and
absorption processes.  This enhancement could, in principle, compete
with the smallness of the coupling constant.  A finite photon mass
suppresses these effects and, as a consequence, may become a major
alteration of the theory.  On the other hand, the mass may remain to be
merely a stabilizing factor in the calculations if the effective
dynamics turns out to be dominated by a limited range of photon
wavelengths so that the momenta order $\delta$ or smaller do not matter.
This is certainly expected to happen in the finite size neutral bound
states.

A fixed value of $\mu_\delta$ leads to the conclusion in perturbation
theory that the photon eigenstates have masses equal to $\mu_\delta$
when the charge $e$ approaches $0$.  Therefore, we will be forced to
consider the limit $\mu_\delta \rightarrow 0$ in order to discuss
physical photons to order $e^2$.  Also, the nonzero mass squared term
for photons leads to additional divergences when $\delta \rightarrow 0$
and the limit of $\mu_\delta \rightarrow 0$ removes those.

The infrared finiteness of QED suggests that physical results may be
independent of $\mu_\delta$ when it is sufficiently small.  But we do
not provide an explanation of how this comes about in the light-front
hamiltonian approach in general.  We introduce the photon mass
$\mu_\delta$ and investigate the limit $\mu_\delta \rightarrow 0$.
The lowest order calculations in this paper lead to results which are
independent of $\mu^2_\delta$ when it tends to zero.

\vskip.3in
\centerline {\bf Photon mass squared}
\vskip.1in

The same procedure from Section 2 which led to Eqs. (3.4) and (3.5) in
the case of mesons in Yukawa theory, leads in QED to

$$ {\cal G}_{1\, photon \,\lambda} = \sum_\sigma \int[k] { k^{\perp 2} +
\mu^2_\lambda \over k^+ } a^\dagger_{k\sigma} a_{k\sigma} \, . \eqno
(3.85)
$$

\noindent No correction arises to the term $k^{\perp 2} /k^+$
because our regularization preserves the kinematical symmetries of
light-front dynamics.

A new feature in comparison to the Yukawa theory is the polarization of
photons.  With the kinematical symmetries explicitly preserved, only
terms diagonal in the photon polarization emerge.  For example, terms
proportional to $k^i \varepsilon^i_{\sigma_1} \, k^j
\varepsilon^j_{\sigma_2}$ with $\sigma_1 \neq \sigma_2$ cannot appear
because the regularization and similarity factors do not introduce
dependence on the photon momentum.  Note that such terms are allowed by
the power counting.  \cite{W3}

The net result of the photon self-interaction is an effective photon
mass squared term which is independent of the photon momentum but varies
with the effective hamiltonian width $\lambda$.  One obtains more
complicated results for the effective photon free energy if the
regularization or similarity factors violate kinematical symmetries of
light-front dynamics. \cite{W3}

The dependence of $\mu^2_\lambda$ on $\lambda$ is determined to order
$e^2$ by the equation

$$ {d \mu^2_\lambda \over d\lambda} \delta^{\sigma_1 \sigma_2} = e^2
\int[x\kappa] {d f^2(z^2_\lambda) \over d\lambda} { Tr
\not\!\varepsilon^*_{k \sigma_1} (\not\!k_{1m} + m) \not\!\varepsilon
_{k \sigma_1} (\not\!k_{2m} - m) \over {\cal M}^2 - \mu^2_\delta}
r_{\epsilon }(x,\kappa) \,  \eqno (3.86) $$

\noindent and the initial condition at $\lambda = \infty$.  However, it
is also sufficient to know the effective mass squared at any single
value of $\lambda$ to determine its value at other values of $\lambda$
using Eq.  (3.86).  The initial condition at $\lambda = \infty$ is
distinguished only because it provides connection with standard
approaches based on the local lagrangian for electrodynamics.

In Eq.  (3.86), ${\cal M}^2 = (\kappa^{\perp \, 2} + m^2)/x(1-x)$,
$\Delta {\cal M}^2 = {\cal M}^2 - \mu^2_\delta$ and $\Sigma {\cal M}^2 =
{\cal M}^2 + \mu^2_\delta$ so that $z_\lambda = ({\cal M}^2 -
\mu^2_\delta) /({\cal M}^2 + \mu^2_\delta + \lambda^2)$.  In the limit
of Eq.  (3.8), we have $f^2(z^2_\lambda) = \theta(\lambda^2 +
3\mu^2_\delta - {\cal M}^2)$.  In fact, Eq.  (3.86) is free from
infrared singularities and we could skip the introduction of
$\mu^2_\delta$ by letting it go to zero at this point.  However, the
systematic approach defines the hamiltonian of QED including the
infrared regulator mass for photons and we can keep it here for
illustration.  The regularization factor $r_\epsilon (x,\kappa)$ in Eq.
(3.86) is the same as in Eq.  (3.9) in Yukawa theory because only
fermion regularization factors enter Eq.  (3.86), according to Eqs.
(2.47) to (2.49), and these factors are the same in both theories.

Evaluation of the spin factor gives

$$ {d \mu^2_\lambda \over d\lambda} = e^2 \int[x\kappa] \, {d
f^2(z^2_\lambda) \over d\lambda} \, { 2 {\cal M}^2 - 4 \kappa^{\perp \,
2} \over {\cal M}^2 - \mu^2_\delta} \, r_{\epsilon }(x,\kappa) , \eqno
(3.87) $$

\noindent which is the QED analog of Eq.  (3.6) from Yukawa theory.

Integration of Eq.  (3.87) is carried out through the same steps as in
the Yukawa theory.  We can use Eq.  (3.22) to calculate the effective
photon mass squared $\mu^2_\lambda$ knowing its value $\mu^2_0$ at some
value of $\lambda = \lambda_0$.

The value of $\mu^2_0$ is found by requesting that the effective
hamiltonian eigenvalues for photon states contain the physical photon
mass $\tilde \mu$, expected to be equal 0. However, solving the
eigenvalue equation to second order in the coupling constant $e$ through
the same steps as in the case of mesons in Yukawa theory in Eqs.  (3.23)
to (3.25), leads to the physical photon mass ${\tilde \mu} =
\mu_\delta$.  $\mu_\delta$ is small but finite and it is considered, in
terms of powers of $e$, to be of order $e^0 = 1$ when $e \rightarrow 0$.

In the $\theta$-function limit for the similarity function $f$ with $u_0
= 1/4$,

$$ \mu^2_0 = \mu^2_\delta + {\alpha \over 4\pi} \int_0^1dx \int_0^\infty
d\kappa^2 { 2 {\cal M}^2 - 4 \kappa^{\perp \, 2} \over x(1-x){\cal M}^2
- \mu^2_\delta} \theta(\lambda^2_0 + 3\mu^2_\delta -{\cal M}^2) + o(e^4)
\, , \eqno(3.88) $$

\noindent Thus, at other values of $\lambda$, the effective photon mass
squared is

$$ \mu^2_\lambda = \mu^2_\delta + {\alpha \over 4\pi} \int_0^1dx
\int_0^\infty d\kappa^2 { 2 {\cal M}^2 - 4 \kappa^{\perp \, 2} \over
x(1-x){\cal M}^2 - \mu^2_\delta} \theta(\lambda^2 + 3\mu^2_\delta -{\cal
M}^2) + o(e^4) \, . \eqno(3.89) $$

\noindent This result naturally depends on the infrared regularization
parameter $\mu^2_\delta$ but no singularity appears when this parameter
is set equal to zero.  For $ \lambda^2 + 3\mu^2_\delta \leq 4m^2$, where
$4m^2$ is the lowest possible free invariant mass squared for the two
intermediate fermions, the photon mass is independent of the hamiltonian
width $\lambda$ and it equals $\mu^2_\delta$.  For larger values of
$\lambda^2$, the effective photon mass grows with the width $\lambda$ so
that its larger value compensates effects of the interactions which
become active for the larger width.  The net result is that the photon
eigenstates have eigenvalues with masses squared equal $\mu^2_\delta$
independently of $\lambda$.  Finally, the result favored by experimental
data is obtained in the limit $\mu^2_\delta \rightarrow 0$ at the end of
the calculation.

\vskip.3in
\centerline {\bf Electron mass squared}
\vskip.1in

Electron and positron self-interactions through emission and
reabsorbtion of transverse photons lead to the fermion free energy terms
of the form exactly the same in QED as in Eq.  (3.27) in Yukawa theory.
However, the effective mass of electrons and positrons depends on the
width differently than in the case of Yukawa theory.  Instead of Eq.
(3.28), one obtains now

$$ {d m^2_\lambda \over d\lambda} = e^2 \sum_{\tilde \sigma}
\int[x\kappa] {df^2(z^2_\lambda) \over d\lambda}
{ {\bar u}_{m\sigma k} \not\!\varepsilon_{\tilde k \tilde \sigma}
(\not\!p_m + m)
\not\!\varepsilon^*_{\tilde k \tilde \sigma} u_{m\sigma k} \over
{\cal M}^2 - m^2} \, r_{\epsilon \delta}(x,\kappa) \, ,\eqno (3.90) $$

\noindent where ${\tilde k} = (k - p)_0$, $p^+ = xk^+$, $p^\perp =
xk^\perp + \kappa^\perp$, ${\cal M}^2 = (m^2 + \kappa^2)/ x  +
(\mu_\delta^2 + \kappa^2)/(1-x)$, $\Delta {\cal M}^2 = {\cal M}^2 -
m^2$, $\Sigma {\cal M}^2 = {\cal M}^2 + m^2$ and $z_\lambda = \Delta
{\cal M}^2 /(\Sigma {\cal M}^2 + \lambda^2)$.  The regularization factor
of Eq.  (2.49) for the intermediate particles and the infrared regulator
for the intermediate photon, as given at the end of Section 2.b, imply

$$ r_{\epsilon \delta}(x,\kappa) = \left[ 1+{\epsilon\over\Lambda^2}
{\cal M}^2 + \left({\epsilon\over\Lambda^2}\right)^2{\kappa^2 +m^2 \over
x}{\kappa^2 +\mu^2_\delta \over 1-x}\right]^{-2} \left(1 + {\delta \over
1-x} \right)^{-2} \, .\eqno(3.91) $$

\noindent The sum over photon polarizations in Eq. (3.90) produces
the well known expression

$$ \sum_{\tilde \sigma} \varepsilon^\alpha_{\tilde k \tilde \sigma}
\varepsilon^{* \beta}_{\tilde k \tilde \sigma} = -g^{\alpha \beta} +
{{\tilde k}^\alpha g^{+\beta} + g^{+\alpha} {\tilde
k}^\beta \over {\tilde k}^+} \,\, , \eqno(3.92) $$

\noindent and the spin factor in Eq. (3.90) is

$$ {\bar u}_{m\sigma k} \gamma_\alpha (\not\!p_m + m) \gamma_\beta
u_{m\sigma k} \left[-g^{\alpha \beta} + {{\tilde k}^\alpha g^{+\beta} +
g^{+\alpha} {\tilde k}^\beta \over {\tilde k}^+} \right] = {2\over
x}\left[ (1-x)^2 m^2 + \kappa^2 {1 + x^2 \over (1-x)^2}\right] \, .
\eqno(3.93)$$

\noindent The new feature of this expression, in comparison to the
Yukawa theory, is the divergence for $x \rightarrow 1$, i.e. where the
photon longitudinal momentum approaches 0.

The rate of change of the electron mass term versus the effective
hamiltonian width in the $\theta$-function limit for the similarity
function with $u_0 = 1/4$ is following.

$$ {d m^2_\lambda \over d\lambda^2} = {\alpha \over 4\pi} \int_0^1 dx
\int_0^\infty du \,\, \delta(3m^2 + \lambda^2 - {\cal M}^2) \, {m^2
{2(1-x)^2 \over x} + u {2(1 + x^2) \over 1-x} \over {\cal M}^2 - m^2} \,
r_{\epsilon \delta}(x,\kappa) \, ,\eqno (3.94.a) $$

\noindent where $u = \kappa^2/x(1-x)$,

$$ r_{\epsilon \delta}(x,\kappa) = \left[ 1+{\epsilon\over\Lambda^2}
{\cal M}^2 + \left({\epsilon\over\Lambda^2}\right)^2 \left[(1-x)u + {m^2
\over x}\right]\left[x u + {\mu^2_\delta \over 1-x}\right]\right]^{-2}
\left(1 + {\delta \over 1-x} \right)^{-2} \, ,\eqno(3.94.b) $$

\noindent and

$$ {\cal M}^2 = u + {m^2 \over x} + {\mu_\delta^2 \over 1-x}
\, .\eqno(3.94.c) $$

The divergence structure of the effective electron mass in Eq.  (3.94.a)
is obscured by the fact that the whole effective mass term is merely a
number dependent on $\lambda$ while three cutoff and regularization
parameters appear in the integral:  $\epsilon$, $ \delta$ and
$\mu_\delta$.  The only available condition is that the effective
electron mass should have a limit when $\epsilon \rightarrow 0$.
However, this condition has to be satisfied without generating
divergences in the physical electron mass (i.e. in the electron
eigenvalue energy) when we remove the infrared regularization.  Since
other contributions to the physical electron mass may diverge as
$\delta$ or $\mu_\delta$ tend to 0, and only the sum is finite in the
limit, one needs to keep track of the infrared structure in defining the
$\epsilon$-independent (i.e. ultraviolet finite) part of the
counterterm.

The divergences due to $\epsilon \rightarrow 0$, $\delta \rightarrow 0$
and $\mu_\delta \rightarrow 0$, are not resolved in the single mass
constant.  Many elements of a complete analysis overlap in producing the
final answer and many simplifications are possible.  We will proceed in
this Section with a simplified analysis.  A more extended analysis will
be required for other hamiltonian terms where the outcome of the
procedure is not reduced to finding only one number in the effective
interaction.  For example, in the electron-positron interaction term,
the external momenta of fermions introduce a whole range of additional
parameters. That case will be illustrated in the next Section.

The $\delta$-function under the integral on the right-hand side of Eq.
(3.94.a) forces ${\cal M}^2$ to be equal $\lambda^2 + 3m^2$.  The
smallest possible value of $\cal M$ is $m + \mu_\delta$ and the negative
value of $\lambda^2$, namely, $(m + \mu_\delta)^2 - 3m^2 = - m^2 - ( m +
\mu_\delta )^2 + 2[ (m + \mu_\delta)^2 - m^2]$, is required to reach
this lower bound (see the comments about Eq.  (2.20)).  Below this bound
the right-hand side of Eq.  (3.94.a) vanishes, no emission or absorption
of photons by electrons is possible and the effective electron mass
stays constant.  The smallest possible value of $\lambda^2$ allowed by
Eq.  (2.18) is $ - m^2 - ( m + \mu_\delta)^2$.  The difference between
these bounds vanishes when the photon mass goes to zero.

In the next Section, we will also consider the case of the infinitesimal
$u_0$ (see the discussion below Eq.  (2.20)), which leads to
$\delta(\tilde \lambda^2 + m^2 - {\cal M}^2)$ in Eq.  (3.94.a), instead
of $\delta(3m^2 + \lambda^2 - {\cal M}^2)$.

For $\lambda^2 > ( m + \mu_\delta)^2 - 3m^2 $, the right-hand side of
Eq.  (3.94.a) is positive.  Therefore, the effective electron mass
squared term grows together with the width of the hamiltonian.  We will
see below that this growth combines with the growing negative
contributions of the corresponding effective transverse photon emission
and reabsorption, so that the physical electron mass is independent of
the hamiltonian width.

For $\lambda^2$ close to the lower bound, $u$ must be close to 0 and $ x
\sim 1 - \mu_\delta/(m + \mu_\delta)$.  But ${\cal M}^2$ is limited and
determined by the value of $\lambda$.  Quite generally, as long as $\cal
M$ remains limited, $u$ and $\mu^2_\delta/(1-x)$ are limited.  The
invariant mass denominator in Eq.  (3.94.a) equals $\lambda^2 + 2m^2$.
The denominator is small only when $\lambda^2$ is close to the lower
bound.  Then, the integration range is small too.  The denominator, when
expressed in terms of $\lambda$, can be pulled out and put in front of
the integral.  The integration over $u$ sets $u = \tilde u(x) =
\lambda^2 + 3m^2 - m^2/x -\mu^2_\delta/(1-x)$ and forces the condition
$\tilde u(x) > 0$.  This condition implies the following limits on the
integration over $x$, provided $ \lambda^2 + 3m^2 > (m + \mu_\delta)^2 $
since otherwise the integral is 0.

$$ x_0 - \Delta x < x < x_0 + \Delta x \, , \eqno(3.95.a) $$

$$ x_0 = {1\over 2}\left( 1 + {m^2 - \mu_\delta^2 \over \lambda^2 +
3m^2}\right) \, , \eqno(3.95.b) $$

$$ \Delta x = \sqrt{ x_0^2 - {m^2\over \lambda^2 + 3m^2}} \, .
\eqno(3.95.c) $$

\noindent Within these limits, $\tilde u(x)$ varies from the minimal
value of 0 at the lower bound $x_0 - \Delta x$ through a maximum of
$\lambda^2 + 3m^2 - (m + \mu_\delta)^2$ at $x = m/(m + \mu_\delta)$ to
the minimal value of 0 again at the upper bound $x_0 + \Delta x$. In the
case of infinitesimal $u_0$, one replaces $\lambda^2 + 2 m^2$ in these
formulae by $\tilde \lambda^2$.

Eq. (3.94.a) reads

$$ {d m^2_\lambda \over d\lambda^2} = {\alpha \theta [3m^2 + \lambda^2 -
(m + \mu_\delta)^2] \over 4\pi(\lambda^2 + 2m^2)} \int_{x_0 - \Delta
x}^{x_0 + \Delta x} dx \left[m^2 {2(1-x)^2\over x} + {\tilde u}(x) {2(1
-+ x^2) \over 1-x}\right] \, r_{\epsilon \delta}(x,\tilde u(x)) \,
, \eqno (3.96.a) $$

\noindent where

$$ r_{\epsilon \delta}(x, \tilde u(x) ) = \left[
1+{\epsilon\over\Lambda^2} (3m^2 + \lambda^2) +
\left({\epsilon\over\Lambda^2}\right)^2 \left[(1-x) \tilde u(x) + {m^2 \over
x}\right] \left[x \tilde u(x) + {\mu^2_\delta \over 1-x}\right] \right]^{-2}$$
$$ \times \left(1 + {\delta \over 1-x} \right)^{-2} \, .\eqno(3.96.b) $$

The upper limit of integration over $x$ for $\mu^2_\delta$ much smaller
than $m^2$ and $ 2 m^2 + \lambda^2$ , equals $ 1 - \mu^2_\delta/(2m^2 +
\lambda^2)$ and approaches 1 when $\mu_\delta \rightarrow 0$.  For $x$
close to 1, the factor $(1-x)^{-1}$ in the square bracket of the
integrand is large and leads to a logarithmic dependence of the integral
on the upper integration limit.  The logarithm would become infinite for
$\mu_\delta \rightarrow 0$ if $\delta$ were equal 0. Therefore, the
limit $\mu_\delta \rightarrow 0$ is sensitive to the presence of the
regularization factor with $\delta \neq 0$.  For the finite
$\mu_\delta$, the region of $ x \rightarrow 1$ is regulated by
$\epsilon$ and can be considered an ultraviolet limit.  A counterterm to
the diverging $\epsilon$ dependence could remove the divergence due to
$x \rightarrow 1$.  Then, a separate cutoff parameter $\delta$ would not
be needed but the resulting terms would diverge for $\mu^2_\delta
\rightarrow 0$.  For finite $\mu^2_\delta$, the derivative of the
electron mass with respect to $\lambda$ is finite.  For $\mu^2_\delta =
0$, the integrand in the region $x \sim 1$ is regulated solely by the
infrared regularization factor $[1 +\delta/(1-x)]^{-2}$ since the upper
limit of integration over $x$ is equal 1.

For finite $\lambda^2$, all three terms in ${\cal M}^2$, i.e.  $\tilde u
(x)$, $m^2/x$ and $\mu^2_\delta/(1-x)$, are limited.  Therefore, for
finite $\mu_\delta$, one can take the limit $\epsilon \rightarrow 0$
in the integrand.  The factor $[1 - \delta/(1-x)]^{-2}$ remains and
additionally cuts off the integration region at $x \sim 1 - \delta$.
The ratio of $\delta$ to $\mu^2_\delta/(2m^2 +\lambda^2)$ determines the
size of contributions obtained from the upper range of integration over
$x$.  For finite fixed values of $\mu_\delta$, one can take the limit
$\delta \rightarrow 0$ and $\log{[(2m^2 +\lambda^2)/\mu^2_\delta]}$
appears in the answer.

The right-hand side of Eq.  (3.96.a) contains terms which behave for
large $\lambda$ as a constant, as $\lambda^{-2}$ with factors of
logarithms of $\lambda$ and as functions vanishing faster than
$\lambda^{-2}$.  One integrates Eq.  (3.96.a) over $\lambda$ from
$\lambda_0$ to infinity in order to express the effective electron mass
squared term at $\lambda_0$, denoted $m^2_0$, in terms of the initial
$m^2_\epsilon$.  Clearly, the integration over $\lambda$ would diverge
without the regularization factor which depends on $\epsilon$.  The
integration produces terms behaving as $\epsilon^{-1}$, $\log \epsilon$
and terms convergent in the limit $\epsilon \rightarrow 0$.
$m^2_\epsilon$ in the initial hamiltonian must be supplied with a
counterterm to subtract the diverging $\epsilon$-dependent terms in the
effective hamiltonians.

In summary, the infrared divergence due to $\mu_\delta \rightarrow 0$
and $\delta \rightarrow 0$ appears in the derivative of $m^2_\lambda$
with respect to $\lambda$.  Therefore, even if one requests that the
electron mass term is finite at some value of $\lambda$, the effective
masses of electrons in the neighbouring hamiltonians with even slightly
different widths will diverge when $\mu_\delta$ and $\delta$ approach 0.
We have to abandon the requirement that the effective electron mass term
at any value of $\lambda$ remains finite when the infrared
regularization is removed.  The effective masses diverge in the limit
$\mu_\delta \rightarrow 0$ and $\delta \rightarrow 0$.  The only
condition we can fulfill through the ultraviolet renormalization is that
the effective electron masses for finite $\lambda$ are independent of
$\epsilon$.

Mathematical details of the effective electron mass term calculation are
more complicated than in the Yukawa theory because the infrared
regularization parameters are present.  Otherwise, the calculation is
essentially the same and we skip the description here.  We only stress
that the counterterm and the effective masses of electrons and positrons
depend on the infrared cutoffs and they diverge when the cutoffs are
being removed.

Thus, the effective electron mass squared term in the limit $\epsilon
\rightarrow 0$ is

$$ m^2_\lambda = m^2_0 + e^2 \int [x\kappa] \left[f^2(z^2_\lambda) -
f^2(z^2_{\lambda_0})\right] { m^2 {2(1-x)^2\over x} + {\kappa^2 \over
x(1-x)} {2(1 + x^2) \over 1-x} \over {\cal M}^2 - m^2} \left(1+{\delta
\over 1-x}\right)^{-2}  + o(e^4) \, .\eqno (3.97) $$

\noindent The finite term $m^2_0$ has a limit when $\epsilon \rightarrow
0$.  Its dependence on the infrared regularization is not displayed.
$m^2_0$ is found from a suitable renormalization condition.

The natural condition to be satisfied by $m^2_0$ is that the effective
hamiltonian of some width $\lambda$ has the electron eigenstates with
eigenvalues equal $(p^{\perp \, 2} + {\tilde m}^2)/p^+$, where $p$
denotes the electron momentum and $\tilde m$ is the physical electron
mass.  The problems of measuring the physical electron mass and
comparing the result with the eigenvalue of the hamiltonian, are not
considered here.

The eigenvalue equation for electrons can be solved in perturbation
theory in the same way as for bosons in Yukawa theory in Eqs.  (3.23) -
(3.26) and fermions in Eqs.  (3.35) - (3.39), or in QED for photons
in Eqs.  (3.88) - (3.89).

One obtains the condition

$$ {\tilde m}^2 = m^2_0 - e^2 \int [x\kappa]
f^2(z^2_{\lambda_0}) { m^2 {2(1-x)^2\over x} + {\kappa^2 \over
x(1-x)} {2(1 + x^2) \over 1-x} \over {\cal M}^2 - m^2} \left(1+{\delta
\over 1-x}\right)^{-2}  + o(e^4) \, ,\eqno (3.98) $$

\noindent and one can calculate $m_0$ from this condition.
Consequently,

$$ m^2_\lambda = {\tilde m}^2 + e^2 \int [x\kappa] f^2(z^2_\lambda) {
m^2 {2(1-x)^2\over x} + {\kappa^2 \over x(1-x)} {2(1 + x^2) \over 1-x}
\over {\cal M}^2 - m^2} \left(1+{\delta \over 1-x}\right)^{-2} + o(e^4)
\, ,\eqno (3.99) $$

\noindent and $m^2 = {\tilde m}^2 + o(e^2)$.

The physical electron mass is independent of the infrared regularization
because the regularization dependent $m^2_\lambda$ and the effective
emission and absorption of photons combine to the regularization
independent result.  Note that the effective mass squared term for
electrons depends on the infrared regularization parameters and it
diverges when the infrared regularization is removed. Higher order
calculations are required to discuss contributions of the infrared
photons in the experimental determination of the electron mass.

\vskip.3in
\centerline {\bf Electron-positron interaction}
\vskip.1in

Calculation of the effective electron-positron interaction to order
$e^2$ is of general interest as a fundamental force in electrodynamics -
this interaction is responsible for the formation of positronium.  Also,
effective interactions between quarks and antiquarks in lowest order QCD
effective hamiltonian will have similar features and the present
calculation of the electron-positron interaction in QED provides an
introduction to the QCD case.

Generally speaking, the QED calculation of the effective
electron-positron interaction proceeds in the same way as in the case of
fermion-antifermion interaction in Yukawa theory except for three new
elements.

The first is that photons have polarization vectors which enter in the
vertex factors and introduce additional dependence on the exchanged
photon momentum.  This dependence leads to infrared divergences for
small longitudinal momenta of exchanged photons.

The second feature is that the infrared divergences require additional
regularization factors.  We use the parameter $\delta$ and the photon
mass squared $\mu_\delta \neq 0$.  The limits $\delta \rightarrow 0$ and
$\mu_\delta \rightarrow 0$ are generally understood as to be taken at
the end of a calculation of observables and not in the effective
hamiltonian itself.  However, it may also be possible to take the limits
in the matrix elements of the hamiltonian between the states which do
not induce infrared divergences, i.e. do not involve small $x$ photons.

The third feature is that the one-photon exchange interaction needs to
be combined with the 5th term from Eq.  (3.84) to obtain the standard
results for the electron-positron scattering in the Born approximation.
The 5th term from Eq.  (3.84) provides the initial condition for the
renormalization group flow of the effective hamiltonians.  In order
$e^2$, this term is only supplied with the outer similarity factor by
the operation $F_\lambda$.  It does not change in the flow otherwise
because it is order $e^2$ itself.  The initial condition provides an
important contribution which leads to the Coulomb interaction.  The
one-photon exchange by itself is not sufficient to produce the effective
Coulomb potential.

This is a different situation than in Yukawa theory where no
four-fermion seagull interactions appeared and the one-meson exchange
interaction was sufficient to produce the Yukawa potential in the
effective hamiltonians of small widths.

The effective electron-positron interaction term has the analogous
structure as the fermion-antifermion interaction in Eq.  (3.73), i.e.

$$ {\cal G}_{1\bar 1 \bar 1 1 \lambda}
= \int [P] {1 \over P^+} \sum_{\sigma_1 \sigma_2 \sigma_3 \sigma_4}
\int [x\kappa][y\rho] \, g_{1\bar 1 \bar 1 1 \lambda} \,
b^\dagger_{xP+\kappa \sigma_1} \, d^\dagger_{(1-x)P-\kappa \sigma_3}
\, d_{(1-y)P-\rho \sigma_4} \, b_{yP+\rho \sigma_2}
\, . \eqno(3.100)$$

\noindent The coefficient function $g_{1 \bar 1 \bar 1 1 \lambda}$
of order $e^2$ satisfies the differential equation

$$ { d g_{1\bar 1 \bar 1 1 \lambda} \over d \lambda} = S_1
\sum_{\sigma_5} e\bar u_1 \not\!\varepsilon_{k_5 \sigma_5} v_3 \, e \bar
v_4 \not\!\varepsilon^*_{k_5 \sigma_5} u_2 \, r_{11} r_{13} r_{14} r_{12}
{1 \over P^+} $$
$$ - S_2 r_{21} r_{2512} r_{24} r_{2534} r_{3/5} r_{2/5}
 {\theta(y-x) \over (y-x)P^+}
\sum_{\sigma_5} e \bar v_4 \not\!\varepsilon_{k_5 \sigma_5} v_3
e \bar u_1 \not\!\varepsilon^*_{k_5 \sigma_5} u_2 \, $$
$$ - S_3 r_{32} r_{3512} r_{33} r_{3534} r_{1/5} r_{4/5}
{\theta(x-y) \over (x-y)P^+}
\sum_{\sigma_5} e \bar u_1 \not\!\varepsilon_{k_5 \sigma_5} u_2 \,
e \bar v_4 \not\!\varepsilon^*_{k_5 \sigma_5} v_3
\, , \eqno(3.101) $$

\noindent which is the QED analog of Eq.  (3.74) from Yukawa theory.
Notation is the same as in Eqs.  (3.74) to (3.81) with the exception
that $\mu^2$ is replaced by $\mu^2_\delta$.  The new elements are the
infrared regularization factors of Section 2.b, i.e.  $r_{3/5} = r(
k_3^+ \delta /k_5^+)$, $r_{2/5} = r( k_2^+ \delta /k_5^+)$, $r_{1/5} =
r( k_1^+ \delta /k_5^+)$, $r_{4/5} = r( k_4^+ \delta /k_5^+)$, and the
photon polarization vectors.  The sum over the photon polarizations
gives

$$ \sum_{\sigma_5} \varepsilon_{k_5 \sigma_5}^\alpha
\varepsilon^{* \beta}_{k_5 \sigma_5} =
- g^{\alpha \beta} + {k^\alpha_{50} g^{\beta +} + g^{\alpha +}
k^\beta_{50} \over k^+_\beta}
\, . \eqno(3.102)$$

\noindent The terms proportional to the four-vector $k_{50}$ can be
rewritten using the Dirac equation for free fermions of mass $m$ into a
simpler form.  For example, in the second term on the right-hand side of
Eq.  (3.101), we have

$$ \bar u_1 \not\!k_{50} u_2 \, = \bar u_1 \left[ \not\!k_{2m}
- \not\!k_{1m} + {1\over 2} \gamma^+ [ (k_2 - k_1)^-_0 - k_{2m}^- +
k_{1m}^-] \right] u_2 \, = \bar u_1 \gamma^+ u_2 \, { -(k_{2m}-k_{1m})^2 \over
2(k_2^+ - k_1^+)} \, . \eqno(3.103) $$

\noindent Using similar relations for all vertex factors involved one
obtains

$$ { d g_{1\bar 1 \bar 1 1 \lambda} \over d \lambda} = S_1 \left[
-g^{\mu \nu} - g^{\mu +} g^{\nu +} {s_{13} + s_{24} \over 2 P^{+\,2} }
\right] e\bar u_1 \gamma_\mu v_3 \, e \bar v_4 \gamma_\nu u_2 \, r_{11}
r_{13} r_{14} r_{12} {1 \over P^+} $$
$$ - \left\{
     S_2 r_{21} r_{2512} r_{24} r_{2534}r_{3/5}
     r_{2/5}{\theta(y-x) \over (y-x)P^+}
  + S_3 r_{32} r_{3512} r_{33} r_{3534}r_{1/5}
     r_{4/5} {\theta(x-y) \over (x-y)P^+}
     \right\} $$
$$  \times \left[ -g^{\mu \nu} - g^{\mu +} g^{\nu +} {q^2_{12} +
q^2_{34}
     \over 2 (x-y)^2 P^{+\,2} } \right]
     e \bar u_1 \gamma_\mu u_2 \,  e \bar v_4 \gamma_\nu v_3 \,
\, , \eqno(3.104) $$

\noindent We use the notation $ s_{ij} = (k_i + k_j)^2 $ and
$ q^2_{ij} = (k_i - k_j)^2 $.

Integration of Eq.  (3.104) proceeds in the same way as in the case of
Eq.  (3.74).  The initial condition at $\lambda = \infty$ includes the
seagull term.

$$ H^{seagull}
= \int [P] {1 \over P^+} \sum_{\sigma_1 \sigma_2 \sigma_3 \sigma_4}
\int [x\kappa][y\rho] \, g^{seagull}_{1\bar 1 \bar 1 1} \,
b^\dagger_{xP+\kappa \sigma_1} \, d^\dagger_{(1-x)P-\kappa \sigma_3}
\, d_{(1-y)P-\rho \sigma_4} \, b_{yP+\rho \sigma_2}
\, , \eqno(3.105.a)$$

\noindent where

$$ g^{seagull}_{1\bar 1 \bar 1 1} =
    - \left[
           r_{21} r_{2512} r_{24} r_{2534} r_{3/5}
  r_{2/5} { \theta(y-x) \over (y - x)^2}  \right. $$
$$ \left. + r_{32} r_{3512} r_{33} r_{3534} r_{1/5}
           r_{4/5} { \theta(x-y) \over (x-y)^2} \right]
  e \bar u_1 \gamma_\mu u_2 \,  e \bar v_4 \gamma_\nu v_3 \,
  { g^{\mu+} g^{\nu +} \over P^{+\,2}}  $$
$$    + e\bar u_1 \gamma_\mu v_3 \,  \, e \bar v_4 \gamma_\nu u_2 \,
     r_{11} r_{13} r_{14} r_{12}
     { g^{\mu+} g^{\nu +} \over P^{+\,2}}
\, . \eqno(3.105.b)$$

The result of the integration of Eq.  (104) is following.

$$ g_{1\bar 1 \bar 1 1 \lambda} =
   g^{counterterm}_{1\bar 1 \bar 1 1 \epsilon}
   - c_1 e\bar u_1 \gamma^\mu v_3 \,  \, e \bar v_4 \gamma_\mu u_2 \,
         r_{11} r_{13} r_{14} r_{12} $$
$$ +\left[ -{1\over 2} c_1 (s_{13} + s_{24}) + 1 \right]
   {e\bar u_1 \gamma^+ v_3 \,  \, e \bar v_4 \gamma^+ u_2 \, \over P^{+ \, 2}}
         r_{11} r_{13} r_{14} r_{12} $$
$$ - \left[ c_2 r_{21} r_{2512} r_{24} r_{2534}
r_{3/5} r_{2/5} \theta(y-x)
 + c_3 r_{32} r_{3512} r_{33} r_{3534}
r_{1/5} r_{4/5}  \theta(x-y) \right]
     e \bar u_1 \gamma^\mu u_2 \,  e \bar v_4 \gamma_\mu v_3 \,  $$
$$ - \left\{
 \left[ {1 \over 2} c_2 (q_{12}^2 + q_{34}^2) + 1\right]
  r_{21} r_{2512} r_{24} r_{2534}
  r_{3/5}  r_{2/5} \theta(y-x) \right.$$
$$\left.
+  \left[ {1 \over 2} c_3 (q_{12}^2 + q_{34}^2) + 1\right]
r_{32} r_{3512} r_{33} r_{3534}
r_{1/5} r_{4/5}  \theta(x-y) \right\}
{e \bar u_1 \gamma^+ u_2 \,  e \bar v_4 \gamma^+ v_3 \,  \over (x-y)^2
 P^{+\, 2} } \, . \eqno(3.106)$$

\noindent The coefficients $c_1$, $c_2$ and $c_3$ are given by universal
Eqs.  (3.82.b) to (3.82.d) which were derived already in Yukawa theory,
with the replacement of the meson mass by the photon mass, $\mu^2 =
\mu^2_\delta$.

This result will be now analysed term by term for illustration in the
matrix elements between states of equal free energy.  The matrix
elements contribute to the electron-positron scattering in second order
perturbation theory.

The first term in Eq.  (3.71) for the $T$-matrix calculated to order
$e^2$, has matrix elements equal to the matrix elements of the effective
interaction from Eq.  (3.106).  When evaluating the $S$-matrix elements,
one considers configurations where the free energy of incoming fermions
equals the free energy of the outgoing fermions.  This configuration
selects the energy-diagonal part of the effective interaction:  $s_{13}
= s_{24} = s$ and $q^2_{12} = q^2_{34} = q^2 $. Also, the external
similarity factor which appears in the effective haniltonian as an
additional factor to $g_{1\bar 1 \bar 1 1 \lambda}$, equals 1. Thus,
$g_{1\bar 1 \bar 1 1 \lambda}$ in the energy diagonal part can be viewed
as the scattering amplitude.  It simplifies in the energy diagonal part
to the following form.

$$
   g_{1\bar 1 \bar 1 1 \lambda} =
   g^{counterterm}_{1\bar 1 \bar 1 1 \epsilon}
   - c_1 e\bar u_1 \gamma^\mu v_3 \,  \, e \bar v_4 \gamma_\mu u_2 \,
$$
$$
   +\left[ - c_1 s + 1 \right]
   {e\bar u_1 \gamma^+ v_3 \,  \, e \bar v_4 \gamma^+ u_2 \, \over
   P^{+ \, 2}}
$$
$$
- \left[ c_2 r_{2512} r_{2534} r_{3/5} r_{2/5} \theta(y-x)
       + c_3 r_{3512} r_{3534} r_{1/5} r_{4/5} \theta(x-y) \right]
     e \bar u_1 \gamma^\mu u_2 \,  e \bar v_4 \gamma_\mu v_3 \,
$$
$$
- \left\{
  \left[ c_2 q^2 + 1\right] r_{2512} r_{2534} r_{3/5} r_{2/5}
  \theta(y-x)
  \right.
$$
$$
  \left.
+ \left[ c_3 q^2 + 1\right] r_{3512} r_{3534} r_{1/5} r_{4/5}
  \theta(x-y)
  \right\}
{e \bar u_1 \gamma^+ u_2 \,  e \bar v_4 \gamma^+ v_3 \,  \over (x-y)^2
 P^{+\, 2} } \, . \eqno(3.107)$$

\noindent We have removed the regularization factors which equal 1 for
fermions with finite free energy.

In the limit $\mu_\delta \rightarrow 0$ and for the cutoff $\lambda$
close to $-2m^2$, a number of simplifications occur.  We display the
result for the case where the intermediate photon momentum fraction
$|x-y| >> (x,y,1-x,1-y) \delta$, i.e. when the photon momentum is not
negligible in comparison to the fermion momenta.  In this case, the
infrared regularization factors for the photon which are still kept in
Eq.  (3.107), can be set equal 1. The result is following.

$$
   g_{1\bar 1 \bar 1 1 \lambda} =
   g^{counterterm}_{1\bar 1 \bar 1 1 \epsilon}
   - {u_1 \gamma^\mu v_3 \,  \, e \bar v_4 \gamma_\mu u_2 \over s}
$$
$$ + \theta\left[ |q^2| - |x-y|(2m^2
+\lambda^2)/max(x,y,1-x,1-y)\right]
   {e \bar u_1 \gamma^\mu u_2 \,  e \bar v_4 \gamma_\mu v_3 \over q^2}
$$
$$ - \theta\left[ |x-y|(2m^2
+\lambda^2)/max(x,y,1-x,1-y) - |q^2| \right]
{e \bar u_1 \gamma^+ u_2 \,  e \bar v_4 \gamma^+ v_3 \,  \over (x-y)^2
 P^{+\, 2} } \, . \eqno(3.108)$$

The first term is the potentially necessary counterterm which we have
not yet determined.  Since the remaining part of the matrix element is
not sensitive to $\epsilon$, the counterterm matrix element is equal
zero.

The second term is the well known expression for the electron-positron
annihilation channel scattering amplitude in the Born approximation.  No
limits on the fermion momenta appear because the invariant mass squared
of two fermions is larger than $4m^2$ which is the minimal invariant
mass difference in the transition between the two fermions and a one
massless photon state.  Since we assume the width $\lambda^2$ to be
small, the effective hamiltonian contains the full amplitude for
transition through the intermediate photon state.

The third term is equal to the standard second order expression for the
electron-positron scattering amplitude via one-photon exchange, except
for the $\theta$-function factor which forces the momentum transfer to
be sufficiently large.

The meaning of this restriction is visible in the $T$-matrix.  The third
term contributes to the $e^+e^-$-scattering amplitude through the first
term in Eq.  (3.71).  The same contribution would originate from the
second term in Eq.  (3.71) if we were using the initial hamiltonian to
calculate the $T$-matrix.  In contrast, the effective hamiltonian with
the small width $\lambda$ limits the effective photon emissions and
absorptions to small momentum transfers and, therefore, it is not able
to provide this contribution through the second term in Eq.  (3.71).
This contribution is then contained in the effective hamiltonian and
comes through the first term in Eq.  (3.71).

The fourth term is unusual in the sense that it should not appear in the
electron-positron scattering at all.  The fourth term distinguishes the
$z$-axis in its structure and diverges when $x \rightarrow y$.  The
$\theta$-function factor in the fourth term is equal 1 where the
$\theta$-function factor of the third term is equal 0. And vice versa,
the fourth term $\theta$-function equals 0 where the third term
$\theta$-function equals 1.

The need for the fourth term becomes clear when one recalls that the
second term in Eq.  (3.71) also contributes to the electron-positron
scattering amplitude.  The relevant contribution comes through the
effective one-photon exchange which results from the double action of
$H_{I\lambda}$.  $H_{I\lambda}$ is given by the operation $F_\lambda$
applied to the third term of the QED hamiltonian from Eq.  (3.84), (see
Eq.  (2.11)).  The operation $F_\lambda$ multiplies the photon emission
and absorption vertices by the factor $f_\lambda$.  This factor was set
equal to a $\theta$-function in the current example.  Each interaction
provides one factor of the $\theta$-function.  The resulting factor in
the second term of Eq.  (3.71) is the same in QED as in Yukawa theory in
Eq.  (3.72), except for the antisymmetrization effect which leads to the
$\theta$-function which stands in front of the fourth term in Eq.
(3.108).

Now, the second term in Eq.  (3.71), contains the spin factor which is
the same as in the last term in Eq.  (3.104).  The $g^{\mu\nu}$ part
complements the third term in Eq.  (3.108) and produces the full well
known one-photon exchange scattering amplitude which is free from the
$\theta$-function factor.  The remaining part provides the term which
cancels the odd fourth term in Eq.  (3.108).  Thus, the effective
hamiltonian calculated to second order in powers of the charge $e$
contains an odd term and $\theta$-functions which are required to
compensate for the odd contributions to the scattering amplitude from
the effective hamiltonian order $e$ acting twice.  The above analysis of
the energy diagonal part of the second order effective hamiltonian
explains the r\^ole of different terms in Eqs.  (3.106) to (3.108).

The analysis also suggests that apparently infrared diverging terms in
the effective hamiltonian may mutually compensate their diverging
contributions in the scattering amplitude on energy-shell.  In the
current example, we see the interplay between the second-order seagull
term and the double action of the first order emission and absorption of
photons.  The first order hamiltonian matrix elements diverge when the
photon longitudinal momentum approaches zero.  The second order seagull
term compensates this divergence in the on-energy-shell $T$-matrix
elements.

The remaining point to make here is that the result of Eq.  (3.106) with
the counterterm equal 0 leads to the effective light-front hamiltonian
version of the Coulomb force in the limit of small $\lambda^2 + 2m^2 \ll
\alpha m^2$.  The key elements in deriving this conclusion are the outer
similarity factors and the smallness of $\alpha$. The outline of the
derivation is following (c.f.  Ref.  \cite{JPG}).

If only the small energy transfers are allowed by the outer similarity
factor, i.e. transfers much smaller than the electron mass, then, the
wave functions of the lowest mass eigenstates of the effective
hamiltonian are strongly peaked at small relative electron momenta and
they fall off very rapidly as functions of the relative momentum.  This
is not true without the outer similarity factor because the function
$g_{1\bar 1 \bar 1 1 \lambda}$ alone is too large at the large energy
transfers and it would produce singular contributions in the large
relative momentum region, sensitive to the regularization cutoffs.

Below the width scale, the wave functions fall off as dictated by the
eigenvalue equation with small $\alpha$.  Above the width scale, the
fall off is very fast due to the similarity factor which justifies
restriction to momenta much smaller than $m$.  Then, the nonrelativistic
approximation for all factors in Eq.  (3.106) becomes accurate.

In the nonrelativistic approximation, $q_{12}^2 = q_{34}^2 = q^2$ and
Eq.  (3.107) applies.  Further, in Eq.  (3.108), the $\theta$-functions
become effectively equal 1 and 0, respectively.  The last term is not 
leading to important contribution despite its divergent longitudinal structure
because it is canceled by the effective massless photon exchange
as described earlier in this Section.

The dominant contributions are provided by the second and the third
terms from Eq.  (3.108) which are well known to have the right
nonrelativistic structure for predicting positronium properties in the
Schr\"odinger equation.  Now, the outer similarity factor becomes
irrelevant to the spectrum in the leading approximation because the
coupling constant is very small.  In the dominant region, the electron
velocity is order $\alpha$, the nonrelativistic approximation to the
full dynamics produces wave functions with relative momenta order
$\alpha m$ (in the positronium ground state the wave function is
$(k^2 + \alpha^2 m^2 / 4)^{-2}$) and the outer similarity factor in
the effective interaction can be replaced by 1.

We can use the infinitesimal $u_0$ in Eq.  (2.20) and replace $\lambda^2
+ 2m^2$ by $\tilde \lambda^2$ (see the discussion below Eq.  (2.20)).
When $\tilde \lambda$ is order $\alpha m$ and $x-y$ is order $\alpha$
then, the momentum transfer $\vec q ^{\, 2}$ is typically order
$\alpha^2 m^2$ which is much larger than $2 (x-y)\tilde \lambda^2$ in
Eq.  (3.108).  Thus, the $\theta$-function is equal 1 and the second
term in Eq.  (3.108) becomes equal to the standard Coulomb interaction
with the well known Breit-Fermi structure of the spin factors.  This step
completes the derivation of the Coulomb potential.  The derivation
explains the effective nature of the Schr\"odinger equation with the
Coulomb potential in the light-front hamiltonian formulation of QED.

\vskip.3in
{\bf 3.d QCD}
\vskip.1in

The main conceptual complication in hamiltonian calculations in QCD is
confinement, which is not understood. To order $g^2$, the manifestation
of the confinement problem is the lack of a well defined initial
condition for the renormalization group flow of the effective
hamiltonians.  It will require an extended research effort to find the
class of acceptable initial conditions.

For example, the renormalization conditions for the quark and gluon mass
terms are missing in the second order calculations in this paper.  We
do not know the right values of the masses in the initial hamiltonian, 
and we may currently only speculate about their values in the small width
hamiltonians (by comparison with the constituent quark model).  So, we
limit our presentation here to the differential equations themselves,
without solutions.

We stress the urgent need for the higher order calculations by
describing some details of the second order calculation of the $q \bar
q$ effective interaction.  The calculation is similar to the one in QED
above, except for the option for a differnt treatment of the last term in Eq.
(3.108).  Perry suggested that the long distance part of this term may
remain uncanceled because of the gluon non-abelian gauge interactions.
If the last term in Eq.  (3.108) would not be canceled, it could be
claimed to generate confinement in the light-front hamiltonian approach
to QCD. \cite{Perry} We describe the structure of this term in the
present approach since it is different than in Ref.  \cite{Perry}.
The differences result from the diferent definitions of the similarity
transformation.

\vskip.3in
\centerline {\bf Quark and gluon mass terms}
\vskip.1in

Results one obtains in QCD from Eq.  (2.38), in second order in ${\cal
G}_{2\lambda}$, can be illustrated by two equations for the hamiltonian
width dependence of the effective masses of quarks and gluons.  It is
arbitrarily assumed that the equations are relevant to physics although
their structure follows from the second order terms only.  The range of
widths and coupling constants for which these equations can pertain to the
physics of quarks and gluons, are not known yet.

The regularization factors are not written in detail. One can write them
easily using results of Section 2 and previous examples in Section 3.
In the case of effective masses a number of simplifications occur,
as explained in the previous Sections. Let us consider an
infinitesimal $u_0$ in Eq. (2.20) and simplify notation by replacing
$\tilde \lambda$ by $\lambda$ itself. Then, we can write

$$ {d \over d\lambda} {\cal G}_{1\lambda} =
\left[  {\cal G}_{12\lambda}
{d f^2(z_\lambda / \lambda^2) / d\lambda \over  {\cal G}_{1\lambda}
- E_{1\lambda} } {\cal G}_{21\lambda} \right]_{11} \, .\eqno(3.109)$$

\noindent $E_{1\lambda}$ is the eigenvalue of ${\cal G}_{1\lambda}$ 
corresponding to the subscript 11. A set of arguments
$z_\lambda$ is needed. Namely,

$$  z_1 = {\kappa^2 + \mu^2_\lambda \over x(1-x) } - \mu^2_\lambda \, ,
\eqno(3.110) $$

$$ z_2 = {\kappa^2 + m^2_\lambda \over x(1-x) } - \mu^2_\lambda \, ,
\eqno(3.111) $$

$$ z_3 = {\kappa^2 + m^2_\lambda \over x} +
{\kappa^2 + \mu^2_\lambda \over 1-x } - m^2_\lambda \, . \eqno(3.112) $$

\noindent $m_\lambda$ and $\mu_\lambda$ are the effective quark and
gluon masses, respectively. Then,

$$ {d m^2_\lambda \over d\lambda} = \int [x\kappa] g^2_{q\lambda}
 z_3^{-1} {d f^2 (z_3/\lambda^2)\over d\lambda }
\left[\kappa^2[2/x +4/(1-x)^2] + 2 m^2_\lambda (1-x)^2/x \right]
r_{qg\epsilon}(x,\kappa) \eqno(3.113) $$

\noindent and

$$ {d \mu^2_\lambda \over d\lambda} = 3 \int [x\kappa]g^2_{g\lambda}
z_1^{-1} {d f^2 (z_1/\lambda^2)\over d\lambda }
\kappa^2[4/x^2 +2]  r_{gg\epsilon}(x,\kappa)  $$
$$ + \int [x\kappa]g^2_{q\lambda}
z_2^{-1} {d f^2 (z_2/\lambda^2)\over d\lambda }
\left[ {\kappa^2 + m^2_\lambda \over x(1-x)} - 2 \kappa^2 \right] \,
r_{qq\epsilon}(x,\kappa) \, .\eqno(3.114) $$

\noindent The gluon couples to the quark-antiquark pairs and pairs of
gluons while the quark couples only to the quark-gluon pairs.  The
running of the quark and gluon masses is more involved than
in the previous examples.  In Eq.  (3.114), the number of colors is
equal to $3$ and the number of flavors to $1$.

These equations are not further studied here for two major reasons.  The
first one is that we do not know the initial conditions to use for such
study.  The second is that the equations involve two running couplings
which are not known yet.  The third and fourth order calculations are
required to find them, at least approximately, before one will be able
to make contact with data for hadrons using the effective hamiltonians
of small widths.

The importance of the effective mass issue for quarks and gluons is
illustrated below by the calculation of the small energy transfer
effective forces between quarks and antiquarks.

\vskip.3in
\centerline {\bf Quark-antiquark effective interaction}
\vskip.1in

The second order results are similar to QED.  For heavy quarkonia, one
can directly look at Eq.  (3.108).  The only change required is the
color $SU(3)$ matrices sandwiched between color vectors of quarks and
summed over colors of the exchanged gluons.  The counterterm is set to
0. The second term gives the annihilation channel potential but the
color matrix is traceless and this excludes the single gluon annihilation 
channel from the dynamics of color singlet $Q \bar Q$ states.  The third 
term leads to the color Coulomb potential with the well known Breit-Fermi 
spin factors.

About the last term in Eq.  (3.108), it was suggested \cite{Perry} 
that a term of this kind may become a seed for confining interactions 
if it is not canceled.  
The cancelation occurs when one evaluates the model interaction in
the $Q\bar Q$ sector in the explicit expansion in powers of $g$ to
second order using the operation $R$ and Eq.  (1.3) in the
nonrelativistic approximation for the quark relative momentum, exactly
the same way as for electrons in QED with massless photons.  

However, there are reasons for gluons to effectively acquire a mass,
such as the unknown finite part of the mass squared counterterm or the
interaction energy due to the non-abelian terms which are absent 
in QED.  Then, the cancelation of the seagull term can be questioned.  

One can assume that the gluon energies in the $Q\bar Q g$ sector
may be lifted up so that the gluons cannot
contribute to the model $Q\bar Q$ interaction in the way the abelian
massless photons can in the model electron-positron interaction in QED.  
As a result of this assumption, one obtains the last term in Eq.  (3.108)
acting in the effective $Q\bar Q$ sector. 

In the nonrelativistic approximation, the Coulomb term and the term in
question are (see Eq. (3.108))

$$ -\, \theta\left[ {\vec q}^{\,2} - |x-y|\,2\tilde \lambda^2 \right] \,\,
{g^2 4 m^2 \over \vec q^{\,2}}\quad - \quad \theta\left[ |x-y|\,2\tilde
\lambda^2 - {\vec q}^{\,2} \right] \,\, {g^2 \over (x-y)^2 } \,\, .
\eqno(3.115)$$

\noindent The regularization factors are the same in both terms and they
are not displayed.  The only effect of their presence which matters here
is that $|x-y|$ is limited from below by about $\delta/2$.  The initial
infrared regularization parameter $\delta$ comes in through the initial
condition in the renormalization group flow of the seagull term.  The
flow is limited in the second order calculation to the dependence of the
outer similarity factor on $\tilde \lambda$.  The factor $1/2$ results
from $|x-y|/x_{quark}$ being limited from below by $\delta$ and the
quarks having $x_{quark} \sim 1/2$.  In fact, the lower bound on
$|x-y|$ is given by $ 2 \delta \, max(x,1-x,y,1-y)$. This is different
from Ref. \cite{Perry} where instead of the ratios of the $+$-momentum
fractions a separate scale for $+$-momentum is introduced. 

The third component of the exchanged gluon momentum is $q_3 = (x-y) 2m$.
Thus, we see that the uncanceled singular term is represented by the
potential which is analogous to the Coulomb potential except for that
the factor $ - 1 /{\vec q}^{\,2}$ is replaced by $ - 1 / q_3^2$ and both
terms have mutually excluding and complementary supports in the momentum
transfer space.

The seagull term $\theta$-function can be rewritten as $ \theta[
\omega^2 - (|q_3| - \omega)^2 - q^{\perp\,2}]$, where $\omega =
\tilde \lambda^2 /2 m$.  The support of this function is two spheres of
radius $\omega$ centered at $q^\perp = 0$ and $q_3 = \pm \omega$.  The
spheres touch each other at the point $q^\perp = q_3 = 0$.  In this
point, $q_3^2$ in the denominator produces a singularity.

Let us initially consider both terms in Eq.  (3.115) as the actual
interaction in the model $Q\bar Q$ sector, i.e. as if they were not
affected by the operation $R$ in Eq.  (1.3).  
The Coulomb term works outside the two spheres in the $\vec q$-space and
the singular seagull term works inside.

In the region of the singularity, both $q^\perp $ and $q_3$ are small in
comparison to $\omega$.  In the rough analysis, one can 
neglect the outer similarity factor $\theta(\tilde \lambda^2 - |k^2
-k'^2|)$ since it is equal 1 when $\vec q = \vec k - \vec k'$
approaches 0. $\vec k$ is the relative momentum of the created $Q\bar Q$
pair and $\vec k'$ is the relative momentum of the annihilated $Q\bar Q$
pair. $\omega = (\tilde \lambda / m) \tilde \lambda /2 \ll
\tilde \lambda/2$ and the spheres have the radius about $\tilde
\lambda / m$ times smaller than the outer similarity factor width in the
length of the quark relative momenta, i.e. the relative size of the
spheres in comparison to the outer similarity factor support approaches 
0 when $\tilde \lambda/m \rightarrow 0$.

Since $q^{\perp \, 2}$ is order $q_3$ in the singular region, the
divergence when $q_3 \rightarrow 0$ is logarithmic.  The lower limit of
integration over $|q_3|$ for a given $x$ is given by $ 2 m\, \delta \,
max(x,1-x)$. However, we assume $x = 1/2 + o(g^2)$ and we neglect terms
of higher order than $g^2$ in the model $Q \bar Q$ hamiltonian.

The potential resulting from the uncanceled seagull term is
given by the following expression (c.f. \cite{Perry}),

$$ V(\vec r) \quad \sim \quad - \,\, \int {d^3 q \over (2\pi)^3}\,\,
\exp{(i\,\, \vec q \, \vec r)} \,\,\, { \theta(2\omega |q_3| - {\vec q}^{\,2})
\,\,\theta (|q_3| - 2 m \delta) \over q_3^2 } \, .\eqno(3.116) $$

\noindent The sign $\sim$ means that the diverging dependence on
$\delta$ is subtracted and the same coefficient stands in front 
of the integral as in
the Coulomb potential term.  The argument for the infrared subtraction
goes as follows.

If the gluons cannot cancel the last term in Eq.  (3.108), they
presumably cannot contribute to the model quark self-energies either,
for the same reason.  Because the size of the quark mass in the
effective hamiltonian is unknown, one may propose that its value is
chosen in the second order calculation in the same way as for 
nucleons in the Yukawa theory in Eq.  (3.37) or electrons in
QED in Eq.  (3.99):  so that a would-be quark eigenstate has a finite
constituent quark mass when the gluons are allowed to contribute at all
momenta in the eigenvalue equation. This setting is
equivalent to the solution Perry proposed for his coupling coherence
condition for the quark self-energies.  \cite{Perry} The argument also
illustrates the urgency of questions concerning the initial conditions
and higher order analysis.

There is nothing wrong with the mass adjustment despite the infrared
divergence.  We have noticed in the previous Section that the arbitrary
finite parts of the ultraviolet counterterms can be infrared divergent.
This time, however, the positive and infrared logarithmically divergent
part of the effective quark mass term in the model eigenvalue equation
for heavy quarkonia remains uncanceled when the transverse gluons with
$2\omega|q_3| - \vec q^{\,2} > 0$ are declared to be
absent from the model dynamics.  The uncanceled part of the effective
quark mass term stands in the eigenvalue problem.  The point is it can
now cancel the diverging $\delta$ dependence in the seagull term
which is not canceled due to the missing of the gluon exchange below
$\tilde \lambda$.  The cancelation occurs in the colorless states. It
is analysed here in the nonrelativistic limit only.

We describe the cancelation mechanism in the case of equal masses of
quarks.  The mechanism is similar but not identical to that in Ref.
\cite{Perry}.  The infrared divergent mass squared term comes into the
quarkonium eigenvalue equation divided by $x(1-x)$.  But in the second
order analysis the mass divergence appears only as a logarithmically
divergent constant.  The $x$-dependence is of higher order.  The same
diverging constant with the opposite sign is generated by the seagull
term.

The infrared divergent terms and their cancelation are not directly
related to the ultraviolet renormalization procedure.  They appear in
the ultraviolet-finite effective small width hamiltonian dynamics.  Note
also that the introduction of the gluon mass $\mu_\delta$ in the
regularization could matter for the lower bound on $|x-y|$ and it could
even eliminate the whole contribution when the upper bound of $\omega$
meets the lower bound of $\mu_\delta$.  We assume here $\mu_\delta = 0$.

The divergent part in Eq.  (3.116) is independent of $\vec r$ and it is
easily removed by subtracting 1 from $\exp{i \vec q \vec r}$.
Evaluation of the integral leads to the answer that for large $r$
the seagull term produces a logarithmic potential of the form

$$ V(\vec r) \quad  =  \quad 
{2 \, \omega \, a(\hat e_r)  \over \pi}\,\, \log{r} \,\, , \eqno
(3.117) $$

\noindent where $a$ is equal 1 for the radial versor $\hat e_r$ along
the $z$-axis and it equals 2 when $\vec r$ is purely transverse.

The resulting potential is confining.  It is also boost invariant.
 But the asymmetry of the potential raises doubts.  It
suggests that an important piece of physics is missing in the reasoning
used to derive it.  The obvious sources of questions are the r\^ole of
the operation $R$, the r\^ole of the nonrelativistic approximation, the
size of the quark and gluon masses and the strong dependence of the term 
on the width $\tilde \lambda$.  The most urgent question is if the 
higher order calculations do lead to the uncanceled seagull term.  The width
dependence must disappear in the eigenvalues.  The coupling constant
dependence on $\tilde \lambda$ is necessary.

\vskip.5in
{\bf 4. CONCLUSION}
\vskip.1in

We have defined and illustrated on a few perturbative examples a general
method of calculating light-front hamiltonians which can be used for the
relativistic description of interacting particles.  The starting point
in the calculation is a field theoretic expression for the bare
hamiltonian density.  This expression is multiply divergent in the
physically interesting cases.  Therefore, the hamiltonian theory
requires renormalization.

In the renormalization process, one calculates the whole family of
effective hamiltonians as functions of the width parameter $\lambda$
which determines the range of the effective interactions on the energy
scale.

An effective hamiltonian of a small width $\lambda$ is much different
from the initial bare hamiltonian.  It couples only those states whose
masses differ by less than a prescribed amount.  Thus, the effective
theory contains only near-neighbour interactions on the energy scale.
No scale is removed in the calculation but the correlations between
dynamics at significantly different energy scales are integrated out.
Therefore, in principle, the effective eigenvalue problem can be solved
scale by scale using standard techniques for finite matrices which
describe dynamics at a single scale.  The remaining problems due to the
small-$x$ massless particles need further investigation.

Our formalism is based on the earlier work on renormalization of 
hamiltonians from Refs.  \cite{GW1}
and \cite{GW2} where the hamiltonians are defined by their matrix elements
in a given set of basis states. Wegner has developed similar 
equations for hamiltonian matrix elements in solid state physics. \cite{WEG} 
The present approach to renormalization of hamiltonians 
introduces the following features.

Our equations are expressed in terms of creation and annihilation
operators.  Consequently, calculations of counterterms in perturbation
theory can be performed without knowing details of the specific Fock
states which are needed to evaluate the matrix elements.  This is useful
because a large number of Fock states needs to be considered.  The
renormalization scheme is free from practical restrictions on the Fock
space sectors.

Expressing the effective hamiltonians in terms of the creation and
annihilation operators and showing that the effective interactions are
connected is a prerequisit to obtain the cluster decomposition property.
\cite{WE1} The effective interactions in our approach do not contain
disconnected terms.  The number of creation and annihilation operators
in a single term is limited in perturbation theory by $2+n(V-2)$ where
$n$ is the order of a perturbation theory and $V$ is the number of
operators in the perturbing term.

The physically motivated assumptions about the model space of effective
states to be included in solving a particular problem are introduced
after the effective hamiltonian is calculated.  The interaction terms in
the effective hamiltonian contain the similarity factors which diminish 
the dynamical significance of the Fock sectors with numbers of effective 
particles considerably different from the number of effective particles 
in the dominant sectors.

The present operator formulation does not introduce spectator dependent
interactions, even in the case where we include the sums of the
invariant masses for incoming and outgoing particles in the similarity
factors.  The sums are useful for estimates of cutoff dependence in
perturbation theory.

The formalism explicitly preserves kinematical symmetries of the
light-front frame.  The structure of counterterms is constrained by
these symmetries, including boost invariance.  Hence, the number of
possible terms is greately limited.  Preserving boost invariance is
particulary important because it is expected to help in understanding
the parton model and constituent quark model in QCD, simultaneously.

It is essential to include the running of the coupling constants in the
calculation of the small width dynamics.  The examples of second order
calculations we described in this article do not include the running
coupling constant effects.  Inclusion of these effects requires higher
order calculations. Examples of related issues are following.
The asymptotic freedom of QCD suggests that the
effective coupling constant grows at small width.  This rise can also
enhance the initially small perturbative terms in the effective
hamiltonians, such as masses of gluons or light quarks.  In the
meson-barion dynamics, the large couplings and narrow form factors are
commonly used but the correlation between the size of the couplings
and the form factor widths is out of the theoretical control. To 
gain some control on this correlation one needs to include the coupling 
constant dependence on the width. In QED, we expect the coupling 
constant to grow with the width.  There, the hamiltonian
calculation of the running coupling effects may help in understanding
the ultraviolet structure of the theory and asymptotic nature of the 
perturbative expansion.

Wegner's equation can be adapted to building an operator approach
similar to what we described in the present article.  The initial
equation which replaces our Eq.  (2.29) when one uses the Wegner
generator of the similarity transformation, is

$$ {d {\cal H}_\lambda \over d\lambda^2} \quad = \quad {-1\over
\lambda^4}\,\, \left[ \, [{\cal H}_{1 \lambda}, {\cal H}_{2 \lambda}],
\, {\cal H}_\lambda \right]\,\,.\eqno(4.1)$$

\noindent However, there is little flexibility left in the equation so
that the widenning of the hamiltonian band is not readily available.

Wilson and the present author have written a class of generalized
equations for the flow of hamiltonian matrix elements and studied them in a
simple numerical model. \cite{GW4} These equations allow widening of the
effective hamiltonian matrix at large energies.  The generalized
equations can also be adapted for the construction of the creation and
annihilation operator calculus.  Namely,

$$ {d {\cal H}_\lambda \over d\lambda} \quad  = \quad
\left[ F\{{\cal H}_{2 \lambda}\}, \, {\cal H}_{\lambda} \right]
\,\, . \eqno(4.2)$$

\noindent These equations require detailed definitions of the similarity
factors generated by the operation $F$ whose description goes beyond
the scope of this article.

In summary, the present formalism for renormalization of hamiltonians in
the light-front Fock space provides a tool for working on a host of
theoretical issues in particle dynamics.  However, it remains to be
verified if the formalism can lead to quantitative improvements in our
description of particles.  The key problems are the lack of
understanding of rotational symmetry beyond the nonrelativistic limit
and the appearance of infrared singularities in gauge theories.  Most
urgent are the calculations of effective hamiltonians in the third and
fourth order perturbation theory.

\vskip.5in
{\bf Acknowledgement}
\vskip.1in

This article includes the content of lectures delivered by the author  
as a Fulbright Scholar at The Ohio State University in winter 1996. 
The author is grateful to Kenneth Wilson for making this visit 
possible and for many discussions. The author would like to thank 
Robert Perry for discussions and hospitality. He would 
also like to thank Billy Jones, Martina Brisudov\'a, Brent Allen, 
Tomek Mas{\l}owski and Marek Wi{\c e}ckowski for numerous discussions.  
Research described in this report has been supported in part by Maria 
Sk{\l}odowska-Curie Foundation under Grant No.  MEN/NSF-94-190.

\end{document}